\documentclass[11pt]{article}
\usepackage{amsmath}
\usepackage{amssymb}
\usepackage{amsfonts}
\usepackage{graphicx}
\usepackage{float}
\usepackage{fancyhdr}

\allowdisplaybreaks
\input{tcilatex}
\begin{document}

\begin{center}
{\LARGE Non-overshooting sliding mode for UAV control}

Xinhua Wang $^{1}$, and Xuerui Mao $^{2}$

$^{1}${\small \ University of Nottingham, Aerospace Engineering, Nottingham,
United Kingdom}

{\small Email: wangxinhua04@gmail.com}

$^{2}${\small \ Beijing Institute of Technology, Advanced Research Institute
of Multidisciplinary Sciences, Beijing, China}
\end{center}

\emph{Abstract:} For a class of uncertain systems, a non-overshooting
sliding mode control is presented to make them globally exponentially stable
and without overshoot. Even when the unknown stochastic disturbance exists,
and the time-variant reference trajectory is required, the strict
non-overshooting stabilization is still achieved. The control law design is
based on a desired second-order sliding mode (2-sliding mode), which
successively includes two bounded-gain subsystems. Non-overshooting
stability requires that the system gains depend on the initial values of
system variables. In order to obtain the global non-overshooting stability,
the first subsystem with non-overshooting reachability compresses the
initial values of the second subsystem to a given bounded range. By
partitioning these initial values, the bounded system gains are determined
to satisfy the robust non-overshooting stability. In order to reject the
chattering in the controller output, a tanh-function-based sliding mode is
developed for the design of smoothed non-overshooting controller. The
proposed method is applied to a UAV trajectory tracking when the
disturbances and uncertainties exist. The control laws are designed to
implement the non-overshooting stabilization in position and attitude.
Finally, the effectiveness of the proposed method is demonstrated by the
flying tests.


\setcounter{page}{1}

\textbf{Nomenclature}

\begin{tabular}{ll}
${\small e(t)}$ & {\small system variable of error form} \\ 
${\small e}^{(n)}{\small (t)}$ & {\small the }${\small n}${\small -th
derivative of }${\small e(t)}$ \\ 
${\small f(\cdot )}$ & {\small system function} \\ 
${\small g(\cdot )}$ & {\small system function} \\ 
${\small k}_{p}$ & {\small proportional gain of PID controller} \\ 
${\small k}_{i}$ & {\small integral gain of PID controller} \\ 
${\small k}_{d}$ & {\small derivative gain of PID controller} \\ 
${\small e}_{1}{\small (t)}$ & {\small sliding variable or error variable}
\\ 
${\small e}_{2}{\small (t)}$ & {\small sliding variable or error variable}
\\ 
${\small k}_{1}$ & {\small parameter of sliding mode or controller parameter}
\\ 
${\small k}_{2}$ & {\small parameter of sliding mode or controller parameter}
\\ 
${\small \rho }$ & {\small parameter of sliding mode or controller parameter}
\\ 
${\small d(t)}$ & {\small disturbance or system uncertainty} \\ 
${\small L}_{d}$ & {\small upper bound of }${\small d(t)}$ \\ 
${\small \sigma }{\small (t)}$ & {\small sliding function} \\ 
${\small x}_{1}{\small (t)}$ & {\small system state} \\ 
${\small x}_{2}{\small (t)}$ & {\small system state}%
\end{tabular}

\begin{tabular}{ll}
${\small u(t)}$ & {\small controller} \\ 
${\small \delta }{\small (t)}$ & {\small system uncertainty or disturbance}%
\end{tabular}

\begin{tabular}{ll}
${\small x}_{d}{\small (t)}$ & {\small reference} \\ 
${\small \dot{x}}_{d}{\small (t)}$ & {\small derivative of reference} \\ 
${\small m}$ & {\small mass of UAV} \\ 
${\small g}$ & {\small gravity of acceleration} \\ 
${\small l}$ & {\small rotor distance to gravity center} \\ 
${\small J}_{\phi }$ & {\small moment of inertia about roll} \\ 
${\small x}$ & {\small position in }${\small x}${\small \ direction} \\ 
${\small y}$ & {\small position in }${\small y}${\small \ direction} \\ 
${\small z}$ & {\small position in }${\small z}${\small \ direction} \\ 
${\small \phi }$ & {\small roll angle} \\ 
${\small \theta }$ & {\small pitch angle} \\ 
${\small \psi }$ & {\small yaw angle} \\ 
${\small J}_{\theta }$ & {\small moment of inertia about pitch} \\ 
${\small J}_{\psi }$ & {\small moment of inertia about yaw} \\ 
${\small b}$ & {\small rotor force coefficient} \\ 
${\small k}$ & {\small rotor torque coefficient} \\ 
${\small F}_{i}$ & {\small thrust force by rotor }${\small i}$ \\ 
${\small Q}_{i}$ & {\small reactive torque of rotor }${\small i}$ \\ 
${\small k}_{x}$ & {\small drag coefficients of UAV in }${\small x}${\small %
\ direction} \\ 
${\small k}_{y}$ & {\small drag coefficients of UAV in }${\small y}${\small %
\ direction} \\ 
${\small k}_{z}$ & {\small drag coefficients of UAV in }${\small z}${\small %
\ direction} \\ 
${\small k}_{\phi }$ & {\small drag coefficients of UAV about roll} \\ 
${\small k}_{\theta }$ & {\small drag coefficients of UAV about pitch} \\ 
${\small k}_{\psi }$ & {\small drag coefficients of UAV about yaw} \\ 
${\small \Delta }_{x}$ & {\small uncertainty in }${\small x}${\small \
direction} \\ 
${\small \Delta }_{y}$ & {\small uncertainty in }${\small y}${\small \
direction} \\ 
${\small \Delta }_{z}$ & {\small uncertainty in }${\small z}${\small \
direction} \\ 
${\small \Delta }_{\phi }$ & {\small uncertainty about roll} \\ 
${\small \Delta }_{\theta }$ & {\small uncertainty about pitch} \\ 
${\small \Delta }_{\psi }$ & {\small uncertainty about yaw}%
\end{tabular}

\section{Introduction}

\markboth{}
{Murray and Balemi: Using the Document Class IEEEtran.cls}This paper
considers robust non-overshooting stabilization for a class of dynamical
systems with stochastic disturbance and application to UAV flight control.
Control without overshoot is very important for many industrial control
systems [1, 2, 3], for example, aircraft safe landing, automated vehicle
safety control, and manufacturing process control, etc. In a control system,
overshoot makes the actual behavior exceed its target, and it may bring the
devastating results. Therefore, in order to guarantee the safety control, a
known reference needs to be tracked without overshoot, i.e.,
non-overshooting stabilization is required. For control systems, i{}n
addition to reduce the overshoot and oscillations in the system outputs, the
effect from the disturbance or uncertainty also needs to be avoided.
Furthermore, a smoothed control law is helpful to improve the system
response and reduce the actuator chattering.

PID control is very popular for many industrial control systems because of
its simplicity and its acceptable control performance [4, 5]. However, PID
control has some disadvantages, for example: sensitive to the disturbance
and uncertainty types, adverse effect by the time-variant references,
overshoot existence because of integral windup or integration saturation.
Theoretically, PID control can completely reject the effect from an unknown
constant disturbance because of the integration term. If the disturbance is
slowly time-varying, through increasing the control gains, the disturbance
effect can be reduced to some extent. For a fast time-varying disturbance or
nonlinear system uncertainty, PID control performance is affected obviously.
In addition, PID control is usually used to control a system with step
reference for good performance. If the time-variant reference is required,
the feedforward term including reference derivatives information should be
added in the PID controller. For PID or PI control, the large initial error
may bring the phenomenon of integral windup, and a relatively long-time
overshoot exits in the system output. There are some methods to reduce
overshoot in system output: the pole-placement or pole-zero configuration
methods [6, 7], the optimization approach to minimize overshoot [8], the
compensation-based method [9, 10], the characteristic ratio assignment
method [11], the iterative technique based on gradient descent-like
procedure [12], the eigenvector placement technique to construct an
invariant set [13]. They are mainly used for the step reference
non-overshooting tracking in the linear systems. However, they still cannot
overcome the effect from the time-varying disturbances or nonlinear system
uncertainties, and these time-varying disturbances may bring overshoot or
oscillations in the system outputs.

Sliding mode is also widely used in many industrial applications due to its
strong robustness against the bounded stochastic disturbances or
uncertainties, especially for aircraft navigation and control: the robust
controller [14], the robust observer [15], and the signal corrector [16].
However, chattering in sliding mode affects its output performance, and
overshoot is inevitable for the usual sliding mode control systems. Some
methods were proposed to reduce overshoot and chattering, and keep the
robustness property of sliding mode control [17, 18, 19, 20]. In [17], for
the systems with the parameter uncertainties and the matched disturbances,
an adaptive sliding mode was designed to create the non-overshooting
responses over the selected output variables. In [18], a cascade sliding
mode-PID control was presented to get the non-overshooting time responses
for a step reference. In [19, 20], for a linear time-invariant systems with
the matched disturbance, the integral sliding mode technique along with the
Moore's eigenstructure assignment was used to make the system stable with
non-overshooting behavior. For the above methods, not only the upper bound
of disturbance should be known, but also the upper bound information of
disturbance derivatives is required. For some cases, the non-overshooting
controllers with the observers were presented for the systems with the
disturbances [21, 22]. The observers were designed to estimate the
disturbances, and the estimations were used for the non-overshooting
stabilization. However, only the step reference was considered, and the
disturbance was assumed to be constant. Furthermore, the overshoot may exist
in the estimation from the observer, and the non-overshooting control
performance is affected adversely.

Recent years, non-overshooting stabilization for the nonlinear systems with
disturbance was developed [23, 24, 25, 26]. In [23], a non-overshooting
control was presented for a class of nonlinear system under the condition
that the initial value of the system output was strictly required below the
initial value of the reference trajectory. The approximately\
non-overshooting performance was achieved by appropriately choosing the
control gains under the deterministic disturbances. In [24], a controller
was designed for a class of nonlinear systems to make the mean of the system
output asymptotically track a given trajectory without overshoot, i.e., the
mean-nonovershooting tracking was achieved. Furthermore, the initial value
of the system output has the same constraint as that in [23]. In [25, 26],
the non-overshooting stabilizers were designed for a class of nonlinear
systems that were input-output linearizable with a full relative degree, and
the matched disturbances were considered. The bounded measurable
disturbances should satisfy the two inequalities including the initial
values of variables, and it was required to be continuous in the system
variables [26]. In addition, the above non-overshooting control methods are
locally stable.

In this paper, for a class of uncertain systems, a method of globally
exponentially stable control without overshoot is proposed. The design of
control law is based on a global non-overshooting 2-sliding mode of error
form. Non-overshooting stability requires that the system gains depend on
the initial values of system variables, making the system locally stable. In
order to achieve the global non-overshooting stability and avoid the
excessively large system gains, the 2-sliding mode consists of two
bounded-gain subsystems. The implementation of global non-overshooting
stability is through the successive connection of the first subsystem with
non-overshooting reachability and the second subsystem with locally
non-overshooting stability. The first subsystem enables the sliding
variables to reach a given bounded range without overshoot within a finite
time. The boundary of this range serves as the initial values for the second
subsystem. Through partitioning these initial values, the sliding variables
are analytically expressed, and in each zone, the bounded system parameters
are determined to achieve the non-overshooting stability. For the second
subsystem, the sliding variables are attracted without overshoot onto a
non-overshooting sliding surface (i.e., the sliding variables are made to
satisfy a linear non-overshooting convergence law). Thus, the sliding
variables converge exponentially to zero, and there is no overshoot for the
first sliding variable. The influence of the bounded stochastic disturbance
or uncertainty is rejected completely due to the sliding mode gain coverage.
The disturbance or uncertainty is only required to be bounded. Also, the
2-sliding mode can separate the measurement noise from the sliding
variables, and it makes the variables smoothed. To implement the trajectory
non-overshooting tracking, it is only required that the second-order
derivative of reference trajectory is bounded. In order to reject the
chattering in the controller output, a tanh-function-based sliding mode is
developed to get the smoothed non-overshooting controller.

The rationality of the proposed non-overshooting sliding mode controller in
this paper incudes: 1) The structure of the two successive subsystems, which
have the non-overshooting reachability and the non-overshooting stability
respectively, can achieve global non-overshooting stability; 2) The initial
value partitioning of the second subsystem is to determine the analytical
expressions for the sliding variables, so that the conditions on
non-overshooting stability can be obtained; 3) The sliding mode gains can be
assigned to completely eliminate the influence of bounded stochastic
disturbance. The advantages of the proposed method are: 1) the robust and
global non-overshooting stability even in the presence of bounded stochastic
disturbance; 2) no restriction on the system initial values; 3) the bounded
and smoothed controller to be easily performed by the actuators.

For flight control, for example, spacecraft, hypersonic vehicle or UAV
control, sliding mode control plays an important role [27,28,29]. However,
overshoot or oscillations exist for the traditional sliding mode control
methods. In the high-speed or high-maneuverability flight conditions,
overshoot or oscillations in attitude control may affect the flight
performance and cause the safety issues. In fact, at a supersonic flying
speed, the angles of attack of the aerodynamic surfaces are usually very
small. The overshoot of the system output will cause these angles of attack
large and irregular, resulting in the unstable flight. Therefore, the
non-overshooting control is necessary for these aircrafts to implement the
safe and maneuvering flight. The robust non-overshooting control proposed in
this paper can overcome the overshoot issue, and the strict non-overshooting
stability can be achieved even in the presence of bounded stochastic
disturbance or system uncertainty. Importantly, the bounded-gain and
smoothed sliding controller is fit for many actuators.

The proposed method is applied to a UAV non-overshooting control. In the UAV
flight test, some adverse situations are considered: only a simple model is
constructed; the system uncertainties and the bounded unknown stochastic
disturbances exist; noise is in the measurements of position and attitude;
and the reference includes the multi-segment trajectory. The control laws
based on the non-overshooting sliding mode are designed to drive the UAV to
achieve the flight mission. The control system can implement the agile and
non-overshooting tracking for the complex reference trajectory. Furthermore,
even when the reference suddenly changes or jumps, i.e., the reference
trajectory is discontinuous, the controller parameters can be updated to
keep the non-overshooting stabilization through the parameter regulation
conditions.

Compared with the research results in the existing relevant literature, the
contributions of this paper include: 1) Even with the presence of stochastic
disturbances and the requirement of time-variant reference trajectories, the
strict non-overshooting stability can still be achieved; 2) The stability is
global, and the system gains are bounded; 3) The stochastic disturbances are
only required to be bounded; 4) There is no restriction on the initial
values of the systems; 5) Due to the filter-corrector property of the
sliding mode surface, the noise can be fully separated from the sliding
variables, even when the frequency bands of the variables and noise overlap;
6) The controller output is bounded and is smoothed, and it is easily
performed by the actuators.

\section{Problem description and analysis}

The problem considered in this paper is for system safety control to
implement non-overshooting stabilization, even the unknown stochastic
disturbance exists, and the time-variant reference is required. Overshoot
means that signal passes over or exceeds its target, and it affects the
safety control adversely.

We know that, the control performance of a system is determined by its
closed-loop error system after a controller is selected. Equivalently
speaking, controller design is to determine a control law to turn the
open-loop error system into a desired stable system of error form. The
control performance is determined by the desired stable system. The
controller is a connection between the open-loop error system and the
desired stable system.

Therefore, when it is difficult to design a controller to make a system
stable and without overshoot, we can construct a desired stable system with
non-overshooting and robust properties. Then, from the relation between the
open-loop error system\ and the desired stable system, the controller is
solved.

A conclusion on controller design based on desired stable system is
introduced as follows.

\bigskip

\emph{2.1 Controller design and desired stable system}

\textbf{Conclusion 2.1} (\emph{Controller design based on desired stable
system}):

A dynamical system of error form is considered as follows:

\begin{equation}
e^{\left( n\right) }\left( t\right) =g\left( e\left( t\right) ,\dot{e}\left(
t\right) ,\cdots ,e^{\left( n-1\right) }\left( t\right) ,t\right) +u\left(
t\right) +d\left( t\right)
\end{equation}%
where, $e\left( t\right) $ is the system error variable, and $\dot{e}\left(
t\right) ,\cdots ,e^{\left( n\right) }\left( t\right) $ are the derivatives
of $e\left( t\right) $; $g\left( \cdot \right) $ is the known function; $%
u\left( t\right) $ is the control input; and $d\left( t\right) $ is the
unknown disturbance or uncertainty of any kind. If the system

\begin{equation}
e^{\left( n\right) }\left( t\right) =f\left( e\left( t\right) ,\dot{e}\left(
t\right) ,\cdots ,e^{\left( n-1\right) }\left( t\right) ,t\right) +d\left(
t\right)
\end{equation}%
is already stable even disturbance $d\left( t\right) $ exists, i.e.,

\begin{equation}
\underset{t\rightarrow \infty }{\lim }e^{\left( i\right) }\left( t\right) =0%
\text{, }i=0,1,\cdots ,n-1
\end{equation}%
where, $f\left( \cdot \right) $ is the known function, then, in order to
make system (1) stable, the controller can be selected as

\begin{equation}
u\left( t\right) =f\left( e\left( t\right) ,\dot{e}\left( t\right) ,\cdots
,e^{\left( n-1\right) }\left( t\right) ,t\right) -g\left( e\left( t\right) ,%
\dot{e}\left( t\right) ,\cdots ,e^{\left( n-1\right) }\left( t\right)
,t\right)
\end{equation}

In fact, system (2) is already stable without control. If we can select
controller $u\left( t\right) $ to turn system (1) into system (2), then, the
system (1) will become stable. Because the left sides of (1) and (2) are
same, we just make the right sides of (1) and (2) equal, i.e.,

\begin{equation}
g\left( e\left( t\right) ,\dot{e}\left( t\right) ,\cdots ,e^{\left(
n-1\right) }\left( t\right) ,t\right) +u\left( t\right) +d\left( t\right)
=f\left( e\left( t\right) ,\dot{e}\left( t\right) ,\cdots ,e^{\left(
n-1\right) }\left( t\right) ,t\right) +d\left( t\right)
\end{equation}%
Then, through solving the equality (5), the disturbance $d\left( t\right) $\
is canceled out, and we get the controller (4).

\bigskip

\textbf{Remark 2.1:} For system (1) with controller (4), system (2)
determines its stability, transient process and robustness. Therefore, for a
dynamical control system, we can use a desired stable system to determine
the controller and analyze the performance of control system. Importantly,
for safety control, if the desired stable system has the non-overshooting
property, then the dynamical control system has the same performance.

\bigskip

\emph{2.2 PID/PI desired stable system}

PID/PI control is popular for many industrial control systems. Ideally, PID
or PI control can completely reject the effect of constant disturbance when
stabilizing a system. The desired stable system in PID form is a third-order
system, and the overshoot often happens. Especially, a large overshoot
exists due to the windup effect or the integration saturation. We have the
following two Lemmas.

\bigskip

\textbf{Lemma 2.1} (\emph{PID desired stable system with unknown constant\
disturbance}): For system%
\begin{equation}
\ddot{e}\left( t\right) =-k_{p}e\left( t\right) -k_{i}\int_{0}^{t}e\left(
\tau \right) d\tau -k_{d}\dot{e}\left( t\right) +d
\end{equation}%
where, $d$ is any unknown constant disturbance, if $k_{p}$, $k_{i}$ and $%
k_{d}$ are selected to make the real parts of all the roots of the
characteristic equation $s^{3}+k_{d}s^{2}+k_{p}s+k_{i}=0$ negative, then,
system (6) is exponentially stable, and

\begin{equation}
\underset{t\rightarrow \infty }{\lim }e\left( t\right) =0\text{, }\underset{%
t\rightarrow \infty }{\lim }\dot{e}\left( t\right) =0\text{, and }\underset{%
t\rightarrow \infty }{\lim }k_{i}\int_{0}^{t}e\left( \tau \right) d\tau =d
\end{equation}

The proof of Lemma 2.1 is presented in Appendix. $\blacksquare $

\textbf{Remark 2.2 (}\emph{PID control for second-order systems}\textbf{):}
Ideally, PID controller is used for the control of a second-order system
with unknown constant disturbance (but not time-varying disturbance). For
the second-order system of error form

\begin{equation}
\ddot{e}\left( t\right) =-g\left( t\right) -u\left( t\right) +d
\end{equation}%
the system (6) is selected as the desired stable system. We select the right
sides of (6) and (8) equal, i.e.,

\begin{equation}
-g\left( t\right) -u\left( t\right) +d=-k_{p}e\left( t\right)
-k_{i}\int_{0}^{t}e\left( \tau \right) d\tau -k_{d}\dot{e}\left( t\right) +d
\end{equation}%
Then, solving the equality (9), the disturbance $d$\ is canceled out, and we
get the controller:

\begin{equation}
u\left( t\right) =k_{p}e\left( t\right) +k_{i}\int_{0}^{t}e\left( \tau
\right) d\tau +k_{d}\dot{e}\left( t\right) -g\left( t\right)
\end{equation}

From (68), the closed-loop error system is a third-order system, the
combination of three roots of $s^{3}+k_{d}s^{2}+k_{p}s+k_{i}=0$ determines
the stability and transient process, and the overshooting response often
happens.

For PID control, if the large initial error of system output variable exist,
then a relatively large long-time overshoot may happen due to the integrator
windup. PID control can only completely reject the effect of constant
disturbance and make $\underset{t\rightarrow \infty }{\lim }e\left( t\right)
=0$ and $\underset{t\rightarrow \infty }{\lim }\dot{e}\left( t\right) =0$.
In addition, for a time-varying disturbance, PID control can only make the
system approximately stable. Even the parameter regulation methods are used
to implement the non-overshooting stabilization, the disturbance is still
assumed to be constant. If disturbance is fast time-varying, then the
control performance becomes worse, and the overshoot and oscillations
deteriorate.

\bigskip

\textbf{Lemma 2.2} (\emph{PI desired stable system with unknown constant\
disturbance}): For system

\begin{equation}
\dot{e}\left( t\right) =-k_{p}e\left( t\right) -k_{i}\int_{0}^{t}e\left(
\tau \right) d\tau +d
\end{equation}%
where, $d$ is any unknown constant disturbance, if $k_{p}$ and $k_{i}$ are
selected to make the real parts of all the roots of the characteristic
equation $s^{2}+k_{p}s+k_{i}=0$ negative, then system is exponentially
stable, and

\begin{equation}
\underset{t\rightarrow \infty }{\lim }e\left( t\right) =0\text{, and }%
\underset{t\rightarrow \infty }{\lim }k_{i}\int_{0}^{t}e\left( \tau \right)
d\tau =d
\end{equation}

The proof of Lemma 2.2 is presented in Appendix. $\blacksquare $

\textbf{Remark 2.3:} Ideally, PI controller is used for the control of a
first-order system with unknown constant disturbance. For the first-order
system of error form

\begin{equation}
\dot{e}\left( t\right) =-g\left( t\right) -u\left( t\right) +d
\end{equation}%
the system (11) is selected as the desired stable system. We select the
right sides of (11) and (13) equal, i.e.,

\begin{equation}
-g\left( t\right) -u\left( t\right) +d=-k_{p}e\left( t\right)
-k_{i}\int_{0}^{t}e\left( \tau \right) d\tau +d
\end{equation}%
Then, solving the equality (14), the disturbance $d$\ is canceled out, and
we get the controller:

\begin{equation}
u\left( t\right) =k_{p}e\left( t\right) +k_{i}\int_{0}^{t}e\left( \tau
\right) d\tau -g\left( t\right)
\end{equation}

From (74), the closed-loop error system is a second-order system, the
combination of two roots of $s^{2}+k_{p}s+k_{i}=0$ determines the stability
and transient process, and sometimes overshooting response happens.
Furthermore, due to the integral windup, the overshoot happens. Strictly, PI
control can only reject the effect of constant disturbance and make $%
\underset{t\rightarrow \infty }{\lim }e\left( t\right) =0$. For a
time-varying disturbance, PI control can only make the system approximately
stable.

\bigskip

\emph{2.3 Performance metrics to evaluate the control methods}

In the view of the problems under consideration, here, we give the
performance metrics for evaluating the proposed control methods: 1) Global
non-overshooting stability; 2) Robustness against the bounded stochastic
disturbances; 3) Precision and maneuverability for reference trajectory
tracking; 4) Non restriction on the system initial values.

\section{Robust non-overshooting 2-sliding mode}

\begin{figure}[tbp]
\begin{center}
\includegraphics[width=3.50in]{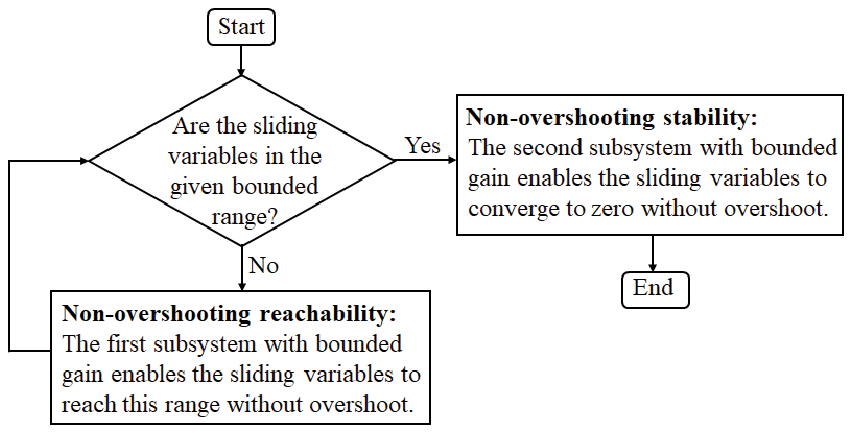}\\[0pt]
{\small (a)}\\[0pt]
\includegraphics[width=3.50in]{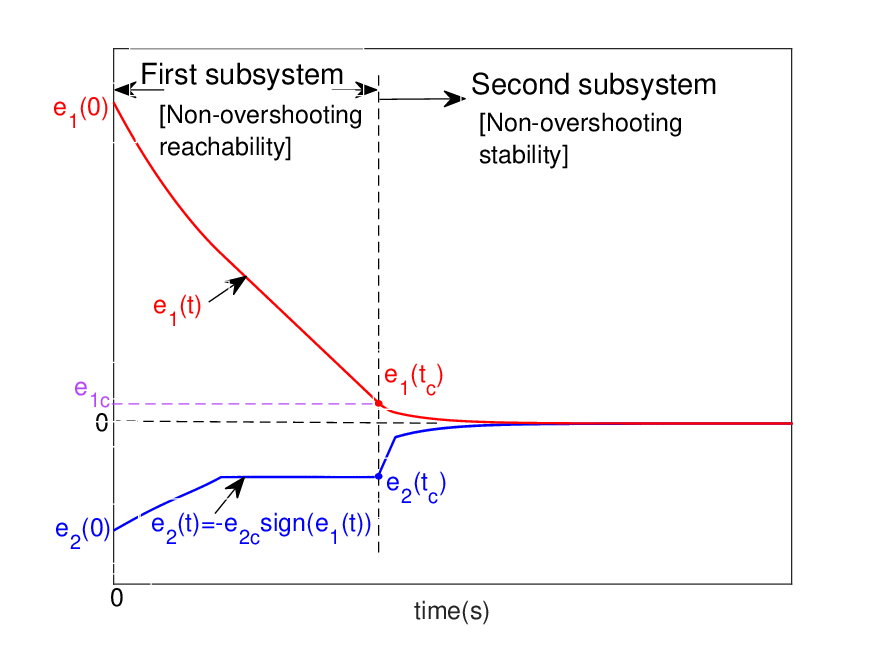}\\[0pt]
{\small (b)}
\end{center}
\caption{Configuration of globally non-overshooting 2-sliding mode. (a) Flow
chart of 2-sliding mode. (b) Convergence process of sliding variables.}
\end{figure}

\emph{3.1 Configuration of robust non-overshooting 2-sliding mode with
global stability}

Before we present the design of non-overshooting control for a class of
uncertain systems, we create a desired non-overshooting 2-sliding mode
system. Non-overshooting stability requires that the system gain depends on
the initial values of the system variables. When the initial value
amplitudes are large, the system gain also becomes large. In order to
achieve the global non-overshooting stability and the bounded system gain,
the 2-sliding mode consists of two successive subsystems. The first
subsystem with non-overshooting reachability enables the sliding variables
to reach a given bounded range. This range boundary serves as the initial
condition of the second subsystem and makes the determined system gain
bounded. The second subsystem has the property of local non-overshooting
stability, such that the sliding variables converge to zero, and no
overshoot exists for the first sliding variable. The flow chart of 2-sliding
mode is shown in Fig.1(a).

The configuration of non-overshooting 2-sliding mode with two subsystems is
expressed by

\begin{eqnarray}
\dot{e}_{1}\left( t\right) &=&e_{2}\left( t\right)  \notag \\
\dot{e}_{2}\left( t\right) &=&\left\{ 
\begin{array}{l}
f_{1}\left[ e_{1}\left( t\right) ,e_{2}\left( t\right) +e_{2c}\text{sign}%
\left( e_{1}\left( t\right) \right) \right] +d(t),\text{ if }\left\vert
e_{1}\left( t\right) \right\vert >e_{1c}; \\ 
f_{2}\left[ e_{1}\left( t\right) ,e_{2}\left( t\right) \right] +d(t),\text{
if }\left\vert e_{1}\left( t\right) \right\vert \leq e_{1c}%
\end{array}%
\right.
\end{eqnarray}%
where, the gains of functions $f_{1}\left[ \cdot \right] $ and $f_{2}\left[
\cdot \right] $ are bounded; $d(t)$ is the bounded disturbance; $e_{1c}>0$
and $e_{2c}>0$ make the system gains bounded. The sliding variables $%
e_{1}\left( t\right) $ and $e_{2}\left( t\right) $ experience the following
convergence process:

\begin{equation}
\left\{ 
\begin{array}{l}
e_{1}\left( t\right) \text{ reaches }e_{1c}\text{sign}\left( e_{1}\left(
t\right) \right) \text{ without overshoot and }e_{2}\left( t\right)
\rightarrow -e_{2c}\text{sign}\left( e_{1}\left( t\right) \right) \text{, if 
}\left\vert e_{1}\left( t\right) \right\vert >e_{1c} \\ 
e_{1}\left( t\right) \rightarrow 0\text{ without overshoot and }e_{2}\left(
t\right) \rightarrow 0\text{, if }\left\vert e_{1}\left( t\right)
\right\vert \leq e_{1c}%
\end{array}%
\right.
\end{equation}

The convergence process of sliding variables is shown in Fig.1(b):

1)\ The first subsystem (\emph{with robust non-overshooting reachability}):
its gain is bounded; it makes $e_{1}\left( t\right) $ reach the\ given range 
$\left\vert e_{1}\left( t\right) \right\vert \leq e_{1c}$ without overshoot,
and $e_{2}\left( t\right) $ approaches to the assigned $-e_{2c}$sign$\left(
e_{1}\left( t\right) \right) $ which is opposite to $e_{1}\left( t\right) $;
the bounded disturbance $d(t)$ can be rejected completely; and then the
first subsystem switches to the second subsystem, which it has the initial
values $e_{1}\left( t_{c}\right) $ and $e_{2}\left( t_{c}\right) $.

2) The second subsystem (\emph{with robust non-overshooting stability}): its
bounded gain is determined from the initial values $e_{1}\left( t_{c}\right) 
$ and $e_{2}\left( t_{c}\right) $; the sliding variables $e_{1}\left(
t\right) $ and $e_{2}\left( t\right) $ converge to zero, and no overshoot
exists for $e_{1}\left( t\right) $; moreover, the bounded disturbance $d(t)$
can be rejected completely.

\bigskip

Two Theorems on the explicit forms of robust non-overshooting 2-sliding mode
are presented as follows.

\emph{3.2 Design of non-overshooting sliding mode system}

\textbf{Theorem 3.1 }(\emph{Robust non-overshooting 2-sliding mode})\textbf{:%
} The 2-sliding mode system is as follows:

\begin{eqnarray}
\dot{e}_{1}\left( t\right) &=&e_{2}\left( t\right)  \notag \\
\dot{e}_{2}\left( t\right) &=&\left\{ 
\begin{array}{l}
-k_{c}\text{sign}\left[ e_{2}\left( t\right) +e_{2c}\text{sign}\left(
e_{1}\left( t\right) \right) \right] +d(t),\text{ if }\left\vert e_{1}\left(
t\right) \right\vert >e_{1c}; \\ 
-k_{2}\text{sign}\left[ e_{2}\left( t\right) +k_{1}e_{1}\left( t\right) %
\right] +d(t),\text{ if }\left\vert e_{1}\left( t\right) \right\vert \leq
e_{1c}%
\end{array}%
\right.
\end{eqnarray}%
where, $e_{1}\left( t\right) $ and $e_{2}\left( t\right) $ are the sliding
variables; the bounded unknown disturbance $d(t)$ satisfies $\sup_{t\in %
\left[ 0,\infty \right) }\left\vert d(t)\right\vert \leq L_{d}<\infty $; $%
k_{c}>L_{d}$; $e_{1c}\in \left( 0,k_{2M}-L_{d}\right) $, $e_{2c}\in \left(
e_{1c},\sqrt{\left( k_{2M}-L_{d}\right) e_{1c}}\right] $, and $k_{2M}>L_{d}$
is the up-bound of $k_{2}$ from the system gain limitation; $e_{1}\left(
t_{c}\right) $ and $e_{2}\left( t_{c}\right) $ are the initial values of $%
e_{1}\left( t\right) $ and $e_{2}\left( t\right) $ respectively when $%
\left\vert e_{1}\left( t\right) \right\vert \leq e_{1c}$; and

\begin{eqnarray}
k_{1} &\in &\left\{ 
\begin{array}{l}
\left( 0,\frac{\left\vert e_{2}\left( t_{c}\right) \right\vert }{\left\vert
e_{1}\left( t_{c}\right) \right\vert }\right) ,\text{ if }e_{1}\left(
t_{c}\right) e_{2}\left( t_{c}\right) <0\text{ and }\left\vert e_{1}\left(
t_{c}\right) \right\vert <\left\vert e_{2}\left( t_{c}\right) \right\vert ;
\\ 
\left( \frac{\left\vert e_{2}\left( t_{c}\right) \right\vert }{\left\vert
e_{1}\left( t_{c}\right) \right\vert },\infty \right) ,\text{ if }%
e_{1}\left( t_{c}\right) e_{2}\left( t_{c}\right) <0\text{ and }\left\vert
e_{1}\left( t_{c}\right) \right\vert \geq \left\vert e_{2}\left(
t_{c}\right) \right\vert ; \\ 
\left( 0,\infty \right) ,\text{ if others}%
\end{array}%
\right. \\
k_{2} &>&\left\{ 
\begin{array}{l}
\max \left\{ k_{1}\left\vert e_{2}\left( t_{c}\right) \right\vert +L_{d},%
\frac{e_{2}^{2}\left( t_{c}\right) }{2\left\vert e_{1}\left( t_{c}\right)
\right\vert }+L_{d}\right\} ,\text{ if }e_{1}\left( t_{c}\right) e_{2}\left(
t_{c}\right) <0\text{ and }\left\vert e_{1}\left( t_{c}\right) \right\vert
<\left\vert e_{2}\left( t_{c}\right) \right\vert ; \\ 
\max \left\{ k_{1}\left\vert e_{2}\left( t_{c}\right) \right\vert +L_{d},%
\frac{k_{1}^{2}}{3}\left( \left\vert e_{1}\left( t_{c}\right) \right\vert +%
\sqrt{e_{1}^{2}\left( t_{c}\right) +3\left( \frac{e_{2}\left( t_{c}\right) }{%
k_{1}}\right) ^{2}}\right) +L_{d}\right\} ,\text{ if others}%
\end{array}%
\right.
\end{eqnarray}%
Then, we get the linear convergence law $\dot{e}_{1}\left( t\right)
=-k_{1}e_{1}\left( t\right) $ (i.e., sliding surface $e_{2}\left( t\right)
+k_{1}e_{1}\left( t\right) =0$), i.e., the system (18) is globally
exponentially stable, and

\begin{equation}
\underset{t\rightarrow \infty }{\lim }e_{1}\left( t\right) =0\text{ and }%
\underset{t\rightarrow \infty }{\lim }e_{2}\left( t\right) =0
\end{equation}%
In addition, the convergence of variable $e_{1}\left( t\right) $ is
non-overshooting.

The proof of Theorem 3.1 is presented in Appendix. $\blacksquare $

\bigskip

\textbf{Remark 3.1} (\emph{Performance analysis of sliding mode (18) with
conditions (19) and (20)}):

\emph{1) Globally exponential stability and no overshoot.} During the whole
transient process, both $e_{1}\left( t\right) $ and $e_{2}\left( t\right) $
converge to zero, and no overshoot exists in $e_{1}\left( t\right) $:

i) Firstly, there exists a finite time $t_{c}>0$, for $t\geq t_{c}$, the
first subsystem enables $e_{1}\left( t\right) $ to reach $\left\vert
e_{1}\left( t\right) \right\vert \leq e_{1c}$ without overshoot; meanwhile, $%
e_{2}\left( t\right) $ gets to $e_{2c}$sign$\left( e_{1}\left( t\right)
\right) $; then, the first subsystem switches to the second subsystem.

ii) Secondly, for the second subsystem within the range $\left\vert
e_{1}\left( t\right) \right\vert \leq e_{1c}$, from its initial values $%
e_{1}\left( t_{c}\right) $ and $e_{2}\left( t_{c}\right) $, there exists a
finite time $t_{s}>0$, for $t\geq t_{c}+t_{s}$, the variables $e_{1}\left(
t\right) $ and $e_{2}\left( t\right) $\ are attracted without overshoot onto
the sliding surface $\sigma \left( t\right) =e_{2}\left( t\right)
+k_{1}e_{1}\left( t\right) =0$, and the sliding function $\sigma \left(
t\right) $ and the sliding variable $e_{1}\left( t\right) $ are
non-overshooting during $t\in \left[ t_{c},t_{c}+t_{s}\right) $; because $%
\dot{e}_{1}\left( t\right) =e_{2}\left( t\right) $, the linear convergence
law $\dot{e}_{1}\left( t\right) =-k_{1}e_{1}\left( t\right) $ is achieved;
therefore, for $t\geq t_{c}+t_{s}$, $e_{1}\left( t\right) $ and $e_{2}\left(
t\right) $ are exponentially convergent, and no overshoot exists in $%
e_{1}\left( t\right) $. Therefore, it is a non-overshooting convergence
process.

\emph{2) Complete rejection of bounded stochastic disturbance/uncertainty.}
Disturbances\ or uncertainties exist in many dynamic systems. For example,
in a UAV flight, the aerodynamic disturbance exists from the crosswind, and
the influences of unmodelled dynamic uncertainties are not avoidable in
modeling. They perform to be stochastic and bounded. Therefore, we can
define the disturbance or uncertainty $d(t)$ to satisfy $\sup_{t\in \left[
0,\infty \right) }\left\vert d(t)\right\vert \leq L_{d}<\infty $. For the
sliding mode (18), even\ the bounded disturbance $d\left( t\right) $ exists,
the strict non-overshooting and exponential stability is still achieved, and 
$\underset{t\rightarrow \infty }{\lim }e_{1}\left( t\right) =0$ and $%
\underset{t\rightarrow \infty }{\lim }e_{2}\left( t\right) =0$. In fact,
from (86)$\sim $(88) in the proof of Theorem 3.1, due to $k_{c}>L_{d}$ and $%
k_{2}>L_{d}$ in the sliding mode, the influence of bounded disturbances or
uncertainties can be completely rejected. Furthermore, for this sliding
mode, the condition on the stochastic disturbance $d\left( t\right) $ is
relax, and $d\left( t\right) $ is only required to be bounded.

\emph{3) Bounded gain of the sliding mode.} From (20), the selection of $%
e_{1c}\in \left( 0,k_{2M}-L_{d}\right) $ (i.e., $e_{1c}=\left\vert
e_{1}\left( t_{c}\right) \right\vert $) and $e_{2c}\in \left( e_{1c},\sqrt{%
\left( k_{2M}-L_{d}\right) e_{1c}}\right] $ (i.e., $e_{2c}=\left\vert
e_{2}\left( t_{c}\right) \right\vert $) makes $\left\vert e_{1}\left(
t_{c}\right) \right\vert $, $\left\vert e_{2}\left( t_{c}\right) \right\vert 
$ and $\frac{e_{2}^{2}\left( t_{c}\right) }{2\left\vert e_{1}\left(
t_{c}\right) \right\vert }$ all bounded. Therefore, $k_{2}$ is bounded, and
the non-overshooting performance is guaranteed.

\emph{4) Chattering phenomenon in the sliding variable }$e_{2}\left(
t\right) $\emph{.} For the sliding mode (18), due to the switching functions
in the $e_{2}\left( t\right) $ dynamic equation, chattering may exist in $%
e_{2}\left( t\right) $.

\bigskip

\textbf{Remark 3.2} (\emph{Parameter conditions that cause slow convergence})%
\textbf{:} For the selection of $k_{1}$ and $k_{2}$, we can use the
partitioning $e_{1}\left( t_{c}\right) e_{2}\left( t_{c}\right) <0$ and $%
e_{1}\left( t_{c}\right) e_{2}\left( t_{c}\right) \geq 0$, and we get

\begin{eqnarray}
k_{1} &\in &\left\{ 
\begin{array}{l}
\left( 0,\frac{\left\vert e_{2}\left( t_{c}\right) \right\vert }{\left\vert
e_{1}\left( t_{c}\right) \right\vert }\right) ,\text{ if }e_{1}\left(
t_{c}\right) e_{2}\left( t_{c}\right) <0; \\ 
\left( 0,\infty \right) ,\text{ if others}%
\end{array}%
\right. \\
k_{2} &>&\left\{ 
\begin{array}{l}
\max \left\{ k_{1}\left\vert e_{2}\left( t_{c}\right) \right\vert +L_{d},%
\frac{e_{2}^{2}\left( t_{c}\right) }{2\left\vert e_{1}\left( t_{c}\right)
\right\vert }+L_{d}\right\} ,\text{ if }e_{1}\left( t_{c}\right) e_{2}\left(
t_{c}\right) <0; \\ 
\max \left\{ k_{1}\left\vert e_{2}\left( t_{c}\right) \right\vert +L_{d},%
\frac{k_{1}^{2}}{3}\left( \left\vert e_{1}\left( t_{c}\right) \right\vert +%
\sqrt{e_{1}^{2}\left( t_{c}\right) +3\left( \frac{e_{2}\left( t_{c}\right) }{%
k_{1}}\right) ^{2}}\right) +L_{d}\right\} ,\text{ if others}%
\end{array}%
\right.
\end{eqnarray}%
Thus, the system (18) is also globally exponentially stable, and no
overshoot exists in $e_{1}\left( t\right) $.

However, for the second subsystem, when the initial values $e_{1}\left(
t_{c}\right) $\ and $e_{2}\left( t_{c}\right) $ of sliding variables are in
the zones II-1 and IV-1, i.e., in \{$\left( e_{1}\left( t_{c}\right)
,e_{2}\left( t_{c}\right) \right) |e_{1}\left( t_{c}\right) e_{2}\left(
t_{c}\right) <0\ $and $\left\vert e_{1}\left( t_{c}\right) \right\vert \geq
\left\vert e_{2}\left( t_{c}\right) \right\vert $\}, the slow convergence
exists\ for sliding mode (18) with conditions (22) and (23), but it does not
happen for sliding mode (18) with conditions (19) and (20). In fact:

1) Slow convergence for sliding mode (18) with conditions (22) and (23): For
the second subsystem, when the initial values of sliding variables are in
the zones II-1 and IV-1, from (22), we know that $k_{1}\in \left( 0,\frac{%
\left\vert e_{2}\left( t_{c}\right) \right\vert }{\left\vert e_{1}\left(
t_{c}\right) \right\vert }\right) $. $k_{1}\in \left( 0,1\right) $ should be
satisfied when $\left\vert e_{1}\left( t_{c}\right) \right\vert \geq
\left\vert e_{2}\left( t_{c}\right) \right\vert $. Therefore, the $%
e_{1}\left( t\right) $ convergence may be slow due to the convergence law $%
\dot{e}_{1}\left( t\right) =-k_{1}e_{1}\left( t\right) $.

2) Fast convergence for sliding mode (18) with conditions (19) and (20): For
the second subsystem, when the initial values of sliding variables are in
the zones II-1 and IV-1, from (19), we know that $k_{1}\in \left( \frac{%
\left\vert e_{2}\left( t_{c}\right) \right\vert }{\left\vert e_{1}\left(
t_{c}\right) \right\vert },\infty \right) $. Because $\left\vert e_{1}\left(
t_{c}\right) \right\vert \geq \left\vert e_{2}\left( t_{c}\right)
\right\vert $, we need to select $k_{1}\geq 1$ for the convergence law $\dot{%
e}_{1}\left( t\right) =-k_{1}e_{1}\left( t\right) $ to get a fast
convergence.

\bigskip

\emph{3.3 Design of smoothed non-overshooting sliding mode system}

In order to avoid $e_{2}\left( t\right) $ chattering in the sliding mode,
the continuous functions are used in the sliding mode, and the following
Theorem is presented.

\textbf{Theorem 3.2 }(\emph{Smoothed non-overshooting sliding mode})\textbf{:%
} The tanh-function-based 2-sliding mode system is as follows:

\begin{eqnarray}
\dot{e}_{1}\left( t\right) &=&e_{2}\left( t\right)  \notag \\
\dot{e}_{2}\left( t\right) &=&\left\{ 
\begin{array}{l}
-k_{c}\text{tanh}\left[ \rho _{c}\left( e_{2}\left( t\right) +e_{2c}\text{%
sign}\left( e_{1}\left( t\right) \right) \right) \right] +d(t),\text{ if }%
\left\vert e_{1}\left( t\right) \right\vert >e_{1c}; \\ 
-k_{2}\text{tanh}\left[ \rho (e_{2}\left( t\right) +k_{1}e_{1}\left(
t\right) )\right] +d(t),\text{ if }\left\vert e_{1}\left( t\right)
\right\vert \leq e_{1c}%
\end{array}%
\right.
\end{eqnarray}%
where, $e_{1}\left( t\right) $ and $e_{2}\left( t\right) $ are the sliding
variables; the bounded unknown disturbance $d(t)$ satisfies $\sup_{t\in %
\left[ 0,\infty \right) }\left\vert d(t)\right\vert \leq L_{d}<\infty $; $%
e_{1c}\in \left( 0,k_{2M}-L_{d}\right) $, $e_{2c}\in \left( e_{1c},\sqrt{%
\left( k_{2M}-L_{d}\right) e_{1c}}\right] $, and $k_{2M}>L_{d}$ is the
up-bound of $k_{2}$ from the system gain limitation; function

\begin{equation}
\text{tanh}\left( \rho \cdot x\right) =\frac{\text{e}^{\rho \cdot x}-\text{e}%
^{-\rho \cdot x}}{\text{e}^{\rho \cdot x}+\text{e}^{-\rho \cdot x}}=1-\frac{2%
}{\text{e}^{2\rho \cdot x}+1}
\end{equation}

\begin{equation}
e_{1c}>0,\text{ }e_{2c}>0,\text{ }k_{c}>L_{d},\text{ }\rho _{c}\gg \frac{1}{2%
}\ln \frac{k_{c}+L_{d}}{k_{c}-L_{d}}
\end{equation}%
$e_{1}\left( t_{c}\right) $ and $e_{2}\left( t_{c}\right) $ are the initial
values of $e_{1}\left( t\right) $ and $e_{2}\left( t\right) $ respectively
when $\left\vert e_{1}\left( t\right) \right\vert \leq e_{1c}$; and

\begin{eqnarray}
k_{1} &\in &\left\{ 
\begin{array}{l}
\left( 0,\frac{\left\vert e_{2}\left( t_{c}\right) \right\vert }{\left\vert
e_{1}\left( t_{c}\right) \right\vert }\right) ,\text{ if }e_{1}\left(
t_{c}\right) e_{2}\left( t_{c}\right) <0\text{ and }\left\vert e_{1}\left(
t_{c}\right) \right\vert <\left\vert e_{2}\left( t_{c}\right) \right\vert ;
\\ 
\left( \frac{\left\vert e_{2}\left( t_{c}\right) \right\vert }{\left\vert
e_{1}\left( t_{c}\right) \right\vert },\infty \right) ,\text{ if }%
e_{1}\left( t_{c}\right) e_{2}\left( t_{c}\right) <0\text{ and }\left\vert
e_{1}\left( t_{c}\right) \right\vert \geq \left\vert e_{2}\left(
t_{c}\right) \right\vert ; \\ 
\left( 0,\infty \right) ,\text{ if others}%
\end{array}%
\right. \\
k_{2} &>&\left\{ 
\begin{array}{l}
\max \left\{ k_{1}\left\vert e_{2}\left( t_{c}\right) \right\vert +L_{d},%
\frac{e_{2}^{2}\left( t_{c}\right) }{2\left\vert e_{1}\left( t_{c}\right)
\right\vert }+L_{d}\right\} ,\text{ if }e_{1}\left( t_{c}\right) e_{2}\left(
t_{c}\right) <0\text{ and }\left\vert e_{1}\left( t_{c}\right) \right\vert
<\left\vert e_{2}\left( t_{c}\right) \right\vert ; \\ 
\max \left\{ k_{1}\left\vert e_{2}\left( t_{c}\right) \right\vert +L_{d},%
\frac{k_{1}^{2}}{3}\left( \left\vert e_{1}\left( t_{c}\right) \right\vert +%
\sqrt{e_{1}^{2}\left( t_{c}\right) +3\left( \frac{e_{2}\left( t_{c}\right) }{%
k_{1}}\right) ^{2}}\right) +L_{d}\right\} ,\text{ if others}%
\end{array}%
\right.
\end{eqnarray}

\begin{equation}
\rho \gg \max \left\{ \frac{1}{2k_{1}},1\right\} \ln \frac{%
k_{2}+k_{1}e_{2\max }+L_{d}}{k_{2}-k_{1}e_{2\max }-L_{d}}
\end{equation}%
and

\begin{equation}
e_{2\max }=\max \left\{ \left\vert e_{2}\left( t_{c}\right) \right\vert ,%
\frac{k_{1}}{3}\left[ \left\vert e_{1}\left( t_{c}\right) \right\vert +\sqrt{%
e_{1}^{2}\left( t_{c}\right) +3\left( \frac{e_{2}\left( t_{c}\right) }{k_{1}}%
\right) ^{2}}\right] \right\}
\end{equation}%
Then:

(i) The effect of disturbances is rejected, and the variables $e_{1}\left(
t\right) $ and $e_{2}\left( t\right) $ of system (24) are in the bounds as
follows:

\begin{equation}
\underset{t\rightarrow \infty }{\lim }\left\vert e_{1}\left( t\right)
\right\vert \leq \frac{1}{2\rho k_{1}}\ln \frac{k_{2}+k_{1}e_{2\max }+L_{d}}{%
k_{2}-k_{1}e_{2\max }-L_{d}}\text{, and }\underset{t\rightarrow \infty }{%
\lim }\left\vert e_{2}\left( t\right) \right\vert \leq \frac{1}{\rho }\ln 
\frac{k_{2}+k_{1}e_{2\max }+L_{d}}{k_{2}-k_{1}e_{2\max }-L_{d}}
\end{equation}%
In addition, the convergence of variable $e_{1}\left( t\right) $ is
non-overshooting.

(ii) Specially, if $\rho $\ is large enough, i.e., $\rho \rightarrow +\infty 
$, then the system (24) becomes the ideal sliding mode (18), and we get

\begin{equation}
\underset{t\rightarrow \infty }{\lim }\underset{\rho \rightarrow +\infty }{%
\lim }e_{1}\left( t\right) =0\text{ and }\underset{t\rightarrow \infty }{%
\lim }\underset{\rho \rightarrow +\infty }{\lim }e_{2}\left( t\right) =0
\end{equation}

The proof of Theorem 3.2 is presented in Appendix. $\blacksquare $

\bigskip

\textbf{Remark 3.3 }(\emph{Smoothed and non-overshooting\ convergence for
sliding mode (24) with (25)}$\sim $\emph{(30)}):

In addition to the non-overshooting convergence, the smoothed sliding mode
system (24) has the following properties:

1) Universal approximation: Because $\underset{\rho \rightarrow +\infty }{%
\lim }$tanh$\left( \rho \cdot x\right) =$sign$(x)$, the sliding mode system
(24) is the smoothed approximation of ideal sliding mode system (18).

2) Smoothed outputs of $e_{1}\left( t\right) $ and $e_{2}\left( t\right) $:
Due to continuity in the sliding mode system (24), the outputs of both $%
e_{1}\left( t\right) $ and $e_{2}\left( t\right) $ are smoothed. For $-k_{c}$%
tanh$\left[ \rho _{c}\left( e_{2}\left( t\right) +e_{2c}\text{sign}\left(
e_{1}\left( t\right) \right) \right) \right] +d(t)$ in (24), the function
sign$\left( e_{1}\left( t\right) \right) $ does not change its sign due to
continuity of $e_{1}\left( t\right) $ when $\left\vert e_{1}\left( t\right)
\right\vert >e_{1c}$. Therefore, no chattering happens.

\bigskip

\textbf{Remark 3.4} (\emph{Parameters determination for the sliding mode
system (24)})

For the algorithm calculation, we need to turn the inequality expressions of
the parameters into the corresponding equalities, and the calculated maximum
or minimum values of $k_{1}$ and $k_{2}$ are multiplied by the corresponding
coefficients. The determination steps of $e_{1c}$, $e_{2c}$, $k_{c}$, $\rho
_{c}$, $k_{1}$, $k_{2}$ and $\rho $ are presented as follows.

Step 1: Get the initial errors $e_{1}\left( 0\right) $ and $e_{2}\left(
0\right) $.

Step 2: Select

\begin{eqnarray}
e_{1c} &\in &\left( 0,k_{2M}-L_{d}\right) ,e_{2c}\in \left( e_{1c},\sqrt{%
\left( k_{2M}-L_{d}\right) e_{1c}}\right]  \notag \\
k_{c} &>&L_{d},\text{ }\rho _{c}=\rho _{c0}\frac{1}{2}\ln \frac{k_{c}+L_{d}}{%
k_{c}-L_{d}}
\end{eqnarray}%
where, $k_{2M}>L_{d}$ is the up-bound of $k_{2}$ limited from the system
gain; and $\rho _{c0}>1$

Step 3: Determine the parameters $k_{1}$, $k_{2}$ and $\rho $ through the
following calculations.

\begin{equation}
k_{1}=\left\{ 
\begin{array}{l}
\beta _{11}\frac{\left\vert e_{2}\left( t_{c}\right) \right\vert }{%
\left\vert e_{1}\left( t_{c}\right) \right\vert }\in \left( 0,\frac{%
\left\vert e_{2}\left( t_{c}\right) \right\vert }{\left\vert e_{1}\left(
t_{c}\right) \right\vert }\right) ,\text{ if }e_{1}\left( t_{c}\right)
e_{2}\left( t_{c}\right) <0\text{ and }\left\vert e_{1}\left( t_{c}\right)
\right\vert <\left\vert e_{2}\left( t_{c}\right) \right\vert ; \\ 
\beta _{12}\frac{\left\vert e_{2}\left( t_{c}\right) \right\vert }{%
\left\vert e_{1}\left( t_{c}\right) \right\vert }\in \left( \frac{\left\vert
e_{2}\left( t_{c}\right) \right\vert }{\left\vert e_{1}\left( t_{c}\right)
\right\vert },\infty \right) ,\text{ if }e_{1}\left( t_{c}\right)
e_{2}\left( t_{c}\right) <0\text{ and }\left\vert e_{1}\left( t_{c}\right)
\right\vert \geq \left\vert e_{2}\left( t_{c}\right) \right\vert ; \\ 
\beta _{13}\in \left( 0,\infty \right) ,\text{ if others}%
\end{array}%
\right.
\end{equation}

\begin{equation}
k_{2}=\left\{ 
\begin{array}{l}
\beta _{2}\max \left\{ k_{1}\left\vert e_{2}\left( t_{c}\right) \right\vert
+L_{d},\frac{e_{2}^{2}\left( t_{c}\right) }{2\left\vert e_{1}\left(
t_{c}\right) \right\vert }+L_{d}\right\} ,\text{ if }e_{1}\left(
t_{c}\right) e_{2}\left( t_{c}\right) <0\text{ and }\left\vert e_{1}\left(
t_{c}\right) \right\vert <\left\vert e_{2}\left( t_{c}\right) \right\vert ;
\\ 
\beta _{2}\max \left\{ k_{1}\left\vert e_{2}\left( t_{c}\right) \right\vert
+L_{d},\frac{k_{1}^{2}}{3}\left( \left\vert e_{1}\left( t_{c}\right)
\right\vert +\sqrt{e_{1}^{2}\left( t_{c}\right) +3\left( \frac{e_{2}\left(
t_{c}\right) }{k_{1}}\right) ^{2}}\right) +L_{d}\right\} ,\text{ if others}%
\end{array}%
\right.
\end{equation}%
where, $e_{1}\left( t_{c}\right) $ and $e_{2}\left( t_{c}\right) $ are the
initial values of $e_{1}\left( t\right) $ and $e_{2}\left( t\right) $
respectively when $\left\vert e_{1}\left( t\right) \right\vert \leq e_{1c}$; 
$\beta _{11}\in (0,1)$, $\beta _{12}>1$ and $\beta _{2}>1$.

\emph{Adjustment of }$k_{1}$\emph{:}

\begin{equation}
k_{1}=1\text{ is selected if the calculated }k_{1}\in \left( 0,1\right)
\end{equation}%
[\emph{Note:} If $k_{1}$ is calculated to be $k_{1}\in \left( 0,1\right) $,
we can select $k_{1}=1$ for the convergence law $\dot{e}_{1}\left( t\right)
=-k_{1}e_{1}\left( t\right) $ to get a fast convergence. We find that $%
k_{1}=1$ always holds from the condition (27).]

For $\rho $, we select

\begin{equation}
\rho =\rho _{0}\max \left\{ \frac{1}{2k_{1}},1\right\} \ln \frac{%
k_{2}+k_{1}e_{2\max }+L_{d}}{k_{2}-k_{1}e_{2\max }-L_{d}}
\end{equation}%
where, $\rho _{0}>1$, and

\begin{equation}
e_{2\max }=\max \left\{ \left\vert e_{2}\left( t_{c}\right) \right\vert ,%
\frac{k_{1}}{3}\left[ \left\vert e_{1}\left( t_{c}\right) \right\vert +\sqrt{%
e_{1}^{2}\left( t_{c}\right) +3\left( \frac{e_{2}\left( t_{c}\right) }{k_{1}}%
\right) ^{2}}\right] \right\}
\end{equation}

\section{Simulation examples on non-overshooting sliding mode}

We use two examples to demonstrate the stability of the two non-overshooting
sliding mode systems.

For sliding modes (18) and (24), we suppose:

the initial sliding variables $e_{1}\left( 0\right) =100$, $e_{2}\left(
0\right) =-10$;

the disturbance $d\left( t\right) =3+2\sin \left( 0.3t\right) \sin \left(
1.6t\right) $, and its upper bound $L_{d}=5$.

Suppose the system gain $k_{2}\leq k_{2M}=20$.

\begin{figure}[tbp]
\centering
\includegraphics[width=3.50in]{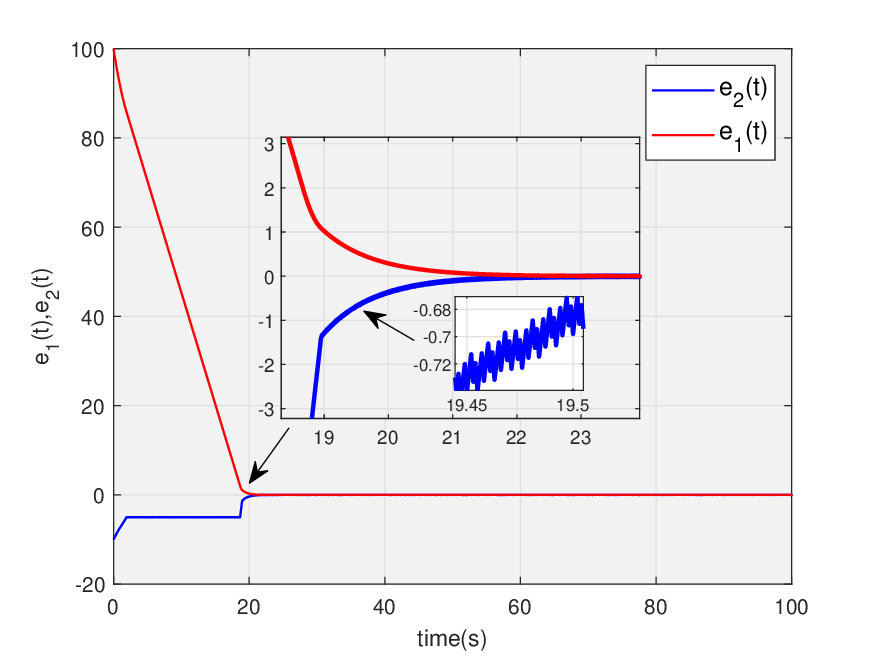}
\caption{Example 4.1 Sliding variables $e_{1}\left( t\right) $ and $%
e_{2}\left( t\right) $ of sliding mode (18).}
\end{figure}

\textbf{Example 4.1:} \emph{(Sliding mode (18) from Theorem 3.1):}

From $e_{1c}\in \left( 0,k_{2M}-L_{d}\right) =\left( 0,15\right) $, we
select $e_{1c}=2$. Then, we get $e_{2c}\in \left( e_{1c},\sqrt{\left(
k_{2M}-L_{d}\right) e_{1c}}\right] =\left( 2,\sqrt{\left( 20-5\right) \times
2}\right] =\left( 2,5.5\right] $. We select $e_{2c}=5$. Select $%
k_{c}=6>L_{d} $.

\emph{Determination of sliding mode parameters }$k_{1}$\emph{\ and }$k_{2}$%
\emph{\ according to the parameter determination steps (34)}$\sim $\emph{%
(36):}

\begin{equation*}
k_{1}=\left\{ 
\begin{array}{l}
0.5\frac{\left\vert e_{2}\left( t_{c}\right) \right\vert }{\left\vert
e_{1}\left( t_{c}\right) \right\vert },\text{ if }e_{1}\left( t_{c}\right)
e_{2}\left( t_{c}\right) <0\text{ and }\left\vert e_{1}\left( t_{c}\right)
\right\vert <\left\vert e_{2}\left( t_{c}\right) \right\vert ; \\ 
2.3\frac{\left\vert e_{2}\left( t_{c}\right) \right\vert }{\left\vert
e_{1}\left( t_{c}\right) \right\vert },\text{ if }e_{1}\left( t_{c}\right)
e_{2}\left( t_{c}\right) <0\text{ and }\left\vert e_{1}\left( t_{c}\right)
\right\vert \geq \left\vert e_{2}\left( t_{c}\right) \right\vert ; \\ 
1\in \left( 0,\infty \right) ,\text{ if others}%
\end{array}%
\right.
\end{equation*}

where, $e_{1}\left( t_{c}\right) $ and $e_{2}\left( t_{c}\right) $ are the
initial values of $e_{1}\left( t\right) $ and $e_{2}\left( t\right) $
respectively when $\left\vert e_{1}\left( t\right) \right\vert \leq e_{1c}$.

\emph{Adjustment of }$k_{1}$\emph{:} $k_{1}=1$ is selected if the calculated 
$k_{1}\in \left( 0,1\right) $; and

\begin{equation*}
k_{2}=\left\{ 
\begin{array}{l}
1.5\max \left\{ k_{1}\left\vert e_{2}\left( t_{c}\right) \right\vert +L_{d},%
\frac{e_{2}^{2}\left( t_{c}\right) }{2\left\vert e_{1}\left( t_{c}\right)
\right\vert }+L_{d}\right\} ,\text{ if }e_{1}\left( t_{c}\right) e_{2}\left(
t_{c}\right) <0\text{ and }\left\vert e_{1}\left( t_{c}\right) \right\vert
<\left\vert e_{2}\left( t_{c}\right) \right\vert ; \\ 
1.5\max \left\{ k_{1}\left\vert e_{2}\left( t_{c}\right) \right\vert +L_{d},%
\frac{k_{1}^{2}}{3}\left( \left\vert e_{1}\left( t_{c}\right) \right\vert +%
\sqrt{e_{1}^{2}\left( t_{c}\right) +3\left( \frac{e_{2}\left( t_{c}\right) }{%
k_{1}}\right) ^{2}}\right) +L_{d}\right\} ,\text{ if others}%
\end{array}%
\right.
\end{equation*}

From the algorithm calculation for the above equations, we can obtain $%
k_{1}=1.25$ and $k_{2}=16.93$.

Figure 2 shows the plots of sliding variables $e_{1}\left( t\right) $ and $%
e_{2}\left( t\right) $. Even the time-varying disturbance exists, the
non-overshooting convergence is implemented: the sign of $e_{1}\left(
t\right) $ is always unchanged, and $\underset{t\rightarrow \infty }{\lim }%
e_{1}\left( t\right) =0$ and $\underset{t\rightarrow \infty }{\lim }%
e_{2}\left( t\right) =0$ hold. Also, we find that, small chattering exists
in $e_{2}\left( t\right) $.

\textbf{Example 4.2: }\emph{(Smoothed sliding mode (24) from Theorem 3.2):}

Similar to Example 4.1, select $e_{1c}=2$, $e_{2c}=5$, and $k_{c}=6>L_{d}$.
Then, we get

\begin{equation*}
\rho _{c}=50\frac{1}{2}\ln \frac{k_{c}+L_{d}}{k_{c}-L_{d}}=59.95
\end{equation*}

\emph{Determination of sliding mode parameters }$k_{1}$\emph{, }$k_{2}$ 
\emph{and }$\rho $ \emph{according to the parameter determination steps (34)}%
$\sim $\emph{(38):}

Firstly, for determination of $k_{1}$\ and $k_{2}$, we use the same
algorithm steps as example 3.1, and we can obtain $k_{1}=1.26$ and $%
k_{2}=16.95$. Secondly, for $\rho $, we have

\begin{equation*}
e_{2\max }=\max \left\{ \left\vert e_{2}\left( t_{c}\right) \right\vert ,%
\frac{k_{1}}{3}\left[ \left\vert e_{1}\left( t_{c}\right) \right\vert +\sqrt{%
e_{1}^{2}\left( t_{c}\right) +3\left( \frac{e_{2}\left( t_{c}\right) }{k_{1}}%
\right) ^{2}}\right] \right\}
\end{equation*}

\begin{equation*}
\rho =20\max \left\{ \frac{1}{2k_{1}},1\right\} \ln \frac{%
k_{2}+k_{1}e_{2\max }+L_{d}}{k_{2}-k_{1}e_{2\max }-L_{d}}
\end{equation*}

From the algorithm calculation for the above equations, we can read $\rho
=12.19$.

\begin{figure}[tbp]
\centering
\includegraphics[width=3.50in]{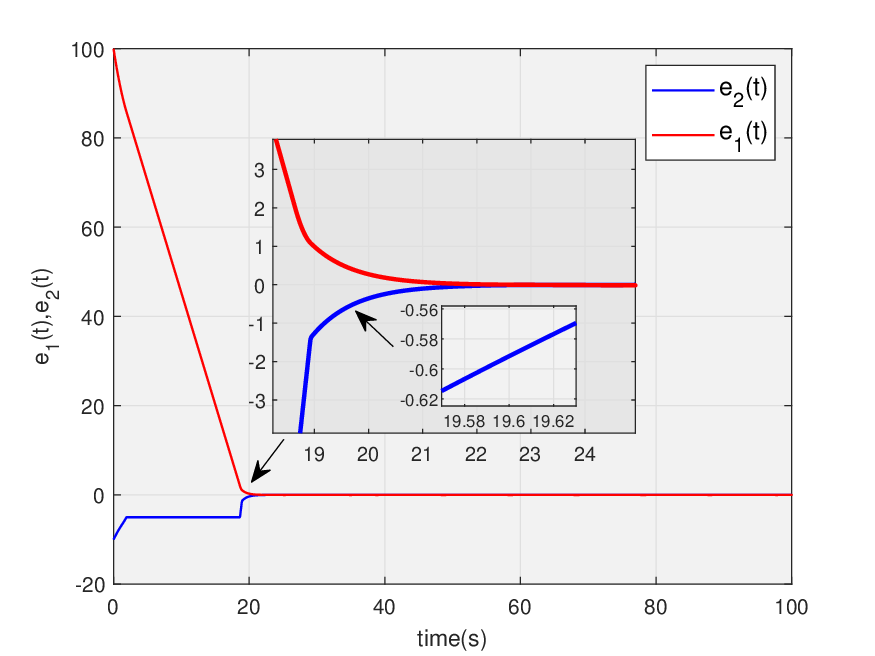}
\caption{Example 4.2 Sliding variables $e_{1}\left( t\right) $ and $%
e_{2}\left( t\right) $ of sliding mode (24).}
\end{figure}

Figure 3 describes the plots of sliding variables $e_{1}\left( t\right) $
and $e_{2}\left( t\right) $. Even the time-varying disturbance exists, the
sliding variables $e_{1}\left( t\right) $ and $e_{2}\left( t\right) $
converge to zero, and $e_{1}\left( t\right) $ convergence is
non-overshooting. In addition, both $e_{1}\left( t\right) $ and $e_{2}\left(
t\right) $ are smoothed.

\section{Non-overshooting control for uncertain systems}

\emph{5.1 Model of uncertain systems}

The following uncertain system has a minimum number of states and inputs but
retains the essential features that must be considered when designing
control laws for many dynamical systems (e.g., UAV dynamics):

\begin{eqnarray}
\dot{x}_{1}\left( t\right) &=&x_{2}\left( t\right)  \notag \\
\dot{x}_{2}\left( t\right) &=&h(t)+u(t)-\delta \left( t\right)
\end{eqnarray}%
where, $x_{1}\left( t\right) $ and $x_{2}\left( t\right) $ the system
states; $h(t)$ is the known function; $u(t)$ is the control input; and $%
\delta \left( t\right) $ is the unknown time-varying disturbance or system
uncertainty, and $\sup_{t\in \left[ 0,\infty \right) }\left\vert \delta
(t)\right\vert \leq L_{\delta }<\infty $. We consider to design a controller
for the uncertain system (39) when the reference is time variant, and $%
x_{1}\left( t\right) $ tracking the reference is required to be
non-overshooting.

\bigskip

\emph{5.2 Non-overshooting control for uncertain systems}

\textbf{Theorem 5.1 (}\emph{Non-overshooting control based on sliding mode
(18)}\textbf{):} For the uncertain system (39) with the time-variant
reference $x_{d}\left( t\right) $, if the controller is selected as

\begin{equation}
u(t)=\left\{ 
\begin{array}{l}
k_{c}\text{sign}\left[ e_{2}\left( t\right) +e_{2c}\text{sign}\left(
e_{1}\left( t\right) \right) \right] -h(t),\text{ if }\left\vert e_{1}\left(
t\right) \right\vert >e_{1c} \\ 
k_{2}\text{sign}\left[ e_{2}\left( t\right) +k_{1}e_{1}\left( t\right) %
\right] -h(t),\text{ if }\left\vert e_{1}\left( t\right) \right\vert \leq
e_{1c}%
\end{array}%
\right.
\end{equation}%
then, the system is globally exponentially stable, $x_{1}\left( t\right) $
tracking $x_{d}\left( t\right) $ is non-overshooting, and

\begin{equation}
\underset{t\rightarrow \infty }{\lim }x_{1}\left( t\right) =x_{d}\left(
t\right) \text{ and}\underset{t\rightarrow \infty }{\lim }x_{2}\left(
t\right) =\dot{x}_{d}\left( t\right)
\end{equation}%
where, $e_{1}\left( t\right) =x_{d}\left( t\right) -x_{1}\left( t\right) $
and $e_{2}\left( t\right) =\dot{x}_{d}\left( t\right) -x_{2}\left( t\right) $%
; $\sup_{t\in \left[ 0,\infty \right) }\left\vert \ddot{x}_{d}\left(
t\right) \right\vert \leq L_{x}<\infty $, and $\sup_{t\in \left[ 0,\infty
\right) }\left\vert \delta (t)\right\vert +\sup_{t\in \left[ 0,\infty
\right) }\left\vert \ddot{x}_{d}\left( t\right) \right\vert \leq
L_{d}<\infty $; $k_{c}>L_{d}$; $e_{1c}\in \left( 0,k_{2M}-L_{d}\right) $, $%
e_{2c}\in \left( e_{1c},\sqrt{\left( k_{2M}-L_{d}\right) e_{1c}}\right] $,
and $k_{2M}>L_{d}$ is the up-bound of $k_{2}$ from the system gain
limitation; $e_{1}\left( t_{c}\right) $ and $e_{2}\left( t_{c}\right) $ are
the initial values of $e_{1}\left( t\right) $ and $e_{2}\left( t\right) $
respectively when $\left\vert e_{1}\left( t\right) \right\vert \leq e_{1c}$;
and

\begin{eqnarray}
k_{1} &\in &\left\{ 
\begin{array}{l}
\left( 0,\frac{\left\vert e_{2}\left( t_{c}\right) \right\vert }{\left\vert
e_{1}\left( t_{c}\right) \right\vert }\right) ,\text{ if }e_{1}\left(
t_{c}\right) e_{2}\left( t_{c}\right) <0\text{ and }\left\vert e_{1}\left(
t_{c}\right) \right\vert <\left\vert e_{2}\left( t_{c}\right) \right\vert ;
\\ 
\left( \frac{\left\vert e_{2}\left( t_{c}\right) \right\vert }{\left\vert
e_{1}\left( t_{c}\right) \right\vert },\infty \right) ,\text{ if }%
e_{1}\left( t_{c}\right) e_{2}\left( t_{c}\right) <0\text{ and }\left\vert
e_{1}\left( t_{c}\right) \right\vert \geq \left\vert e_{2}\left(
t_{c}\right) \right\vert ; \\ 
\left( 0,\infty \right) ,\text{ if others}%
\end{array}%
\right. \\
k_{2} &>&\left\{ 
\begin{array}{l}
\max \left\{ k_{1}\left\vert e_{2}\left( t_{c}\right) \right\vert +L_{d},%
\frac{e_{2}^{2}\left( t_{c}\right) }{2\left\vert e_{1}\left( t_{c}\right)
\right\vert }+L_{d}\right\} ,\text{ if }e_{1}\left( t_{c}\right) e_{2}\left(
t_{c}\right) <0\text{ and }\left\vert e_{1}\left( t_{c}\right) \right\vert
<\left\vert e_{2}\left( t_{c}\right) \right\vert ; \\ 
\max \left\{ k_{1}\left\vert e_{2}\left( t_{c}\right) \right\vert +L_{d},%
\frac{k_{1}^{2}}{3}\left( \left\vert e_{1}\left( t_{c}\right) \right\vert +%
\sqrt{e_{1}^{2}\left( t_{c}\right) +3\left( \frac{e_{2}\left( t_{c}\right) }{%
k_{1}}\right) ^{2}}\right) +L_{d}\right\} ,\text{ if others}%
\end{array}%
\right.
\end{eqnarray}
The proof of Theorem 5.1 is presented in Appendix. $\blacksquare $

\bigskip

\textbf{Remark 5.1} (\emph{Non-overshooting control (40)})\textbf{:}

\emph{1) Non-overshooting of }$x_{1}\left( t\right) $\emph{\ tracking }$%
x_{d}\left( t\right) $\emph{:} The sliding mode (18) is the desired stable
system of error form, therefore, $e_{1}\left( t\right) =x_{d}\left( t\right)
-x_{1}\left( t\right) $ converges to zero without overshoot. Thus, $%
x_{1}\left( t\right) $ tracking $x_{d}\left( t\right) $ is non-overshooting.

\emph{2) Complete rejection of the influence from bounded disturbance and
time-variant reference:} For the uncertain system (39), the disturbance $%
\delta (t)$ and the second-order derivative of reference $x_{d}(t)$\ are
bounded, and $\sup_{t\in \left[ 0,\infty \right) }\left\vert \delta
(t)\right\vert +\sup_{t\in \left[ 0,\infty \right) }\left\vert \ddot{x}%
_{d}(t)\right\vert \leq L_{d}<\infty $ are satisfied. The parameters of
controller (40) satisfy $k_{c}>L_{d}$ and (43). When selecting the
controller (40), from (173) and (174) in the proof of Theorem 5.1, the
closed-loop error system for (39) is the robust non-overshooting 2-sliding
mode (18). Therefore, $\underset{t\rightarrow \infty }{\lim }x_{1}\left(
t\right) =x_{d}\left( t\right) $ and $\underset{t\rightarrow \infty }{\lim }%
x_{2}\left( t\right) =\dot{x}_{d}\left( t\right) $, and there is no
overshoot \ for $x_{1}\left( t\right) $ tracking $x_{d}\left( t\right) $.

\emph{3) Smoothed }$x_{1}(t)$\emph{:} Due to the integral-chain structure of
second-order sliding mode (18), $e_{1}\left( t\right) $ is smoothed.
Therefore, for the control system, $x_{1}\left( t\right) $ is smoothed.

\emph{4) Chattering in the variable }$x_{2}\left( t\right) $\emph{\ and the
controller }$u\left( t\right) $\emph{:} For system (39), due to the
controller of switching function exists in the $x_{2}\left( t\right) $
dynamic equation, chattering happens in $x_{2}\left( t\right) $. Also, the
controller $u\left( t\right) $ in (40) is discontinuous, therefore,
chattering exists in the controller output. The chattering in controller
output will affect actuator performance adversely.

\bigskip

\emph{5.3 Smoothed non-overshooting control for uncertain systems}

In order to reject chattering in the outputs of controller $u\left( t\right) 
$ and and variable $x_{2}\left( t\right) $, we present a smoothed control
based on the sliding mode (24), and a Theorem is presented as follows.

\textbf{Theorem 5.2 (}\emph{Non-overshooting control based on smoothed
sliding mode (24)}\textbf{):} For the uncertain system (39) with the
time-variant reference $x_{d}\left( t\right) $, if the controller is
selected as

\begin{equation}
u(t)=\left\{ 
\begin{array}{l}
k_{c}\text{tanh}\left[ \rho _{c}\left( e_{2}\left( t\right) +e_{2c}\text{sign%
}\left( e_{1}\left( t\right) \right) \right) \right] -h(t),\text{ if }%
\left\vert e_{1}\left( t\right) \right\vert >e_{1c}; \\ 
k_{2}\text{tanh}\left[ \rho (e_{2}\left( t\right) +k_{1}e_{1}\left( t\right)
)\right] -h(t),\text{ if }\left\vert e_{1}\left( t\right) \right\vert \leq
e_{1c}%
\end{array}%
\right.
\end{equation}%
then, the system is globally exponentially stable, $x_{1}\left( t\right) $
tracking $x_{d}\left( t\right) $ is non-overshooting, and

\begin{equation}
\underset{t\rightarrow \infty }{\lim }\left\vert e_{1}\left( t\right)
\right\vert \leq \frac{1}{2\rho k_{1}}\ln \frac{k_{2}+k_{1}e_{2\max }+L_{d}}{%
k_{2}-k_{1}e_{2\max }-L_{d}}\text{, and }\underset{t\rightarrow \infty }{%
\lim }\left\vert e_{2}\left( t\right) \right\vert \leq \frac{1}{\rho }\ln 
\frac{k_{2}+k_{1}e_{2\max }+L_{d}}{k_{2}-k_{1}e_{2\max }-L_{d}}
\end{equation}%
where, $e_{1}\left( t\right) =x_{d}\left( t\right) -x_{1}\left( t\right) $
and $e_{2}\left( t\right) =\dot{x}_{d}\left( t\right) -x_{2}\left( t\right) $%
; $\sup_{t\in \left[ 0,\infty \right) }\left\vert \ddot{x}_{d}\left(
t\right) \right\vert \leq L_{x}<\infty $, and $\sup_{t\in \left[ 0,\infty
\right) }\left\vert \delta (t)\right\vert +\sup_{t\in \left[ 0,\infty
\right) }\left\vert \ddot{x}_{d}\left( t\right) \right\vert \leq
L_{d}<\infty $;$e_{1c}\in \left( 0,k_{2M}-L_{d}\right) $, $e_{2c}\in \left(
e_{1c},\sqrt{\left( k_{2M}-L_{d}\right) e_{1c}}\right] $, and $k_{2M}>L_{d}$
is the up-bound of $k_{2}$ from the system gain limitation;

\begin{equation}
k_{c}>L_{d},\text{ }\rho _{c}\gg \frac{1}{2}\ln \frac{k_{c}+L_{d}}{%
k_{c}-L_{d}}
\end{equation}%
$e_{1}\left( t_{c}\right) $ and $e_{2}\left( t_{c}\right) $ are the initial
values of $e_{1}\left( t\right) $ and $e_{2}\left( t\right) $ respectively
when $\left\vert e_{1}\left( t\right) \right\vert \leq e_{1c}$; and

\begin{eqnarray}
k_{1} &\in &\left\{ 
\begin{array}{l}
\left( 0,\frac{\left\vert e_{2}\left( t_{c}\right) \right\vert }{\left\vert
e_{1}\left( t_{c}\right) \right\vert }\right) ,\text{ if }e_{1}\left(
t_{c}\right) e_{2}\left( t_{c}\right) <0\text{ and }\left\vert e_{1}\left(
t_{c}\right) \right\vert <\left\vert e_{2}\left( t_{c}\right) \right\vert ;
\\ 
\left( \frac{\left\vert e_{2}\left( t_{c}\right) \right\vert }{\left\vert
e_{1}\left( t_{c}\right) \right\vert },\infty \right) ,\text{ if }%
e_{1}\left( t_{c}\right) e_{2}\left( t_{c}\right) <0\text{ and }\left\vert
e_{1}\left( t_{c}\right) \right\vert \geq \left\vert e_{2}\left(
t_{c}\right) \right\vert ; \\ 
\left( 0,\infty \right) ,\text{ if others}%
\end{array}%
\right. \\
k_{2} &>&\left\{ 
\begin{array}{l}
\max \left\{ k_{1}\left\vert e_{2}\left( t_{c}\right) \right\vert +L_{d},%
\frac{e_{2}^{2}\left( t_{c}\right) }{2\left\vert e_{1}\left( t_{c}\right)
\right\vert }+L_{d}\right\} ,\text{ if }e_{1}\left( t_{c}\right) e_{2}\left(
t_{c}\right) <0\text{ and }\left\vert e_{1}\left( t_{c}\right) \right\vert
<\left\vert e_{2}\left( t_{c}\right) \right\vert ; \\ 
\max \left\{ k_{1}\left\vert e_{2}\left( t_{c}\right) \right\vert +L_{d},%
\frac{k_{1}^{2}}{3}\left( \left\vert e_{1}\left( t_{c}\right) \right\vert +%
\sqrt{e_{1}^{2}\left( t_{c}\right) +3\left( \frac{e_{2}\left( t_{c}\right) }{%
k_{1}}\right) ^{2}}\right) +L_{d}\right\} ,\text{ if others}%
\end{array}%
\right.
\end{eqnarray}

\begin{equation}
\rho \gg \max \left\{ \frac{1}{2k_{1}},1\right\} \ln \frac{%
k_{2}+k_{1}e_{2\max }+L_{d}}{k_{2}-k_{1}e_{2\max }-L_{d}}
\end{equation}%
with

\begin{equation}
e_{2\max }=\max \left\{ \left\vert e_{2}\left( t_{c}\right) \right\vert ,%
\frac{k_{1}}{3}\left[ \left\vert e_{1}\left( t_{c}\right) \right\vert +\sqrt{%
e_{1}^{2}\left( t_{c}\right) +3\left( \frac{e_{2}\left( t_{c}\right) }{k_{1}}%
\right) ^{2}}\right] \right\}
\end{equation}%
The proof of Theorem 5.2 is presented in Appendix. $\blacksquare $

\bigskip

\textbf{Remark 5.2} (\emph{Smoothed non-overshooting control (44)})\textbf{:}

1) Non-overshooting of $x_{1}\left( t\right) $ tracking $x_{d}\left(
t\right) $: The sliding mode (24) is the desired stable system of error
form, therefore, no overshoot happens in $e_{1}\left( t\right) =x_{d}\left(
t\right) -x_{1}\left( t\right) $. Thus, $x_{1}\left( t\right) $ tracking $%
x_{d}\left( t\right) $ is non-overshooting.

2) Smoothed $x_{1}\left( t\right) $ and $x_{2}\left( t\right) $: Due to the
use of continuous functions in sliding mode (24), both $e_{1}\left( t\right) 
$ and $e_{2}\left( t\right) $ are smoothed. Therefore, for the control
system, $x_{1}\left( t\right) $ and $x_{2}\left( t\right) $ are smoothed.

3) Smoothed controller $u(t)$: For system (39), the controller $u(t)$ in
(44) is smoothed, and it is fit for the implementation by many actuators.

\bigskip

\textbf{Remark 5.3} (\emph{Rejection of measurement noise}):

For system (39) with the controller (40) or (44), the measurement noise is
rejected on the sliding surface because of its filter-corrector property.
Even when the frequency bands of the variables and noise overlap, the
outputs of sliding surface are smoothed and are accurate.

In fact, suppose noise $n_{1}\left( t\right) $ and $n_{2}\left( t\right) $
exist in the measurements of $x_{1}\left( t\right) $ and $x_{2}\left(
t\right) $, respectively. Then, for the sliding surface, we get

\begin{equation}
\dot{x}_{d}\left( t\right) -x_{2}\left( t\right) -n_{2}\left( t\right) +k_{1}%
\left[ x_{d}\left( t\right) -x_{1}\left( t\right) -n_{1}\left( t\right) %
\right] =0
\end{equation}%
i.e.,

\begin{equation}
e_{2}\left( t\right) +k_{1}e_{1}\left( t\right) =k_{1}n_{1}\left( t\right)
+n_{2}\left( t\right)
\end{equation}

Define the Laplace transforms $E_{1}\left( s\right) =L\left[ e_{1}\left(
t\right) \right] $, $E_{2}\left( s\right) =L\left[ e_{2}\left( t\right) %
\right] $, $N_{1}\left( s\right) =L\left[ n_{1}\left( t\right) \right] $,
and $N_{2}\left( s\right) =L\left[ n_{2}\left( t\right) \right] $. Taking
Laplace transform for (52), we get

\begin{equation}
sE_{1}\left( s\right) +k_{1}E_{1}\left( s\right) =k_{1}N_{1}\left( s\right)
+N_{2}\left( s\right)
\end{equation}%
i.e.,

\begin{equation}
E_{1}\left( s\right) =\frac{k_{1}}{s+k_{1}}\left( N_{1}\left( s\right) +%
\frac{1}{k_{1}}N_{2}\left( s\right) \right)
\end{equation}%
where, $\frac{k_{1}}{s+k_{1}}$ is the form of first-order filter, and $k_{1}$
is the cut-off frequency of the filter. For the filter, the input is the
noise, and the output is the system error. As long as $k_{1}$ is less than
the minimum frequency of the noise, the noise will be rejected. From $k_{1}$
selection conditions, $k_{1}$ can be neither too large nor too small, for
example, $k_{1}=1$ or $k_{1}=2$. Therefore, the noise $N_{1}\left( s\right) +%
\frac{1}{k_{1}}N_{2}\left( s\right) $ is rejected sufficiently, and the
system error is reduced to be small enough.

\bigskip

\textbf{Remark 5.4} (\emph{Parameters regulation of the controller}):

1) The parameter selection conditions (42) and (43) (i.e., (47) and (48))
make the system non-overshooting stable.

2) $k_{c}>L_{d}$. If $\left\vert e_{2}\left( 0\right) \right\vert $ is
large, $k_{c}$ should increase to reduce $\left\vert e_{2}\left( t\right)
\right\vert $ effectively.

2) $k_{1}$ determines the convergence rate of linear convergence law $\dot{e}%
_{1}\left( t\right) =-k_{1}e_{1}\left( t\right) $; from (54), $k_{1}$ also
determines the filtering frequency band of the sliding mode. Therefore, $%
k_{1}$ should not be too small to keep a convergence rate of linear
convergence law; and $k_{1}$ should not be too large to get a suitable
frequency band for noise rejection.

3) $k_{2}$ determines the convergence rate of the second subsystem; $k_{2}$
also keeps the signs of sliding function $\sigma \left( t\right)
=e_{2}\left( t\right) +k_{1}e_{1}\left( t\right) $ and sliding variable $%
e_{1}\left( t\right) $ unchanged for $t\in \left[ t_{c},\infty \right) $;
also, $k_{1}$ affects the selection of $k_{2}$.

4) Because $\left\vert e_{1}\left( t_{c}\right) \right\vert =e_{1c}$ and $%
\left\vert e_{2}\left( t_{c}\right) \right\vert =e_{2c}$ when $t=t_{c}$, the
selection of $e_{1c}\in \left( 0,k_{2M}-L_{d}\right) $ and $e_{2c}\in \left(
e_{1c},\sqrt{\left( k_{2M}-L_{d}\right) e_{1c}}\right] $ makes $\left\vert
e_{1}\left( t_{c}\right) \right\vert $, $\left\vert e_{2}\left( t_{c}\right)
\right\vert $, $\frac{\left\vert e_{2}\left( t_{c}\right) \right\vert }{%
\left\vert e_{1}\left( t_{c}\right) \right\vert }$ and $\frac{%
e_{2}^{2}\left( t_{c}\right) }{2\left\vert e_{1}\left( t_{c}\right)
\right\vert }$ are all bounded. Then, the bounded $k_{1}$ and $k_{2}$ are
determined from (47) and (48).

5) The selection of $\rho $ from (49) affects the smoothness of system
variables and controller output. Also, from (45), $\rho $ affects the
control precision. Therefore, the selection of $\rho $ should balance the
smoothness of variables and controller and the precision of control
performance.

\begin{figure}[tbp]
\centering
\includegraphics[width=4.00in]{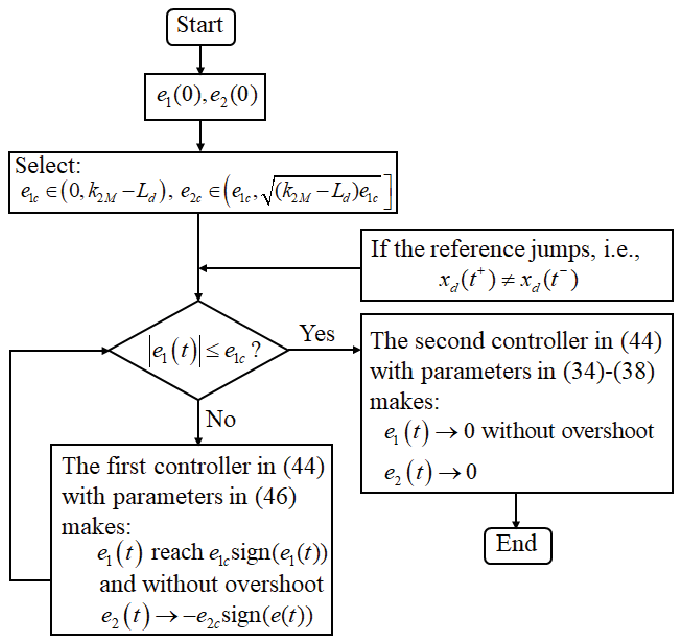}
\caption{Flow chart of non-overshooting controller design.}
\end{figure}

\textbf{Remark 5.5} (\emph{Steps on determination of non-overshooting
controller })

For the system (39), the flow chart of controller design is explained in
Fig. 4. Furthermore, the the steps on controller determination are described
as follows.

Step 1: Measure the system initial states, and get the initial errors $%
e_{1}\left( 0\right) $ and $e_{2}\left( 0\right) $.

Step 2: Select $e_{1c}\in \left( 0,k_{2M}-L_{d}\right) $ and $e_{2c}\in
\left( e_{1c},\sqrt{\left( k_{2M}-L_{d}\right) e_{1c}}\right] $,
respectively, and

\begin{equation}
k_{c}>L_{d},\text{ }\rho _{c}=\rho _{c0}\frac{1}{2}\ln \frac{k_{c}+L_{d}}{%
k_{c}-L_{d}}
\end{equation}%
where, $k_{2M}>L_{d}$ is the up-bound of $k_{2}$ from the system gain
limitation; $\rho _{c0}>1$.

Step 3: Determine the controller parameters $k_{1}$, $k_{2}$ and $\rho $
through the calculations in Equations (34)$\sim $(38). [Note: When the known
reference jumps or suddenly changes, the parameters $k_{1}$, $k_{2}$ and $%
\rho $ are updated]

Step 4: Controller output:

\begin{equation}
u(t)=\left\{ 
\begin{array}{l}
k_{c}\text{tanh}\left[ \rho _{c}\left( e_{2}\left( t\right) +e_{2c}\text{sign%
}\left( e_{1}\left( t\right) \right) \right) \right] -h(t),\text{ if }%
\left\vert e_{1}\left( t\right) \right\vert >e_{1c}; \\ 
k_{2}\text{tanh}\left[ \rho (e_{2}\left( t\right) +k_{1}e_{1}\left( t\right)
)\right] -h(t),\text{ if }\left\vert e_{1}\left( t\right) \right\vert \leq
e_{1c}%
\end{array}%
\right.
\end{equation}

\section{Simulation examples on non-overshooting control of uncertain systems%
}

We use two examples to illustrate the non-overshooting control presented in
Theorems 5.1 and 5.2, respectively. Consider the uncertain system:

\begin{subequations}
\begin{eqnarray*}
\dot{x}_{1}\left( t\right) &=&x_{2}\left( t\right) \\
\dot{x}_{2}\left( t\right) &=&h(t)+u(t)-\delta \left( t\right)
\end{eqnarray*}

where, $h(t)=5x_{1}^{\frac{1}{3}}\sin \left( 0.5t\right) $, and the unknown
disturbance or uncertainty $\delta \left( t\right) =1+0.3\sin (0.3t)\sin
(1.6t)$.

The initial conditions of states: $x_{1}\left( 0\right) =10$, $x_{2}\left(
0\right) =-1$

The reference: $x_{d}\left( t\right) =2+0.5\sin (0.8t)$. Therefore, we know
that $\dot{x}_{d}\left( t\right) =0.4\cos (0.8t)$ and $\ddot{x}_{d}\left(
t\right) =-0.32\sin (0.8t)$; $x_{d}\left( 0\right) =2$, and $\dot{x}%
_{d}\left( 0\right) =0.4$.

The upper bound of disturbance/uncertainty: $\sup_{t\in \left[ 0,\infty
\right) }\left\vert \delta \left( t\right) \right\vert +\sup_{t\in \left[
0,\infty \right) }\left\vert \ddot{x}_{d}\left( t\right) \right\vert
=1.3+0.32=1.62$, and we can select $L_{d}=1.62$.

Define system errors $e_{1}\left( t\right) =x_{d}\left( t\right)
-x_{1}\left( t\right) $ and $e_{2}\left( t\right) =\dot{x}_{d}\left(
t\right) -x_{2}\left( t\right) $. Then, the error system is:

\end{subequations}
\begin{subequations}
\begin{eqnarray*}
\dot{e}_{1}\left( t\right) &=&e_{2}\left( t\right) \\
\dot{e}_{2}\left( t\right) &=&-h(t)-u(t)+\ddot{x}_{d}\left( t\right) +\delta
\left( t\right)
\end{eqnarray*}%
and the initial errors are $e_{1}\left( 0\right) =x_{d}\left( 0\right)
-x_{1}\left( 0\right) =-8$, and $e_{2}\left( 0\right) =\dot{x}_{d}\left(
0\right) -x_{2}\left( 0\right) =1.4$.

Suppose the system gain $k_{2}\leq k_{2M}=10$.

\bigskip

\textbf{Example 6.1: }\emph{(Non-overshooting control from Theorem 5.1):}

\emph{1) Selection of the desired stable error system}

The sliding mode (18) is selected as the desired stable error system, and

\end{subequations}
\begin{eqnarray*}
\dot{e}_{1}\left( t\right) &=&e_{2}\left( t\right) \\
\dot{e}_{2}\left( t\right) &=&\left\{ 
\begin{array}{l}
-k_{c}\text{sign}\left[ e_{2}\left( t\right) +e_{2c}\text{sign}\left(
e_{1}\left( t\right) \right) \right] +\ddot{x}_{d}\left( t\right) +\delta
\left( t\right) ,\text{ if }\left\vert e_{1}\left( t\right) \right\vert
>e_{1c}; \\ 
-k_{2}\text{sign}\left[ e_{2}\left( t\right) +k_{1}e_{1}\left( t\right) %
\right] +\ddot{x}_{d}\left( t\right) +\delta \left( t\right) ,\text{ if }%
\left\vert e_{1}\left( t\right) \right\vert \leq e_{1c}%
\end{array}%
\right.
\end{eqnarray*}

\emph{2) Determination of parameters }$k_{1}$\emph{\ and }$k_{2}$

From $e_{1c}\in \left( 0,k_{2M}-L_{d}\right) =\left( 0,8.38\right) $, we
select $e_{1c}=1$. Then, we get $e_{2c}\in \left( e_{1c},\sqrt{\left(
k_{2M}-L_{d}\right) e_{1c}}\right] =\left( 1,\sqrt{\left( 10-1.62\right)
\times 1}\right] =\left( 1,2.9\right] $. We select $e_{2c}=2$. Select $%
k_{c}=2.5>L_{d}$.

According to the parameter determination steps (34)$\sim $(38), we get

\begin{equation*}
k_{1}=\left\{ 
\begin{array}{l}
0.5\frac{\left\vert e_{2}\left( t_{c}\right) \right\vert }{\left\vert
e_{1}\left( t_{c}\right) \right\vert },\text{ if }e_{1}\left( t_{c}\right)
e_{2}\left( t_{c}\right) <0\text{ and }\left\vert e_{1}\left( t_{c}\right)
\right\vert <\left\vert e_{2}\left( t_{c}\right) \right\vert ; \\ 
2.3\frac{\left\vert e_{2}\left( t_{c}\right) \right\vert }{\left\vert
e_{1}\left( t_{c}\right) \right\vert },\text{ if }e_{1}\left( t_{c}\right)
e_{2}\left( t_{c}\right) <0\text{ and }\left\vert e_{1}\left( t_{c}\right)
\right\vert \geq \left\vert e_{2}\left( t_{c}\right) \right\vert ; \\ 
2\in \left( 0,\infty \right) ,\text{ if others}%
\end{array}%
\right.
\end{equation*}

\emph{Adjustment of }$k_{1}$\emph{:} $k_{1}=1$ if the calculated $k_{1}\in
\left( 0,1\right) $; and

\begin{equation*}
k_{2}=\left\{ 
\begin{array}{l}
1.5\max \left\{ k_{1}\left\vert e_{2}\left( t_{c}\right) \right\vert +L_{d},%
\frac{e_{2}^{2}\left( t_{c}\right) }{2\left\vert e_{1}\left( t_{c}\right)
\right\vert }+L_{d}\right\} ,\text{ if }e_{1}\left( t_{c}\right) e_{2}\left(
t_{c}\right) <0\text{ and }\left\vert e_{1}\left( t_{c}\right) \right\vert
<\left\vert e_{2}\left( t_{c}\right) \right\vert ; \\ 
1.5\max \left\{ k_{1}\left\vert e_{2}\left( t_{c}\right) \right\vert +L_{d},%
\frac{k_{1}^{2}}{3}\left( \left\vert e_{1}\left( t_{c}\right) \right\vert +%
\sqrt{e_{1}^{2}\left( t_{c}\right) +3\left( \frac{e_{2}\left( t_{c}\right) }{%
k_{1}}\right) ^{2}}\right) +L_{d}\right\} ,\text{ if others}%
\end{array}%
\right.
\end{equation*}%
where, $e_{1}\left( t_{c}\right) $ and $e_{2}\left( t_{c}\right) $ are the
initial values of $e_{1}\left( t\right) $ and $e_{2}\left( t\right) $
respectively when $\left\vert e_{1}\left( t\right) \right\vert \leq e_{1c}$.
From the algorithm calculation for the above equations, we can read $k_{1}=1$
and $k_{2}=5.44$.

\emph{3) Controller design}

According to the controller (40), we get

\begin{equation*}
u(t)=\left\{ 
\begin{array}{l}
k_{c}\text{sign}\left[ e_{2}\left( t\right) +e_{2c}\text{sign}\left(
e_{1}\left( t\right) \right) \right] -h(t),\text{ if }\left\vert e_{1}\left(
t\right) \right\vert >e_{1c} \\ 
k_{2}\text{sign}\left[ e_{2}\left( t\right) +k_{1}e_{1}\left( t\right) %
\right] -h(t),\text{ if }\left\vert e_{1}\left( t\right) \right\vert \leq
e_{1c}%
\end{array}%
\right.
\end{equation*}%
where, $e_{1c}=1$, $e_{2c}=2$, $k_{c}=2.5$, $k_{1}=1$, $k_{2}=5.44$, and $%
h\left( t\right) =5x_{1}^{\frac{1}{3}}\sin \left( 0.5t\right) $.

\begin{figure}[tbp]
\begin{center}
\includegraphics[width=3.5in]{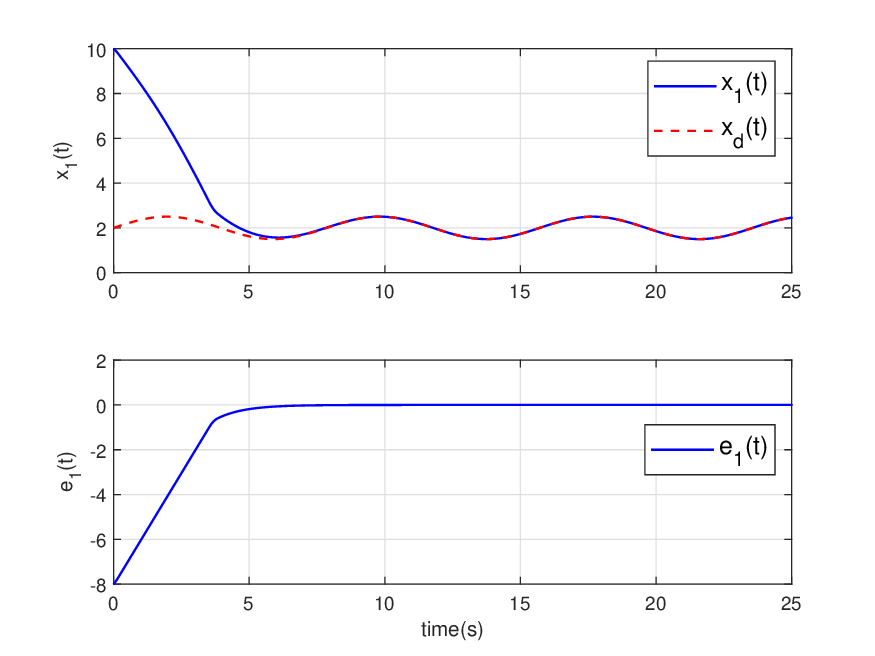}\\[0pt]
{\small (a)}\\[0pt]
\includegraphics[width=3.5in]{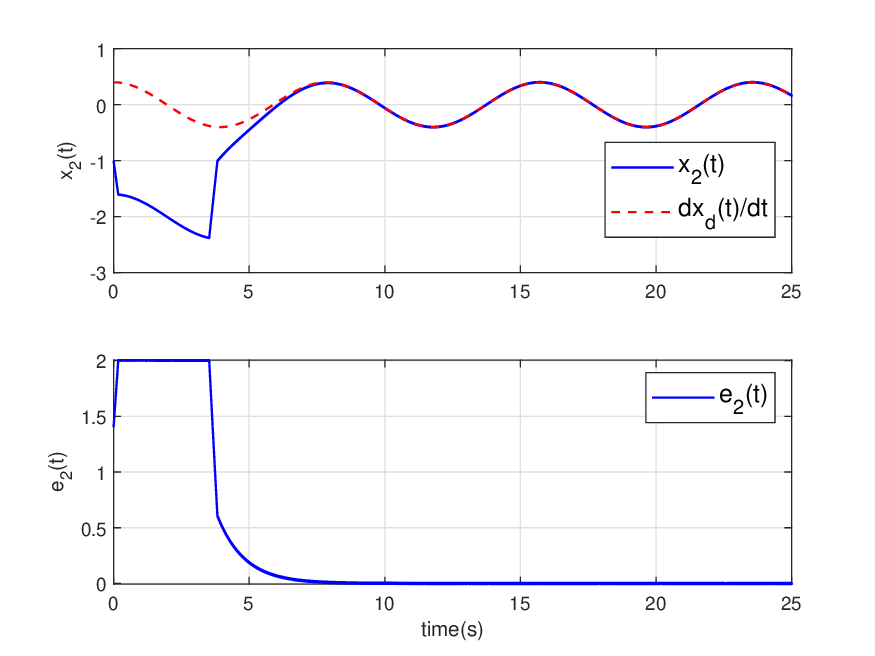}\\[0pt]
{\small (b)}\\[0pt]
\includegraphics[width=3.5in]{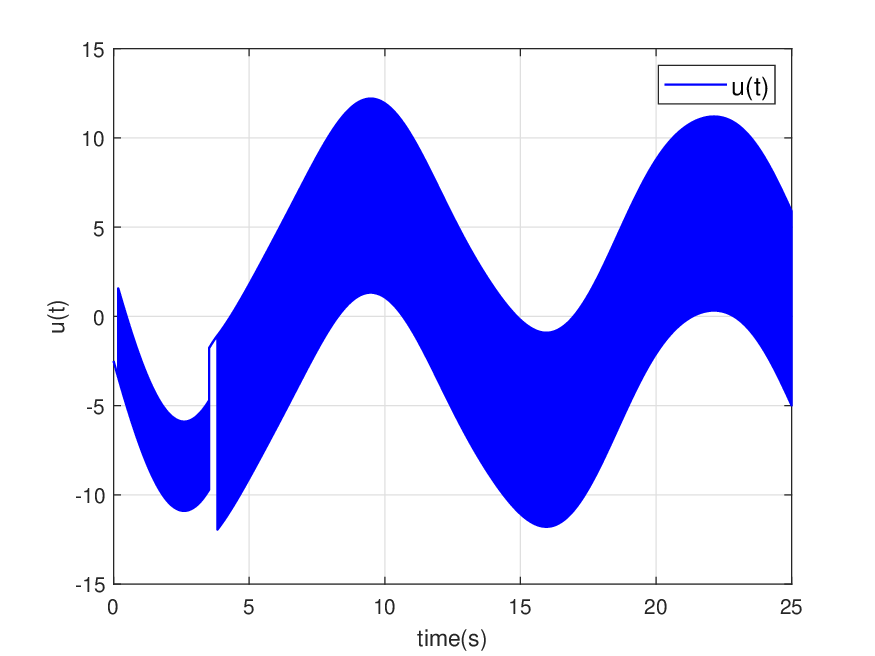}\\[0pt]
{\small (c)}
\end{center}
\caption{Example 6.1 Non-overshooting sliding mode control. (a) $x_{1}$. (b) 
$x_{2}$. (c) Controller $u\left( t\right) $.}
\end{figure}

\begin{figure}[tbp]
\begin{center}
\includegraphics[width=3.5in]{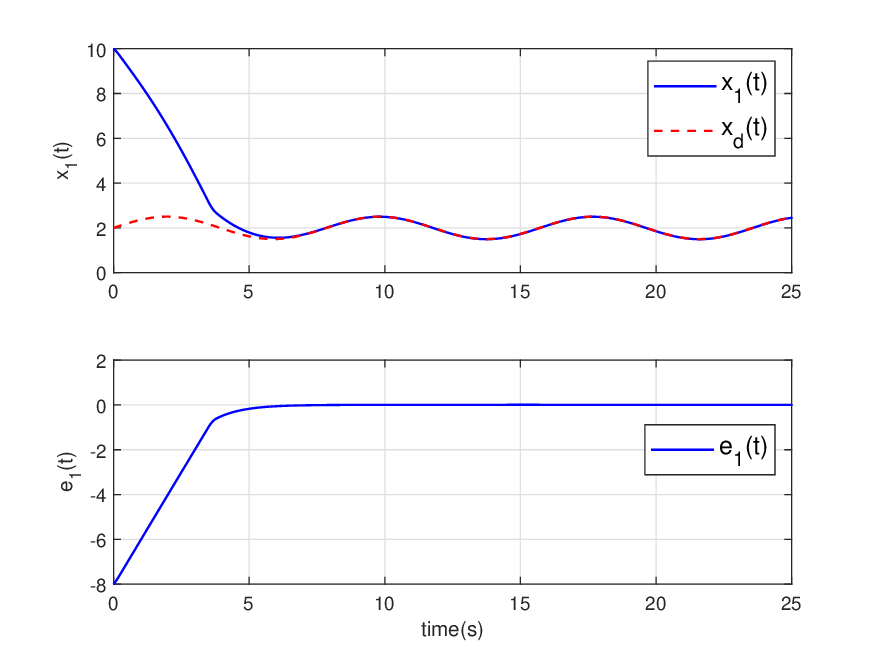}\\[0pt]
{\small (a)}\\[0pt]
\includegraphics[width=3.5in]{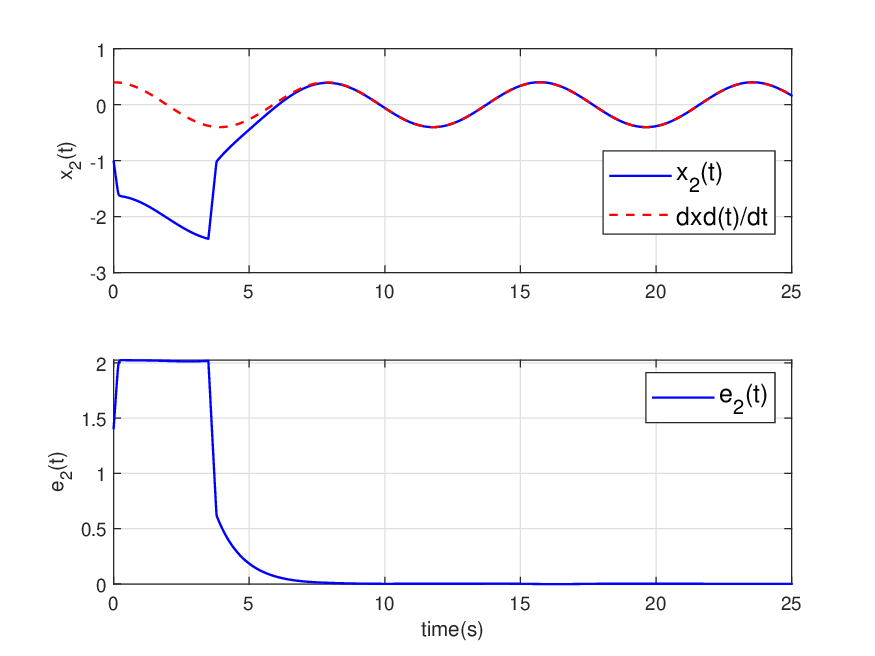}\\[0pt]
{\small (b)}\\[0pt]
\includegraphics[width=3.5in]{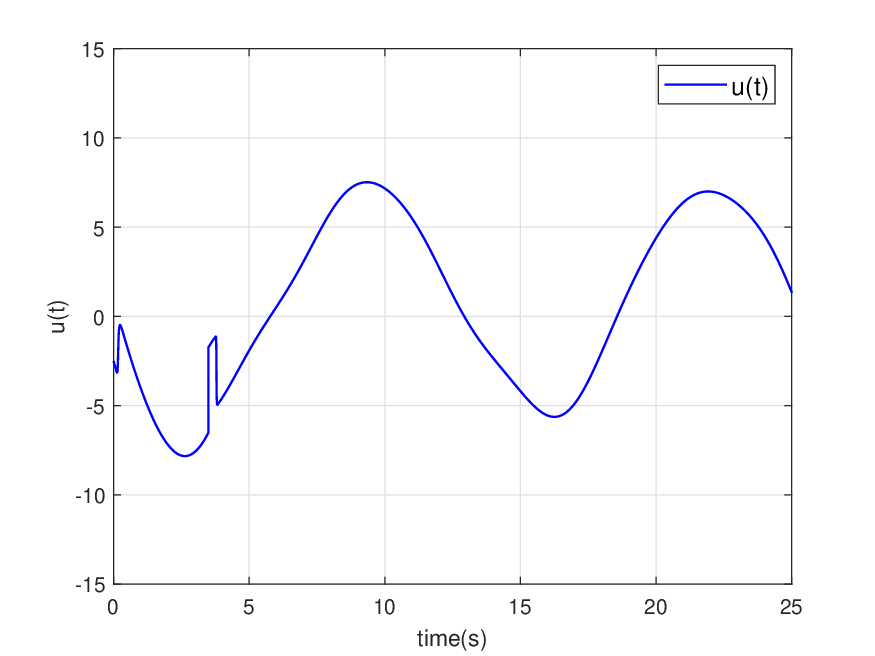}\\[0pt]
{\small (c)}
\end{center}
\caption{Example 6.2 Smoothed non-overshooting sliding mode control. (a) $%
x_{1}$. (b) $x_{2}$. (c) Controller $u\left( t\right) $.}
\end{figure}

\emph{4) Analysis of simulation results}

The control performance based on the ideal sliding mode is presented in Fig.
5. Figure 5(a) describes the system variable $x_{1}$ tracking the reference $%
x_{d}\left( t\right) $. $x_{1}$ tracking $x_{d}\left( t\right) $ is smoothed
and non-overshooting, even the unknown time-varying disturbance exists, and
the reference is also time variant. Figure 5(b) presents the variable $x_{2}$
convergence to the reference derivative $\dot{x}_{d}\left( t\right) $. At
the beginning, $x_{2}$ increases to speed up the finite-time convergence,
then it makes $x_{1}$ in the linear convergence law $\dot{e}_{1}\left(
t\right) =-k_{1}e_{1}\left( t\right) $, and $\underset{t\rightarrow \infty }{%
\lim }x_{1}\left( t\right) =x_{d}\left( t\right) $ and $\underset{%
t\rightarrow \infty }{\lim }x_{2}\left( t\right) =\dot{x}_{d}\left( t\right) 
$. Figure 5(c) shows the controller $u\left( t\right) $ output. Even the
controller can make the system stable and non-overshooting, chattering
happens in the controller output, and it may increase actuator trembling.

\bigskip

\textbf{Example 6.2: }\emph{(Smoothed non-overshooting control from Theorem
5.2):}

\emph{1) Selection of the desired stable error system}

The smoothed sliding mode (24) is selected as the desired stable error
system, and

\begin{eqnarray*}
\dot{e}_{1}\left( t\right) &=&e_{2}\left( t\right) \\
\dot{e}_{2}\left( t\right) &=&\left\{ 
\begin{array}{l}
-k_{c}\text{tanh}\left[ \rho _{c}\left( e_{2}\left( t\right) +e_{2c}\text{%
sign}\left( e_{1}\left( t\right) \right) \right) \right] +\ddot{x}_{d}\left(
t\right) +\delta \left( t\right) ,\text{ if }\left\vert e_{1}\left( t\right)
\right\vert >e_{1c}; \\ 
-k_{2}\text{tanh}\left[ \rho (e_{2}\left( t\right) +k_{1}e_{1}\left(
t\right) )\right] +\ddot{x}_{d}\left( t\right) +\delta \left( t\right) ,%
\text{ if }\left\vert e_{1}\left( t\right) \right\vert \leq e_{1c}%
\end{array}%
\right.
\end{eqnarray*}

\emph{2) Determination of parameters }$k_{1}$\emph{, }$k_{2}$\emph{\ and }$%
\rho $ \emph{according to the parameter determination steps (34)}$\sim $%
\emph{(38):}

Select $e_{1c}=1$, $e_{2c}=2$, and $k_{c}=2.5$. Then, we get $\rho _{c}=20%
\frac{1}{2}\ln \frac{k_{c}+L_{d}}{k_{c}-L_{d}}=15.44$.

Firstly, for determination of $k_{1}$\ and $k_{2}$, we use the same
algorithm steps to example 6.1, and we can read $k_{1}=1.01$ and $k_{2}=5.49$%
.

Secondly, for $\rho $, we have

\begin{equation*}
e_{2\max }=\max \left\{ \left\vert e_{2}\left( t_{c}\right) \right\vert ,%
\frac{k_{1}}{3}\left[ \left\vert e_{1}\left( t_{c}\right) \right\vert +\sqrt{%
e_{1}^{2}\left( t_{c}\right) +3\left( \frac{e_{2}\left( t_{c}\right) }{k_{1}}%
\right) ^{2}}\right] \right\}
\end{equation*}

\begin{equation*}
\rho =20\max \left\{ \frac{1}{2k_{1}},1\right\} \ln \frac{%
k_{2}+k_{1}e_{2\max }+L_{d}}{k_{2}-k_{1}e_{2\max }-L_{d}}
\end{equation*}

From the algorithm calculation for the above equations, we can read $\rho
=32.19$.

\emph{3) Controller design}

According to the smoothed controller (44), we get

\begin{equation*}
u(t)=\left\{ 
\begin{array}{l}
k_{c}\text{tanh}\left[ \rho _{c}\left( e_{2}\left( t\right) +e_{2c}\text{sign%
}\left( e_{1}\left( t\right) \right) \right) \right] -h(t),\text{ if }%
\left\vert e_{1}\left( t\right) \right\vert >e_{1c}; \\ 
k_{2}\text{tanh}\left[ \rho (e_{2}\left( t\right) +k_{1}e_{1}\left( t\right)
)\right] -h(t),\text{ if }\left\vert e_{1}\left( t\right) \right\vert \leq
e_{1c}%
\end{array}%
\right.
\end{equation*}%
where, $e_{1c}=1$, $e_{2c}=2$, $k_{c}=2.5$, $\rho _{c}=15.44$, $k_{1}=1.01$, 
$k_{2}=5.33$, $\rho =32.19$, and $h\left( t\right) =5x_{1}^{\frac{1}{3}}\sin
\left( 0.5t\right) $.

\emph{4) Analysis of simulation results}

Figure 6 presents the control performance based on the smoothed sliding
mode. Figure 6(a) describes $x_{1}$ tracking the reference $x_{d}\left(
t\right) $, and Figure 6(b) presents $x_{2}$ convergence to the reference
derivative $\dot{x}_{d}\left( t\right) $. Even time-varying disturbance
exists, and time-variant reference is required, $x_{1}$ tracking $%
x_{d}\left( t\right) $ is smoothed and non-overshooting. Comparing to $x_{2}$
in Example 6.1, $x_{2}$ in Example 6.2 is smoother. Figure 6(c) shows the
smoothed controller $u\left( t\right) $ output. The smoothed controller
output is beneficial for actuator implementation, and it reduce the actuator
trembling.

\section{UAV control application}

A quadrotor UAV prototype is used [30], which is shown in Fig. 7, and the
forces and torques of UAV are described. The system parameters are
introduced in Table I.

\begin{figure}[tbp]
\centering
\includegraphics[width=3.00in]{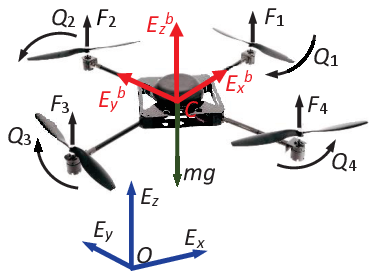}
\caption{Forces and torques in UAV [30].}
\end{figure}

\begin{center}
{\small Table I. UAV Parameters [30]}

\begin{tabular}{lll}
\hline
Symbol & Quantity & Value \\ \hline
$m$ & mass of UAV & $2.01$kg \\ 
$g$ & gravity acceleration & $9.81$m/s$^{2}$ \\ 
$l$ & rotor distance to gravity center & $0.2$m \\ 
$J_{\phi }$ & moment of inertia about roll & $0.25$kg$\cdot $m$^{2}$ \\ 
$J_{\theta }$ & moment of inertia about pitch & $0.25$kg$\cdot $m$^{2}$ \\ 
$J_{\psi }$ & moment of inertia about yaw & $0.5$kg$\cdot $m$^{2}$ \\ 
$b$ & rotor force coefficient & $2.923\times 10^{-3}$ \\ 
$k$ & rotor torque coefficient & $5\times 10^{-4}$ \\ \hline
\end{tabular}
\end{center}

\emph{7.1 Model of UAV flight dynamics [30]}

The inertial and fuselage frames are denoted by $\Xi _{g}=\left(
E_{x},E_{y},E_{z}\right) $ and $\Xi _{b}=\left(
E_{x}^{b},E_{y}^{b},E_{z}^{b}\right) $, respectively; $\psi $, $\theta $ and 
$\phi $ are the yaw, pitch and roll angles, respectively. $F_{i}=b\omega
_{i}^{2}$ is the thrust force by rotor $i$, and its reactive torque is $%
Q_{i}=k\omega _{i}^{2}$. The sum of the four rotor thrusts is $%
F=\sum\limits_{{i=1}}^{{4}}F_{i}$. The motion equations of the UAV flight
dynamics can be expressed by

\begin{eqnarray}
\dot{x}_{\ast 1} &=&x_{\ast 2}  \notag \\
\dot{x}_{\ast 2} &=&h_{\ast }(t)+\bar{u}_{\ast }\left( t\right) +\delta
_{\ast }(t)
\end{eqnarray}%
where, $\ast =x,y,z,\psi ,\theta ,\phi $; $x_{x1}=x$, $x_{y1}=y$, $x_{z1}=z$%
, $x_{\psi 1}=\psi $, $x_{\theta 1}=\theta $, $x_{\phi 1}=\phi $; $%
h_{x}(t)=0 $, $h_{y}(t)=0$, $h_{z}(t)=-g$, $h_{\psi }(t)=0$, $h_{\theta
}(t)=0$, $h_{\phi }(t)=0$; $\delta _{x}(t)=m^{-1}(-k_{x}\dot{x}+\Delta _{x})$%
; $\delta _{y}(t)=m^{-1}(-k_{y}\dot{y}+\Delta _{y})$; $\delta
_{z}(t)=m^{-1}(-k_{z}\dot{z}+\Delta _{z})$; $\delta _{\psi }(t)=J_{\psi
}^{-1}(-k_{\psi }\dot{\psi}+\Delta _{\psi })$; $\delta _{\theta
}(t)=J_{\theta }^{-1}(-lk_{\theta }\dot{\theta}+\Delta _{\theta })$; $\delta
_{\phi }(t)=J_{\phi }^{-1}(-lk_{\phi }\dot{\phi}+\Delta _{\phi })$; $k_{x}$, 
$k_{y}$, $k_{z}$, $k_{\psi }$, $k_{\theta }$ and $k_{\phi }$ are the unknown
drag coefficients; $(\Delta _{x},\Delta _{y},\Delta _{z})$ and $(\Delta
_{\psi },\Delta _{\theta },\Delta _{\phi })$ are the uncertainties in
position and attitude dynamics, respectively; $J=diag\{J_{\psi },J_{\theta
},J_{\phi }\}$ is the matrix of three-axial moment of inertias; $c_{\theta }$
and $s_{\theta }$ are expressed for $\cos \theta $ and $\sin \theta $,
respectively; and

\begin{eqnarray}
\bar{u}_{x}\left( t\right) &=&u_{x}\left( t\right) /m=(c_{\psi }s_{\theta
}c_{\phi }+s_{\psi }s_{\phi })F/m  \notag \\
\bar{u}_{y}\left( t\right) &=&u_{y}\left( t\right) /m=(s_{\psi }s_{\theta
}c_{\phi }-c_{\psi }s_{\phi })F/m  \notag \\
\bar{u}_{z}\left( t\right) &=&u_{z}\left( t\right) /m=c_{\theta }c_{\phi }F/m
\notag \\
\bar{u}_{\psi }\left( t\right) &=&u_{z}\left( t\right) /J_{\psi }=\frac{k}{b}%
\left( \sum\limits_{{i=1}}^{{4}}(-1)^{i+1}F_{i}\right) /J_{\psi }  \notag \\
\bar{u}_{\theta }\left( t\right) &=&u_{\theta }\left( t\right) /J_{\theta
}=(F_{3}-F_{1})l/J_{\theta }  \notag \\
\bar{u}_{\phi }\left( t\right) &=&u_{\phi }\left( t\right) /J_{\phi
}=(F_{2}-F_{4})l/J_{\phi }
\end{eqnarray}

\bigskip

\emph{7.2 Measurements}

A Vicon system provides position and velocity, and a Doppler radar sensor
measures height and vertical velocity. An IMU gives the attitude angle and
angular velocity. The sensor outputs are:

\begin{equation}
y_{\ast 1}\left( t\right) =x_{\ast 1};\text{ }y_{\ast 2}\left( t\right)
=x_{\ast 2}
\end{equation}%
where, $\ast =x,y,z,\psi ,\theta ,\phi $.

\emph{7.3 Controller design}

In this section, the control laws are derived for UAV trajectory tracking
and attitude stabilization.

\emph{1) Error systems}

The control laws are designed to stabilize the UAV flight. For the desired
trajectory ($x_{d}\left( t\right) $, $y_{d}\left( t\right) $, $z_{d}\left(
t\right) $) and attitude angle ($\psi _{d}\left( t\right) $, $\theta
_{d}\left( t\right) $, $\phi _{d}\left( t\right) $), the error systems of
position and attitude dynamics can be expressed, respectively, by

\begin{eqnarray}
\dot{e}_{\ast 1}\left( t\right) &=&\dot{e}_{\ast 2}\left( t\right)  \notag \\
\dot{e}_{\ast 2}\left( t\right) &=&-h_{\ast }(t)-\bar{u}_{\ast }\left(
t\right) +\ddot{\ast}_{d}\left( t\right) -\delta _{\ast }(t)
\end{eqnarray}%
where, $e_{\ast 1}\left( t\right) =\ast _{d}\left( t\right) -x_{\ast 1}$, $%
e_{\ast 2}\left( t\right) =\dot{\ast}_{d}\left( t\right) -x_{\ast 2}$; $\ast
=x,y,z,\psi ,\theta ,\phi $.

\bigskip

\emph{2) Controller design}

From (24), we select the smoothed non-overshooting sliding mode as the
desired stable, i.e.,

\begin{eqnarray}
\dot{e}_{\ast 1}\left( t\right) &=&e_{\ast 2}\left( t\right)  \notag \\
\dot{e}_{\ast 2}\left( t\right) &=&\left\{ 
\begin{array}{l}
-k_{\ast c}\text{tanh}\left[ \rho _{\ast c}\left( e_{\ast 2}\left( t\right)
+e_{\ast 2c}\text{sign}\left( e_{\ast 1}\left( t\right) \right) \right) %
\right] +\ddot{\ast}_{d}\left( t\right) -\delta _{\ast }(t),\text{ if }%
\left\vert e_{\ast 1}\left( t\right) \right\vert >e_{\ast 1c}; \\ 
-k_{\ast 2}\text{tanh}\left[ \rho _{\ast }(e_{\ast 2}\left( t\right)
+k_{\ast 1}e_{\ast 1}\left( t\right) )\right] +\ddot{\ast}_{d}\left(
t\right) -\delta _{\ast }(t),\text{ if }\left\vert e_{\ast 1}\left( t\right)
\right\vert \leq e_{\ast 1c}%
\end{array}%
\right.
\end{eqnarray}%
In order to turn the error system (62) into the desired stable sliding mode
(63), we select

\begin{eqnarray}
&&-h_{\ast }(t)-\bar{u}_{\ast }\left( t\right) +\ddot{\ast}_{d}\left(
t\right) -\delta _{\ast }(t)  \notag \\
&=&\left\{ 
\begin{array}{l}
-k_{\ast c}\text{tanh}\left[ \rho _{\ast c}\left( e_{\ast 2}\left( t\right)
+e_{\ast 2c}\text{sign}\left( e_{\ast 1}\left( t\right) \right) \right) %
\right] +\ddot{\ast}_{d}\left( t\right) -\delta _{\ast }(t),\text{ if }%
\left\vert e_{\ast 1}\left( t\right) \right\vert >e_{\ast 1c}; \\ 
-k_{\ast 2}\text{tanh}\left[ \rho _{\ast }(e_{\ast 2}\left( t\right)
+k_{\ast 1}e_{\ast 1}\left( t\right) )\right] +\ddot{\ast}_{d}\left(
t\right) -\delta _{\ast }(t),\text{ if }\left\vert e_{\ast 1}\left( t\right)
\right\vert \leq e_{\ast 1c}%
\end{array}%
\right.
\end{eqnarray}%
Therefore, we get the controller as follows:

\begin{equation}
\bar{u}_{\ast }\left( t\right) =\left\{ 
\begin{array}{l}
k_{\ast c}\text{tanh}\left[ \rho _{\ast c}\left( e_{\ast 2}\left( t\right)
+e_{\ast 2c}\text{sign}\left( e_{\ast 1}\left( t\right) \right) \right) %
\right] -h_{\ast }(t),\text{ if }\left\vert e_{\ast 1}\left( t\right)
\right\vert >e_{\ast 1c}; \\ 
k_{\ast 2}\text{tanh}\left[ \rho _{\ast }(e_{\ast 2}\left( t\right) +k_{\ast
1}e_{\ast 1}\left( t\right) )\right] -h_{\ast }(t),\text{ if }\left\vert
e_{\ast 1}\left( t\right) \right\vert \leq e_{\ast 1c}%
\end{array}%
\right.
\end{equation}%
where, $\ast =x,y,z,\psi ,\theta ,\phi $. Thus, for the UAV system (59),
when the controller (65) is selected, the system is stable, and the system
variables $x_{\ast 1}$ (where, $\ast =x$, $y$, $z$, $\psi $, $\theta $, $%
\phi $) are non-overshooting.

\section{Experiment on UAV non-overshooting control}

\begin{figure}[tbp]
\centering
\includegraphics[width=4.50in]{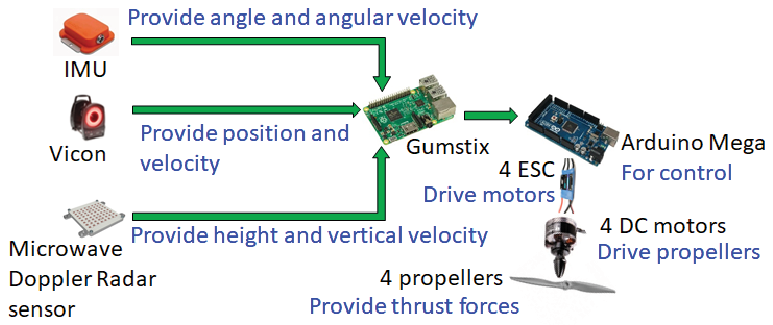}
\caption{Control system hardware.}
\end{figure}

In this section, an experiment on a quadrotor UAV is presented to
demonstrate the proposed non-overshooting control in practice. The UAV
prototype shown in Figure 7 is used for the flight test. The flight control
system implementation on the hardware is shown in Figure 8, whose elements
include: A Gumstix and Arduino Mega 2560 (16MHz) are selected as the driven
boards; Gumstix is to collect data from measurements; Arduino Mega is to run
control algorithm, which has multiple PWM output channels; A XsensMTI AHRS
(10 kHz) provides the 3-axial attitude angles and the angular velocities. A
microwave Doppler radar sensor (24GHz) is to detect the height and its
vertical velocity. The Vicon system provides position and velocity.

\begin{figure}[tbp]
\begin{center}
\includegraphics[width=3.75in]{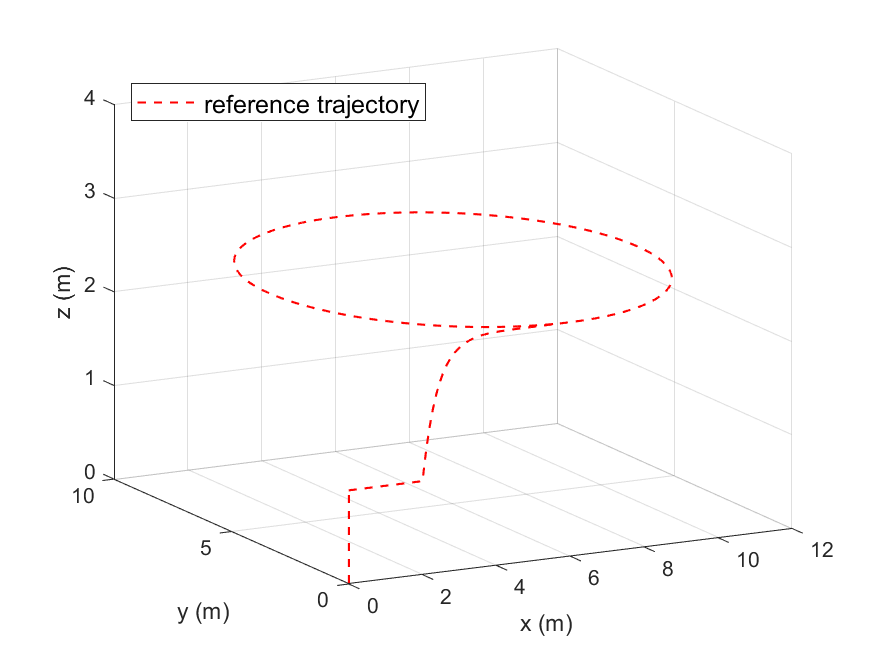}\\[0pt]
{\small (a)}\\[0pt]
\includegraphics[width=3.75in]{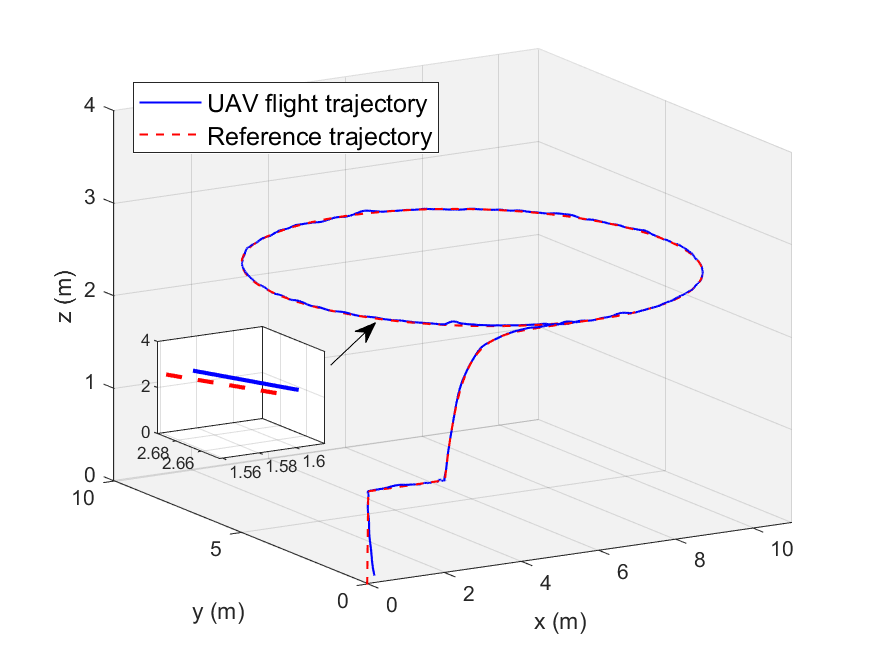}\\[0pt]
{\small (b)}
\end{center}
\caption{UAV 3D flight trajectories. (a) Reference trajectory. (b) Flight
trajectory comparison.}
\end{figure}

\emph{\ Flight reference trajectory:} The UAV reference trajectory includes:
1) take off vertically and hover at the height of 1m; 2) then cruise along a
horizontal line and keep the height; 3) then climb and cruise in a circle
with the radius 5m and the height 2.5m. The 3D reference trajectory is shown
in Fig. 9(a).

In the experiment, considering the disturbance (e.g., the crosswind from a
swinging electric fan) and the modelling uncertainty in the UAV flight
dynamics, the UAV is controlled to track the reference trajectory. The
position and velocity are obtained from the Vicon, the height and its
vertical velocity are detected by the microwave Doppler radar sensor, and
the attitude angle and the angular velocity are measured by the IMU. The
controller (65) drives the UAV to track the reference trajectory.

\begin{figure}[tbp]
\begin{center}
\includegraphics[width=3.5in]{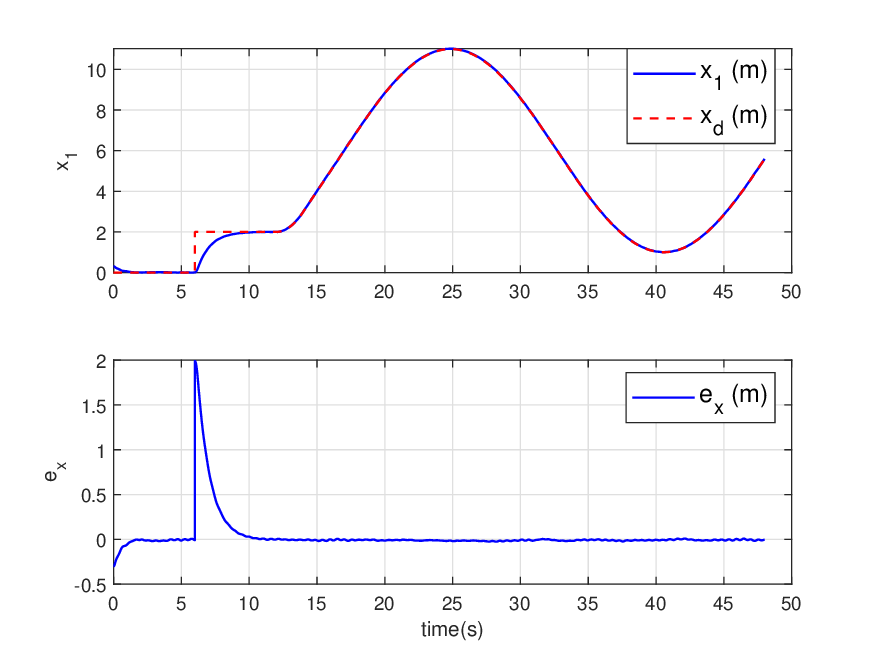}\\[0pt]
{\small (a)}\\[0pt]
\includegraphics[width=3.5in]{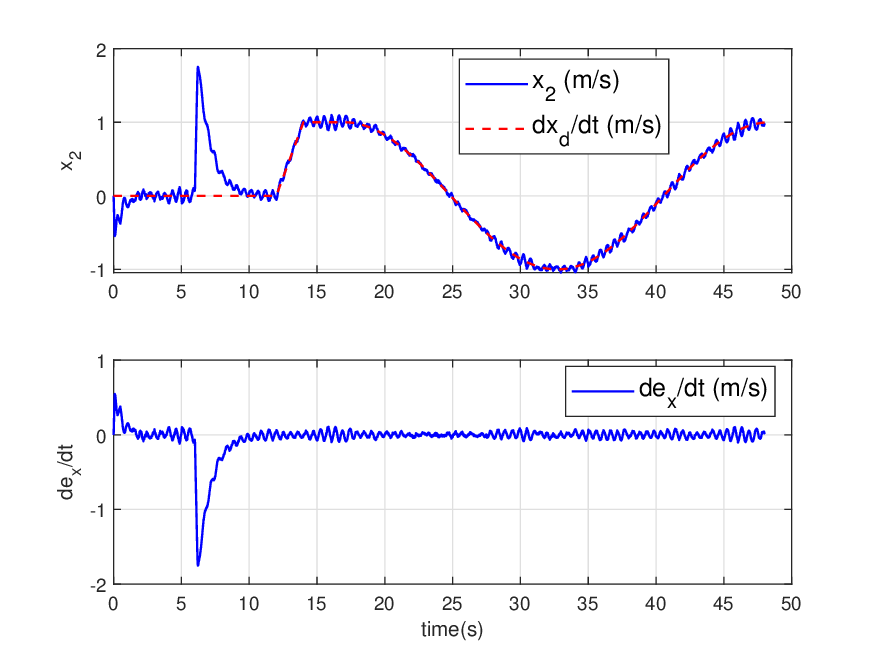}\\[0pt]
{\small (b)}\\[0pt]
\includegraphics[width=3.5in]{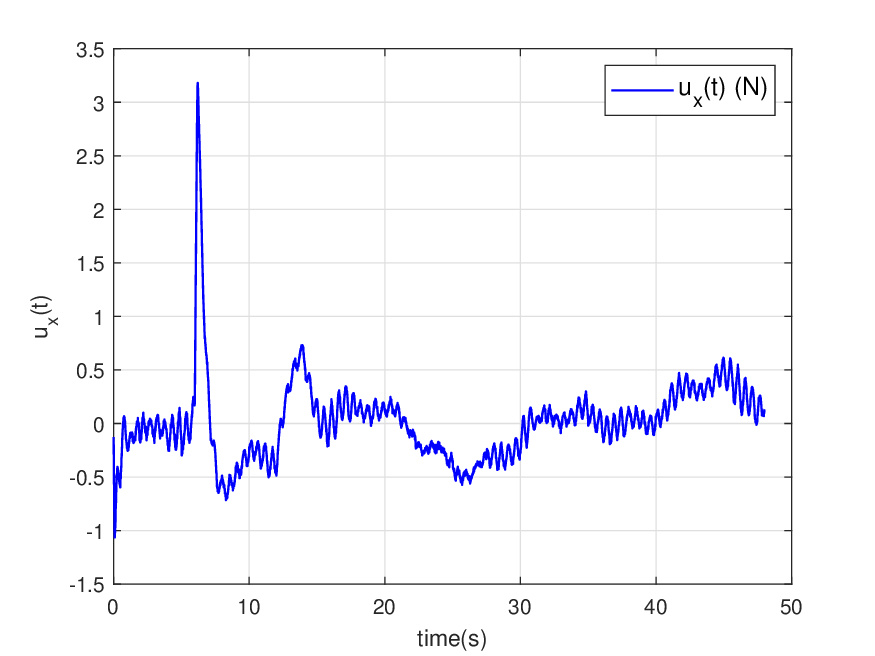}\\[0pt]
{\small (c)}
\end{center}
\caption{Control performance in $x$-direction. (a) $x_{1}$. (b) $x_{2}$. (c)
Controller $u_{x}\left( t\right) $.}
\end{figure}

\begin{figure}[tbp]
\begin{center}
\includegraphics[width=3.5in]{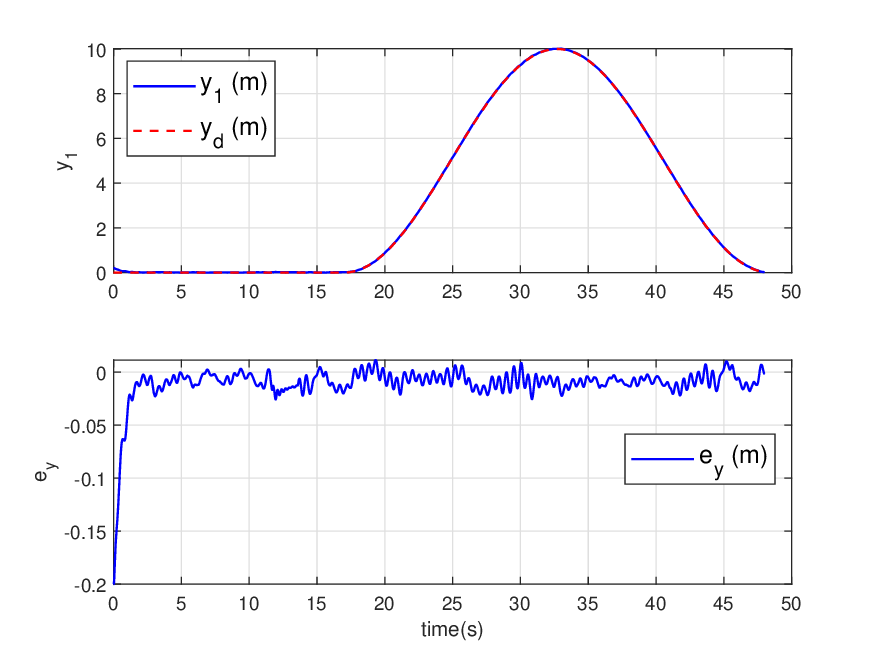}\\[0pt]
{\small (a)}\\[0pt]
\includegraphics[width=3.5in]{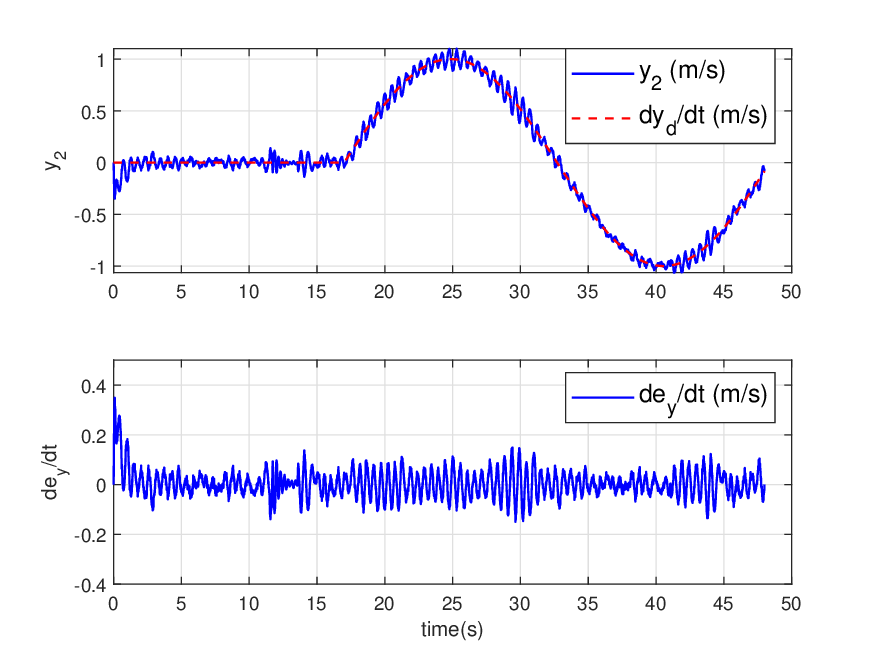}\\[0pt]
{\small (b)}\\[0pt]
\includegraphics[width=3.5in]{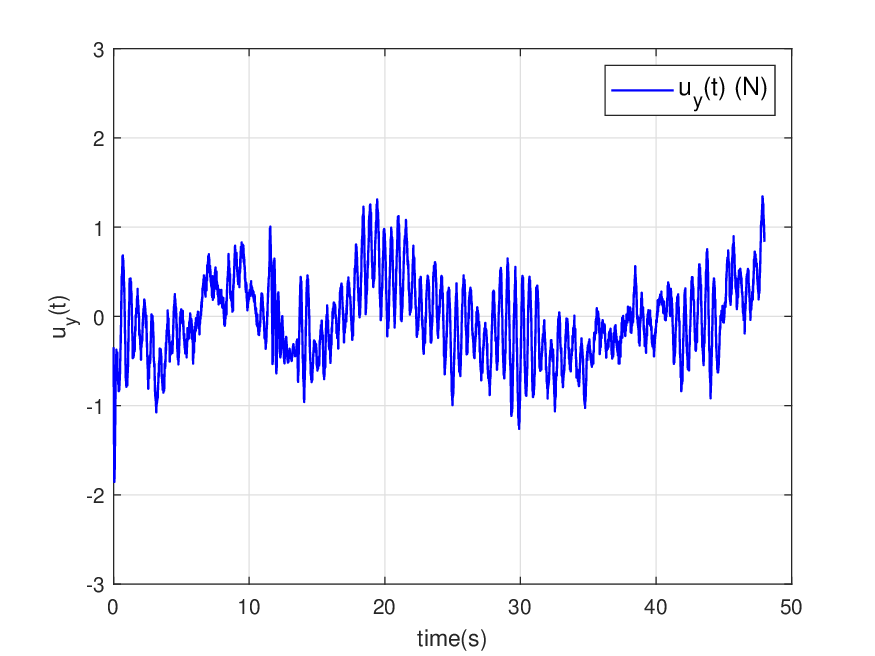}\\[0pt]
{\small (c)}
\end{center}
\caption{Control performance in $y$-direction. (a) $y_{1}$. (b) $y_{2}$. (c)
Controller $u_{y}\left( t\right) $.}
\end{figure}

\begin{figure}[tbp]
\begin{center}
\includegraphics[width=3.5in]{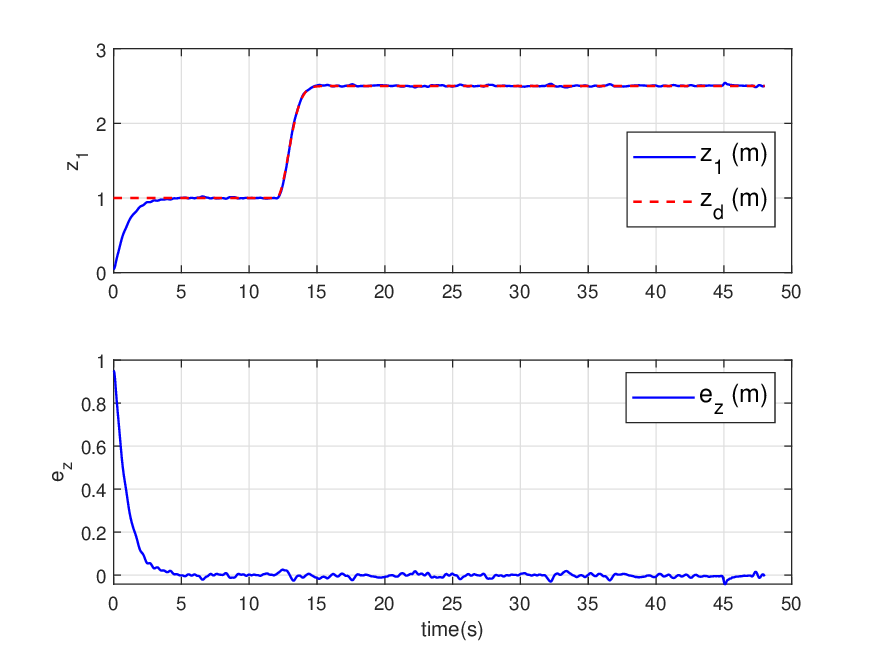}\\[0pt]
{\small (a)}\\[0pt]
\includegraphics[width=3.5in]{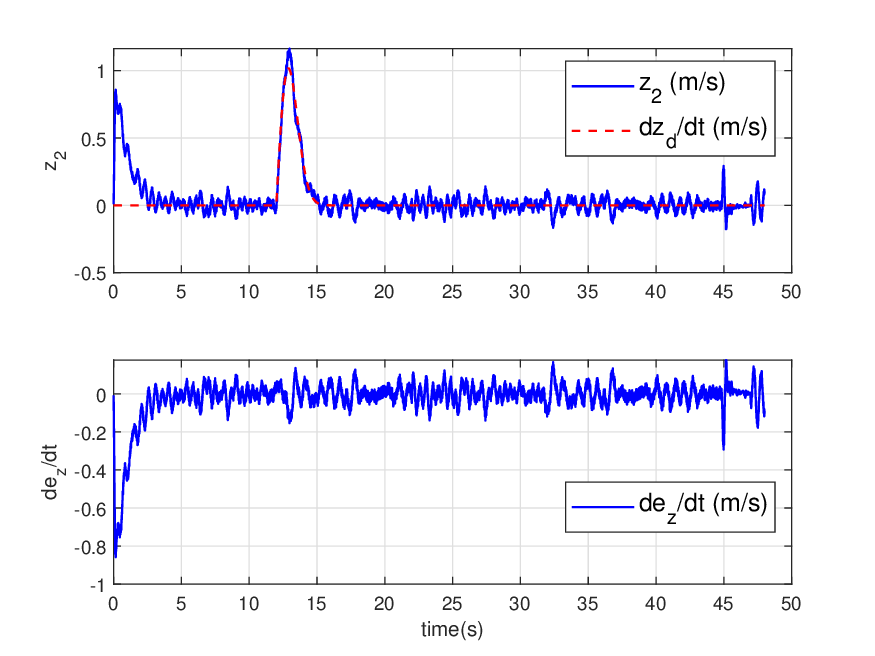}\\[0pt]
{\small (b)}\\[0pt]
\includegraphics[width=3.5in]{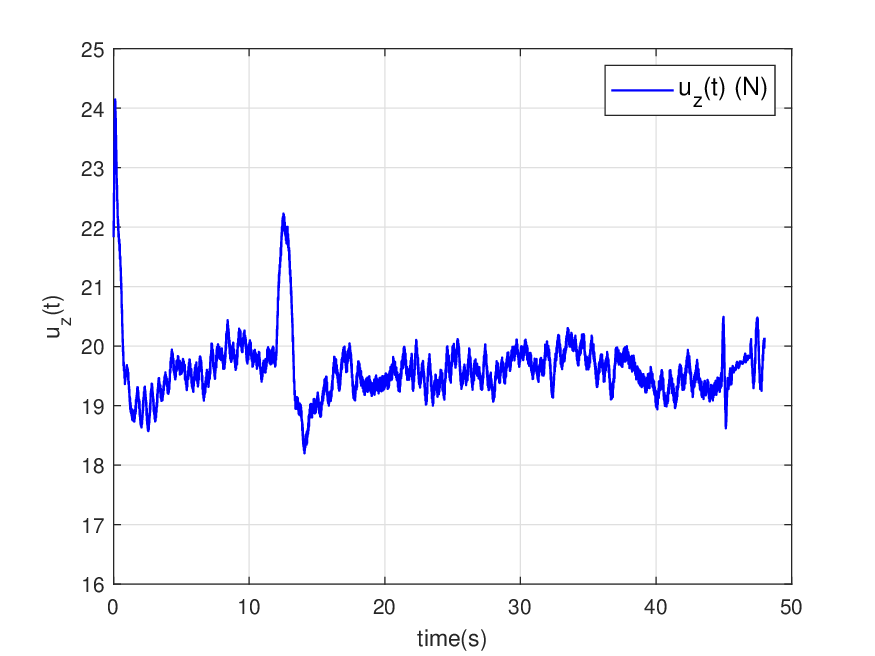}\\[0pt]
{\small (c)}
\end{center}
\caption{Control performance in $z$-direction. (a) $z_{1}$. (b) $z_{2}$. (c)
Controller $u_{z}\left( t\right) $.}
\end{figure}

\bigskip

\emph{8.1 Controller parameters determination:}

Through testing the crosswind from the electric fan, and considering the
system uncertainties, we estimate the upper bound of disturbances and
uncertainties to be within $L_{\delta }=4.5N$. The up-bound of $k_{2}$ is $%
k_{2M}=8N$ through the motor tests.

\emph{Steps on determination of controller parameters according to (34)}$%
\sim $\emph{(38):}

Step 1: Measure the initial states, and determine the initial errors.

the initial states:

$x\left( 0\right) =0.3$, $\dot{x}\left( 0\right) =-0.02$;

$y\left( 0\right) =0.2$, $\dot{y}\left( 0\right) =-0.01$;

$z\left( 0\right) =0.05$, $\dot{z}\left( 0\right) =0.01$;

the initial reference:

$x_{d}\left( 0\right) =0$, $\dot{x}_{d}\left( 0\right) =0$;

$y_{d}\left( 0\right) =0$, $\dot{y}_{d}\left( 0\right) =0$;

$z_{d}\left( 0\right) =1$, $\dot{z}_{d}\left( 0\right) =0$;

then, the initial errors:

$e_{x1}\left( 0\right) =-0.3$, $e_{x2}\left( 0\right) =0.02$;

$e_{y1}\left( 0\right) =-0.2$, $e_{y2}\left( 0\right) =0.01$;

$e_{z1}\left( 0\right) =0.95$, $e_{z2}\left( 0\right) =-0.01$.

Step 2: Select $e_{\ast 1c}$ and $e_{\ast 2c}$ and determine $k_{\ast c}$
and $\rho _{\ast c}$.

From $e_{\ast 1c}\in \left( 0,k_{\ast 2M}-L_{\delta }\right) =\left(
0,3.5\right) $, we select $e_{\ast 1c}=1$. Then, we get $e_{\ast 2c}\in
\left( e_{\ast 1c},\sqrt{\left( k_{\ast 2M}-L_{\delta }\right) e_{\ast 1c}}%
\right] =\left( 1,\sqrt{\left( 8-4.5\right) \times 1}\right] =\left( 1,1.87%
\right] $. We select $e_{\ast 2c}=1.2$. Select $k_{\ast c}=5.5>L_{\delta }$.
Then, $\rho _{\ast c}=6\frac{1}{2}\ln \frac{k_{\ast c}+L_{\ast d}}{k_{\ast
c}-L_{\ast d}}=6\frac{1}{2}\ln \frac{5.5+4.5}{5.5-4.5}=8.83$, where, $\ast
=x,y,z,\psi ,\theta ,\phi $.

Step 3: Determine the controller parameters $k_{\ast 1}$, $k_{\ast 2}$ and $%
\rho _{\ast }$ (where, $\ast =x,y,z,\psi ,\theta ,\phi $) through the
following calculations:

\begin{equation*}
k_{\ast 1}=\left\{ 
\begin{array}{l}
0.5\frac{\left\vert e_{\ast 2}\left( t_{c}\right) \right\vert }{\left\vert
e_{\ast 1}\left( t_{c}\right) \right\vert },\text{ if }e_{\ast 1}\left(
t_{c}\right) e_{\ast 2}\left( t_{c}\right) <0\text{ and }\left\vert e_{\ast
1}\left( t_{c}\right) \right\vert <\left\vert e_{\ast 2}\left( t_{c}\right)
\right\vert ; \\ 
2.3\frac{\left\vert e_{\ast 2}\left( t_{c}\right) \right\vert }{\left\vert
e_{\ast 1}\left( t_{c}\right) \right\vert },\text{ if }e_{\ast 1}\left(
t_{c}\right) e_{\ast 2}\left( t_{c}\right) <0\text{ and }\left\vert e_{\ast
1}\left( t_{c}\right) \right\vert \geq \left\vert e_{\ast 2}\left(
t_{c}\right) \right\vert ; \\ 
2\in \left( 0,\infty \right) ,\text{ if others}%
\end{array}%
\right.
\end{equation*}%
where, $e_{\ast 1}\left( t_{c}\right) $ and $e_{\ast 2}\left( t_{c}\right) $
are the initial values of $e_{\ast 1}\left( t\right) $ and $e_{\ast 2}\left(
t\right) $ respectively when $\left\vert e_{\ast 1}\left( t\right)
\right\vert \leq e_{\ast 1c}$.

\emph{Adjustment of }$k_{\ast 1}$\emph{:} $k_{\ast 1}=1$ if the calculated $%
k_{\ast 1}\in \left( 0,1\right) $;

\begin{equation*}
k_{\ast 2}=\left\{ 
\begin{array}{l}
1.5\max \left\{ k_{\ast 1}\left\vert e_{\ast 2}\left( t_{c}\right)
\right\vert +L_{d},\frac{e_{\ast 2}^{2}\left( t_{c}\right) }{2\left\vert
e_{\ast 1}\left( t_{c}\right) \right\vert }+L_{\ast d}\right\} ,\text{ if }%
e_{\ast 1}\left( t_{c}\right) e_{\ast 2}\left( t_{c}\right) <0\text{ and }%
\left\vert e_{\ast 1}\left( t_{c}\right) \right\vert <\left\vert e_{\ast
2}\left( t_{c}\right) \right\vert ; \\ 
1.5\max \left\{ k_{\ast 1}\left\vert e_{2}\left( t_{c}\right) \right\vert
+L_{\ast d},\frac{k_{\ast 1}^{2}}{3}\left( \left\vert e_{\ast 1}\left(
t_{c}\right) \right\vert +\sqrt{e_{\ast 1}^{2}\left( t_{c}\right) +3\left( 
\frac{e_{\ast 2}\left( t_{c}\right) }{k_{\ast 1}}\right) ^{2}}\right)
+L_{\ast d}\right\} ,\text{ if others}%
\end{array}%
\right.
\end{equation*}

\begin{equation*}
e_{\ast 2\max }=\max \left\{ \left\vert e_{\ast 2}\left( t_{c}\right)
\right\vert ,\frac{k_{\ast 1}}{3}\left[ \left\vert e_{\ast 1}\left(
t_{c}\right) \right\vert +\sqrt{e_{\ast 1}^{2}\left( t_{c}\right) +3\left( 
\frac{e_{\ast 2}\left( t_{c}\right) }{k_{\ast 1}}\right) ^{2}}\right]
\right\}
\end{equation*}

\begin{equation*}
\rho _{\ast }=3\max \left\{ \frac{1}{2k_{\ast 1}},1\right\} \ln \frac{%
k_{\ast 2}+k_{\ast 1}e_{\ast 2\max }+L_{\ast d}}{k_{\ast 2}-k_{\ast
1}e_{\ast 2\max }-L_{\ast d}}
\end{equation*}

Step 4: Controller output (65):

\begin{equation*}
\bar{u}_{\ast }(t)=\left\{ 
\begin{array}{l}
k_{\ast c}\text{tanh}\left[ \rho _{\ast c}\left( e_{\ast 2}\left( t\right)
+e_{\ast 2c}\text{sign}\left( e_{\ast 1}\left( t\right) \right) \right) %
\right] -h_{\ast }(t),\text{ if }\left\vert e_{\ast 1}\left( t\right)
\right\vert >e_{1\ast c}; \\ 
k_{\ast 2}\text{tanh}\left[ \rho _{\ast }(e_{\ast 2}\left( t\right) +k_{\ast
1}e_{\ast 1}\left( t\right) )\right] -h_{\ast }(t),\text{ if }\left\vert
e_{\ast 1}\left( t\right) \right\vert \leq e_{\ast 1c}%
\end{array}%
\right.
\end{equation*}

Because the reference trajectory jumps once at the $6$th second, the
algorithm in Remark 5.5 updates the controller parameters at this time. From
the program algorithm calculation, we can read the two groups of controller
parameters as follows:

1) Parameters for takeoff and hovering for $0\leq t<6$ (sec):

$k_{x1}=1$, $k_{x2}=7.05$, $\rho _{x}=4.83$

$k_{y1}=1$, $k_{y2}=6.95$, $\rho _{y}=4.83$

$k_{z1}=1$, $k_{z2}=7.70$, $\rho _{z}=4.83$

2) Parameters for climbing and cruise in a circle for $t\geq 6$ (sec):

$k_{x1}=1$, $k_{x2}=8.75$, $\rho _{x}=4.83$

$k_{y1}=2$, $k_{y2}=6.81$, $\rho _{y}=4.83$

$k_{z1}=2$, $k_{z2}=6.94$, $\rho _{z}=4.83$

\bigskip

\emph{8.2 Analysis of UAV control performance:}

Figure 9(b) gives the UAV 3D flight trajectory comparison. Figures 10$\sim $%
12 show the UAV flight trajectories, the tracking errors and the controller
outputs in the three directions, respectively: Figure 10 describes the
control performance in $x$-direction; Figure 11 shows the control
performance in $y$-direction; and Figure 12 presents the height control
performance in $z$-direction. From the error outputs, the position errors
were within $0.04$m, and the velocity errors were within $0.2$m/s.

The position ($x,y,z$) of the UAV was controlled to track the reference ($%
x_{d}\left( t\right) $, $y_{d}\left( t\right) $, $z_{d}\left( t\right) $).
From the position trajectory outputs in Figures 10$\sim $12, the smooth and
accurate trajectory tracking was achieved with almost no chattering and no
overshoot. Even in the presence of time-varying disturbance (i.e., the
crosswind from a swinging electric fan), the position and velocity tracking
errors remained very small, and there was no overshoot. In addition, by
reading the program, we found that although the reference trajectory jumped
at the 6th second, the control system re-made the range judgement and the
initial value partitioning, and the control parameters were updated at this
moment. The UAV remained in the safe flight status throughout the flight.

In this indoor flight test, the bounded time-varying crosswind was generated
by a swinging electric fan. Because the up-bound of disturbance was within a
certain range, the influence of disturbance was rejected sufficiently by the
proposed controller. The experimental environment and test equipment are
shown in [30]. Since the proposed control method only requires that the\
magnitude of disturbance\ is within a given range, and it is unrelated to
other aspects of the disturbance. For the outdoor flight tests, as long as
the magnitude of outdoor wind is within the required range, this control
method will be equally effective in suppressing the disturbance.

\bigskip

\emph{8.3 Limitations of the proposed method:}

Through the theoretical analysis (see Theorem 3.1 and Remark 5.3) and the
experimental results, we found that the high-frequency noise in the
measurements and the bounded stochastic disturbances in the system dynamics
almost do not affect the control performance. However, the low-frequency
disturbances or errors in the measurements cannot be suppressed effectively
by the sliding mode system, resulting in the reduced control accuracy. The
potentially effective methods, such as utilizing the signal fusion with
multiple sensors, can be used to obtain the relatively accurate measurements.

\section{Conclusions}

In this paper, a non-overshooting sliding mode control has been presented to
stabilize a class of uncertain systems, and the global non-overshooting
stability has been implemented. Even when the bounded stochastic disturbance
exists, and the time-variant reference is required, the strict
non-overshooting stabilization is still achieved. The performance of the
proposed control method was demonstrated by two simulation examples, and it
was applied successfully to a UAV flying test: 1) The high-precision and
non-overshooting trajectory tracking was performed; 2) The bounded
stochastic disturbances were rejected sufficiently; 3) The bounded and
smoothed controller outputs were easily performed by the actuators; 4) The
UAV trajectory tracking experiments verified the high maneuverability
control capability and non-overshooting performance for the proposed
control. The merits of the control method include its global
non-overshooting stability, strong robustness and no restriction on the
system initial values. Our future work is to optimize the parameters in the
proposed controller.

\bigskip

\textbf{Appendix}

\emph{Proof of Lemma 2.1:}

Define a new variable

\begin{equation}
z\left( t\right) =\int_{0}^{t}e\left( \tau \right) d\tau -\frac{d}{k_{i}}
\end{equation}%
Then, we get

\begin{equation}
\dot{z}\left( t\right) =e\left( t\right) ,\text{ }\ddot{z}\left( t\right) =%
\dot{e}\left( t\right) ,\text{ }\dddot{z}\left( t\right) =\ddot{e}\left(
t\right)
\end{equation}%
Substituting the relations (66) and (67) into system (6), we can rewrite the
system (6) as

\begin{equation}
\dddot{z}\left( t\right) +k_{d}\ddot{z}\left( t\right) +k_{p}\dot{z}\left(
t\right) +k_{i}z\left( t\right) =0
\end{equation}%
The characteristic equation of system (68) can be expressed by

\begin{equation}
s^{3}+k_{d}s^{2}+k_{p}s+k_{i}=0
\end{equation}%
According to the stability conditions, for system (69), if $k_{p}$, $k_{i}$
and $k_{d}$ are selected to make the real parts of all the roots of the
characteristic equation (69) negative, then system is exponentially stable,
and we get

\begin{equation}
\underset{t\rightarrow \infty }{\lim }z\left( t\right) =0\text{, }\underset{%
t\rightarrow \infty }{\lim }\dot{z}\left( t\right) =0\text{, and }\underset{%
t\rightarrow \infty }{\lim }\ddot{z}\left( t\right) =0
\end{equation}%
For (70), from the relations (66) and (67), we get

\begin{equation}
\underset{t\rightarrow \infty }{\lim }\int_{0}^{t}e\left( \tau \right) d\tau
=\frac{d}{k_{i}}\text{, }\underset{t\rightarrow \infty }{\lim }e\left(
t\right) =0\text{, and }\underset{t\rightarrow \infty }{\lim }\dot{e}\left(
t\right) =0
\end{equation}%
This concludes the proof. $\blacksquare $

\bigskip

\emph{Proof of Lemma 2.2:}

Define a new variable

\begin{equation}
z\left( t\right) =\int_{0}^{t}e\left( \tau \right) d\tau -\frac{d}{k_{i}}
\end{equation}%
Then, we get

\begin{equation}
\dot{z}\left( t\right) =e\left( t\right) ,\text{ }\ddot{z}\left( t\right) =%
\dot{e}\left( t\right)
\end{equation}%
Substituting the relations (72) and (73) into system (11), we can rewrite
the system (11) as

\begin{equation}
\ddot{z}\left( t\right) +k_{p}\dot{z}\left( t\right) +k_{i}z\left( t\right)
=0
\end{equation}%
The characteristic equation of system (74) can be expressed by

\begin{equation}
s^{2}+k_{p}s+k_{i}=0
\end{equation}%
According to the stability conditions, for system (74), if $k_{p}$ and $%
k_{i} $ are selected to make the real parts of all the roots of the
characteristic equation (75) negative, then system is exponentially stable,
and we get

\begin{equation}
\underset{t\rightarrow \infty }{\lim }z\left( t\right) =0\text{, and }%
\underset{t\rightarrow \infty }{\lim }\dot{z}\left( t\right) =0
\end{equation}%
For (76), from the relations (72) and (73), we get

\begin{equation}
\underset{t\rightarrow \infty }{\lim }\int_{0}^{t}e\left( \tau \right) d\tau
=\frac{d}{k_{i}}\text{, and }\underset{t\rightarrow \infty }{\lim }e\left(
t\right) =0
\end{equation}%
This concludes the proof. $\blacksquare $

\bigskip

\emph{Proof of Theorem 3.1:}

\emph{Proof introduction.} The proof on the robust and global
non-overshooting stability of the proposed 2-sliding mode is divided into
three steps:

A. Robust non-overshooting reachability of the first subsystem. It proves
that the first subsystem can make the sliding variables reach a given
bounded range and without overshoot. That is to say, the initial values of
the second subsystem are compressed to a given bounded range by the first
subsystem with robust non-overshooting reachability.

B. Robust non-overshooting stability of the second subsystem. It proves
that: conditions on finite-time stability; analytical expressions of sliding
variables through partitioning the initial values; and the determination of
the bounded parameters for the robust non-overshooting stability.

C. Existence and determination of the initial value range for the second
subsystem to obtain the bounded system gains and the non-overshooting
stability.

\bigskip

\textbf{A. Robust non-overshooting reachability of the first subsystem}

Case one: If $e_{1}\left( t\right) >e_{1c}$, for system (18), we get

\begin{equation}
\frac{d\left( e_{2}\left( t\right) +e_{2c}\right) }{dt}=-k_{c}\text{sign}%
\left[ e_{2}\left( t\right) +e_{2c}\right] +d(t)
\end{equation}%
A Lyapunov function candidate is selected as

\begin{equation}
V_{c}=\frac{1}{2}\left( e_{2}\left( t\right) +e_{2c}\right) ^{2}
\end{equation}%
Then, taking the derivative for $V_{c}$, we get

\begin{eqnarray}
\dot{V}_{c} &=&\left( e_{2}\left( t\right) +e_{2c}\right) \left\{ -k_{c}%
\text{sign}\left[ e_{2}\left( t\right) +e_{2c}\right] +d(t)\right\}  \notag
\\
&\leq &-\left( k_{c}-L_{d}\right) \left\vert e_{2}\left( t\right)
+e_{2c}\right\vert =-\sqrt{2}\left( k_{c}-L_{d}\right) V_{c}^{\frac{1}{2}}
\end{eqnarray}%
We know that $k_{c}>L_{d}$. Therefore, there exists a finite time $t_{c1}>0$%
, for $t\geq t_{c1}$, such that $e_{2}\left( t\right) =-e_{2c}$. According
to $\dot{e}_{1}=e_{2}$, we get that $\dot{e}_{1}\left( t\right) =-e_{2c}$
for $t\geq t_{c1}$. Therefore, there exists a finite time $t_{c}>0$, for $%
t\geq t_{c}$, such that $e_{1}\left( t\right) \leq e_{1c}$.

Case two: If $e_{1}\left( t\right) <-e_{1c}$, for system (18), we get

\begin{equation}
\frac{d\left( e_{2}\left( t\right) -e_{2c}\right) }{dt}=-k_{c}\text{sign}%
\left[ e_{2}\left( t\right) -e_{2c}\right] +d(t)
\end{equation}%
Similar method to case $e_{1}\left( t\right) >e_{1c}$, when we select the
Lyapunov function candidate as $V_{c}=\frac{1}{2}\left( e_{2}\left( t\right)
-e_{2c}\right) ^{2}$, we can get that $e_{1}\left( t\right) \geq -e_{1c}$.

Combining cases one and two, we can get that $\left\vert e_{1}\left(
t\right) \right\vert \leq e_{1c}$ for $t\geq t_{c}$.

\bigskip

\textbf{B. Robust non-overshooting stability of the second subsystem}

In the following, we discuss the case $\left\vert e_{1}\left( t\right)
\right\vert \leq e_{1c}$.

\emph{Finite-time stability conditions}

For the sliding mode (18) when $\left\vert e_{1}\left( t\right) \right\vert
\leq e_{1c}$, a Lyapunov function candidate is selected as

\begin{equation}
V_{1}=\frac{1}{2}\sigma \left( t\right) ^{2}
\end{equation}%
where, the sliding function $\sigma \left( t\right) =e_{2}\left( t\right)
+k_{1}e_{1}\left( t\right) $. Then, we get

\begin{eqnarray}
\dot{V}_{1} &=&\left( e_{2}\left( t\right) +k_{1}e_{1}\left( t\right)
\right) \left\{ -k_{2}\text{sign}\left[ e_{2}\left( t\right)
+k_{1}e_{1}\left( t\right) \right] +d(t)+k_{1}e_{2}\left( t\right) \right\} 
\notag \\
&=&-k_{2}\left\vert e_{2}\left( t\right) +k_{1}e_{1}\left( t\right)
\right\vert +\left( k_{1}e_{2}\left( t\right) +d(t)\right) \left(
e_{2}\left( t\right) +k_{1}e_{1}\left( t\right) \right)  \notag \\
&\leq &-k_{2}\left\vert e_{2}\left( t\right) +k_{1}e_{1}\left( t\right)
\right\vert +\left( k_{1}\left\vert e_{2}\left( t\right) \right\vert
+L_{d}\right) \left\vert e_{2}\left( t\right) +k_{1}e_{1}\left( t\right)
\right\vert  \notag \\
&=&-\left( k_{2}-k_{1}\left\vert e_{2}\left( t\right) \right\vert
-L_{d}\right) \left\vert e_{2}\left( t\right) +k_{1}e_{1}\left( t\right)
\right\vert  \notag \\
&=&-\sqrt{2}\left( k_{2}-k_{1}\left\vert e_{2}\left( t\right) \right\vert
-L_{d}\right) V^{\frac{1}{2}}
\end{eqnarray}%
If $k_{2}>k_{1}\left\vert e_{2}\left( t\right) \right\vert +L_{d}$, then the
sliding mode is finite-time stable. That means there exists a finite time $%
t_{s}>0$, for $t\geq t_{c}+t_{s}$, the sliding function $\sigma \left(
t\right) =e_{2}\left( t\right) +k_{1}e_{1}\left( t\right) =0$. Then, we get
the linear convergence law $\dot{e}_{1}\left( t\right) =-k_{1}e_{1}\left(
t\right) $, and $\underset{t\rightarrow \infty }{\lim }e_{1}\left( t\right)
=0$.

\bigskip

\emph{Trajectories of }$e_{1}\left( t\right) $,\emph{\ }$e_{2}\left(
t\right) $\emph{\ and }$\sigma \left( t\right) $ \emph{for }$t\in \left[
t_{c},t_{c}+t_{s}\right) $

For the sliding mode (18), according to the differential inclusion theory,
we get

\begin{equation}
\dot{e}_{2}\left( t\right) \in -\left[ k_{2}-L_{d},k_{2}-L_{d}\right] \text{%
sign}\left[ e_{2}\left( t\right) +k_{1}e_{1}\left( t\right) \right]
\end{equation}%
That is to say, there exists $\bar{k}_{2}\in \left[ k_{2}-L_{d},k_{2}-L_{d}%
\right] $, such that

\begin{equation}
\dot{e}_{2}\left( t\right) =-\bar{k}_{2}\text{sign}\left[ e_{2}\left(
t\right) +k_{1}e_{1}\left( t\right) \right]
\end{equation}%
We will determine $k_{1}$ and $k_{2}$ to make the sign of sliding function $%
\sigma \left( t\right) =e_{2}\left( t\right) +k_{1}e_{1}\left( t\right) $
unchanged for $t\in \left[ t_{c},t_{c}+t_{s}\right) $, i.e., sign$\left[
e_{2}\left( t\right) +k_{1}e_{1}\left( t\right) \right] =$sign$\left[
e_{2}\left( t_{c}\right) +k_{1}e_{1}\left( t_{c}\right) \right] $ for $t\in %
\left[ t_{c},t_{c}+t_{s}\right) $. Then, we can get the simple solution to
(85) for $t\in \left[ t_{c},t_{c}+t_{s}\right) $, as follows:

\begin{equation}
e_{2}\left( t\right) =e_{2}\left( t_{c}\right) -\bar{k}_{2}\text{sign}\left[
e_{2}\left( t\right) +k_{1}e_{1}\left( t\right) \right] \left( t-t_{c}\right)
\end{equation}%
From $\dot{e}_{1}\left( t\right) =e_{2}\left( t\right) $ in sliding mode
(18) and $e_{2}\left( t\right) $ expression in (86), we get

\begin{equation}
e_{1}\left( t\right) =e_{1}\left( t_{c}\right) +e_{2}\left( t_{c}\right)
\left( t-t_{c}\right) -\frac{1}{2}\bar{k}_{2}\text{sign}\left[ e_{2}\left(
t\right) +k_{1}e_{1}\left( t\right) \right] \cdot \left( t-t_{c}\right) ^{2}
\end{equation}%
and

\begin{eqnarray}
\sigma \left( t\right) &=&e_{2}\left( t\right) +k_{1}e_{1}\left( t\right)
=e_{2}\left( t_{c}\right) +k_{1}e_{1}\left( t_{c}\right) -\left( \bar{k}_{2}%
\text{sign}\left[ e_{2}\left( t\right) +k_{1}e_{1}\left( t\right) \right]
-k_{1}e_{2}\left( t_{c}\right) \right) \left( t-t_{c}\right)  \notag \\
&&-\frac{1}{2}k_{1}\bar{k}_{2}\text{sign}\left[ e_{2}\left( t\right)
+k_{1}e_{1}\left( t\right) \right] \cdot \left( t-t_{c}\right) ^{2}
\end{eqnarray}

\bigskip

\emph{Non-overshooting convergence conditions}

1) For the 2-sliding mode (18), we will determine $k_{1}$ and $k_{2}$ to
generate the two convergence laws: 1) the finite-time convergence law to get
the sliding surface $\sigma \left( t\right) =e_{2}\left( t\right)
+k_{1}e_{1}\left( t\right) =0$ after a finite time $t_{c}+t_{s}$; 2) the
linear convergence law $\dot{e}_{1}\left( t\right) =-k_{1}e_{1}\left(
t\right) $ to make $\underset{t\rightarrow \infty }{\lim }e_{1}\left(
t\right) =0$.

2) For the 2-sliding mode (18), in order to get the simple form of sliding
variables assumed in (86)$\sim $(88), we hope that, for $t\in \left[
t_{c},t_{c}+t_{s}\right) $, the sign of sliding function $\sigma \left(
t\right) =e_{2}\left( t\right) +k_{1}e_{1}\left( t\right) $ unchanged as $%
\sigma \left( t_{c}\right) =e_{2}\left( t_{c}\right) +k_{1}e_{1}\left(
t_{c}\right) $. Thus, function sign$\left[ e_{2}\left( t\right)
+k_{1}e_{1}\left( t\right) \right] $ becomes constant $1$ or $-1$ for $t\in %
\left[ t_{c},t_{c}+t_{s}\right) $. Then, we can get $\dot{e}_{2}\left(
t\right) =-\bar{k}_{2}$ or $\dot{e}_{2}\left( t\right) =\bar{k}_{2}$ for $%
t\in \left[ t_{c},t_{c}+t_{s}\right) $.

3) For being non-overshooting, sign$\left[ e_{1}\left( t\right) \right] =$%
sign$\left[ e_{1}\left( t_{c}\right) \right] $ needs always hold until $%
\underset{t\rightarrow \infty }{\lim }e_{1}\left( t\right) =0$.

We know that, the sliding variable $e_{1}\left( t\right) $ is
non-overshooting when it is in the convergence law $\dot{e}_{1}\left(
t\right) =-k_{1}e_{1}\left( t\right) $ for $t\geq t_{c}+t_{s}$. Therefore,
in order to make $e_{1}\left( t\right) $ non-overshooting during the whole
transient process, we only need to guarantee $e_{1}\left( t\right) $
non-overshooting in the finite-time convergence law for $t\in \left[
t_{c},t_{c}+t_{s}\right) $.

\bigskip

\emph{Partitioning of }$e_{1}\left( t_{c}\right) $\emph{\ and }$e_{2}\left(
t_{c}\right) $\emph{\ in coordinate}

\begin{figure}[tbp]
\centering
\includegraphics[width=4.00in]{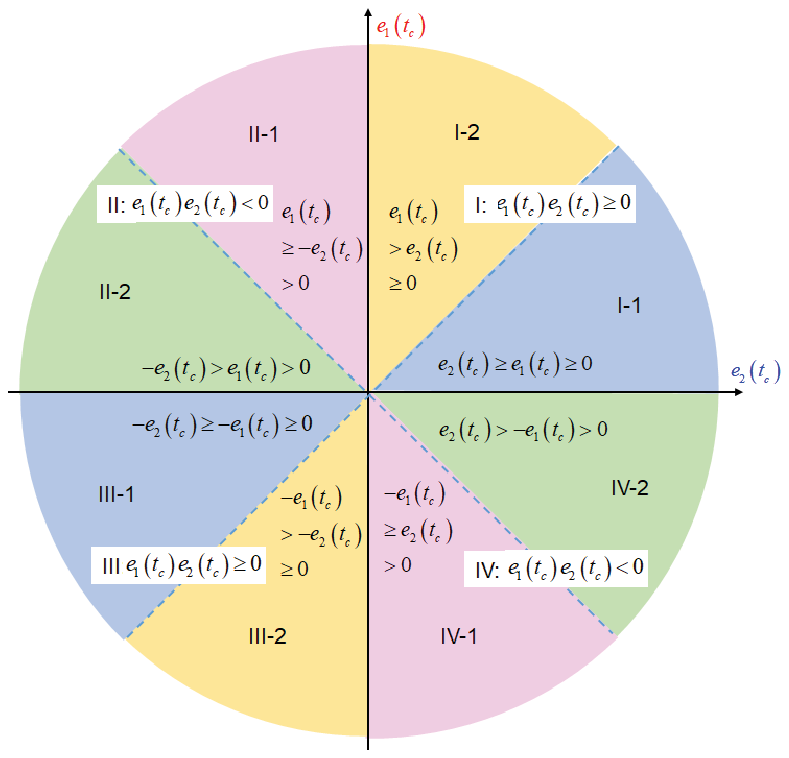}
\caption{Partitioning of $e_{1}\left( t_{c}\right) $ and $e_{2}\left(
t_{c}\right) $ in coordinate.}
\end{figure}

We consider $e_{1}\left( t_{c}\right) $ and $e_{2}\left( t_{c}\right) $ in
the zones shown in Fig. 13:

1) Zones II-2 and IV-2: $e_{1}\left( t_{c}\right) e_{2}\left( t_{c}\right)
<0 $ with $\left\vert e_{1}\left( t_{c}\right) \right\vert <\left\vert
e_{2}\left( t_{c}\right) \right\vert $

2) Zones IV-1 and II-1: $e_{1}\left( t_{c}\right) e_{2}\left( t_{c}\right)
<0 $ with $\left\vert e_{1}\left( t_{c}\right) \right\vert \geq \left\vert
e_{2}\left( t_{c}\right) \right\vert $

3) Zones III-1 and I-1: $e_{1}\left( t_{c}\right) e_{2}\left( t_{c}\right)
\geq 0$ with $\left\vert e_{1}\left( t_{c}\right) \right\vert \leq
\left\vert e_{2}\left( t_{c}\right) \right\vert $

4) Zones III-2 and I-2: $e_{1}\left( t_{c}\right) e_{2}\left( t_{c}\right)
\geq 0$ with $\left\vert e_{1}\left( t_{c}\right) \right\vert >\left\vert
e_{2}\left( t_{c}\right) \right\vert $

The right side of the sliding mode (18) is the odd function about the
origin, therefore, we only consider the convergence performance for the
zones (II-2, IV-1, III-1 and III-2). The corresponding zones (IV-2,II-1, I-1
and I-2) about the origin have the same stability performance to the zones
(II-2, IV-1, III-1 and III-2), respectively.

Therefore, in the following, we will consider the convergence performance
for the following zones:

1) Zone II-2: $-e_{2}\left( t_{c}\right) >e_{1}\left( t_{c}\right) >0$

2) Zone IV-1: $-e_{1}\left( t_{c}\right) \geq e_{2}\left( t_{c}\right) >0$

3) Zone III-1: $-e_{2}\left( t_{c}\right) \geq -e_{1}\left( t_{c}\right)
\geq 0$

4) Zone III-2: $-e_{1}\left( t_{c}\right) >-e_{2}\left( t_{c}\right) \geq 0$

\bigskip

\emph{Analytical expressions of sliding variables\ by assuming the unchanged
sign of }$\sigma \left( t\right) =e_{2}\left( t\right) +k_{1}e_{1}\left(
t\right) $\emph{\ for }$t\in \left[ t_{c},t_{c}+t_{s}\right) $

In the selected zones (II-2, IV-1, III-1 and III-2), we will determine $%
k_{1} $ and $k_{2}$ to make $\sigma \left( t\right) =e_{2}\left( t\right)
+k_{1}e_{1}\left( t\right) <0$ for $t\in \left[ t_{c},t_{c}+t_{s}\right) $.
Then, sign$\left[ e_{2}\left( t\right) +k_{1}e_{1}\left( t\right) \right] =$%
sign$\left[ e_{2}\left( t_{c}\right) +k_{1}e_{1}\left( t_{c}\right) \right]
=-1$. Therefore, from (86), (87) and (88), for $t\in \left[
t_{c},t_{c}+t_{s}\right) $, we get the expressions of variables $e_{2}\left(
t\right) $, $e_{1}\left( t\right) $ and function $\sigma \left( t\right)
=e_{2}\left( t\right) +k_{1}e_{1}\left( t\right) $, respectively, as follows:

\begin{equation}
e_{2}\left( t\right) =e_{2}\left( t_{c}\right) +\bar{k}_{2}\left(
t-t_{c}\right)
\end{equation}

\begin{equation}
e_{1}\left( t\right) =e_{1}\left( t_{c}\right) +e_{2}\left( t_{c}\right)
\left( t-t_{c}\right) +\frac{1}{2}\bar{k}_{2}\left( t-t_{c}\right) ^{2}%
\overset{\text{define}}{=}c_{1}+b_{1}\left( t-t_{c}\right) +a_{1}\left(
t-t_{c}\right) ^{2}
\end{equation}

\begin{eqnarray}
\sigma \left( t\right) &=&e_{2}\left( t\right) +k_{1}e_{1}\left( t\right)
=e_{2}\left( t_{c}\right) +k_{1}e_{1}\left( t_{c}\right) +\left( \bar{k}%
_{2}+k_{1}e_{2}\left( t_{c}\right) \right) \left( t-t_{c}\right) +\frac{1}{2}%
k_{1}\bar{k}_{2}\left( t-t_{c}\right) ^{2}  \notag \\
&&\overset{\text{define}}{=}c+b\left( t-t_{c}\right) +a\left( t-t_{c}\right)
^{2}
\end{eqnarray}%
From (90), $e_{1}\left( t\right) $ is a segment of a\ parabola for $t\in %
\left[ t_{c},t_{c}+t_{s}\right) $, and we get its axis of symmetry:

\begin{equation}
-\frac{b_{1}}{2a_{1}}=-\frac{e_{2}\left( t_{c}\right) }{\bar{k}_{2}}
\end{equation}%
its vertex:

\begin{equation}
\frac{4a_{1}c_{1}-b_{1}^{2}}{4a_{1}}=\frac{2\bar{k}_{2}e_{1}\left(
t_{c}\right) -e_{2}^{2}\left( t_{c}\right) }{2\bar{k}_{2}}
\end{equation}%
and the time instant when $e_{1}\left( t\right) =0$:

\begin{equation}
\left. t\right\vert _{e_{1}\left( t\right) =0}=\frac{-b_{1}+\sqrt{%
b_{1}^{2}-4a_{1}c_{1}}}{2a_{1}}+t_{c}=-\frac{e_{2}\left( t_{c}\right) }{\bar{%
k}_{2}}+\sqrt{\left( \frac{e_{2}\left( t_{c}\right) }{\bar{k}_{2}}\right)
^{2}-\frac{2e_{1}\left( t_{c}\right) }{\bar{k}_{2}}}+t_{c}
\end{equation}%
From (91), function $\sigma \left( t\right) =e_{2}\left( t\right)
+k_{1}e_{1}\left( t\right) $ is a segment of a\ parabola for $t\in \left[
t_{c},t_{c}+t_{s}\right) $, and we get its axis of symmetry:

\begin{equation}
-\frac{b}{2a}=-\frac{\bar{k}_{2}+k_{1}e_{2}\left( t_{c}\right) }{k_{1}\bar{k}%
_{2}}
\end{equation}%
its vertex:

\begin{equation}
\frac{4ac-b^{2}}{4a}=\frac{2k_{1}\bar{k}_{2}\left( e_{2}\left( t_{c}\right)
+k_{1}e_{1}\left( t_{c}\right) \right) -\left( \bar{k}_{2}+k_{1}e_{2}\left(
t_{c}\right) \right) ^{2}}{2k_{1}\bar{k}_{2}}
\end{equation}%
and the time instant when $\sigma \left( t\right) =0$, i.e., the settling
time $t_{c}+t_{s}$:

\begin{equation}
t_{c}+t_{s}=t_{c}+\frac{-b+\sqrt{b^{2}-4ac}}{2a}=t_{c}-\left( \frac{1}{k_{1}}%
+\frac{e_{2}\left( t_{c}\right) }{\bar{k}_{2}}\right) +\sqrt{\frac{1}{%
k_{1}^{2}}+\left( \frac{e_{2}\left( t_{c}\right) }{\bar{k}_{2}}\right) ^{2}-%
\frac{2e_{1}\left( t_{c}\right) }{\bar{k}_{2}}}
\end{equation}

In the following, for any zone of $e_{1}\left( t_{c}\right) $ and $%
e_{2}\left( t_{c}\right) $, we will determine the parameters $k_{1}$ and $%
k_{2}$ to make the sliding mode finite-time stable and $e_{1}\left( t\right) 
$ non-overshooting for $t\in \left[ t_{c},t_{c}+t_{s}\right) $. We know the
linear convergence law $\dot{e}_{1}\left( t\right) =-k_{1}e_{1}\left(
t\right) $ makes the system non-overshooting stable automatically.

\begin{figure}[tbp]
\centering
\includegraphics[width=5.00in]{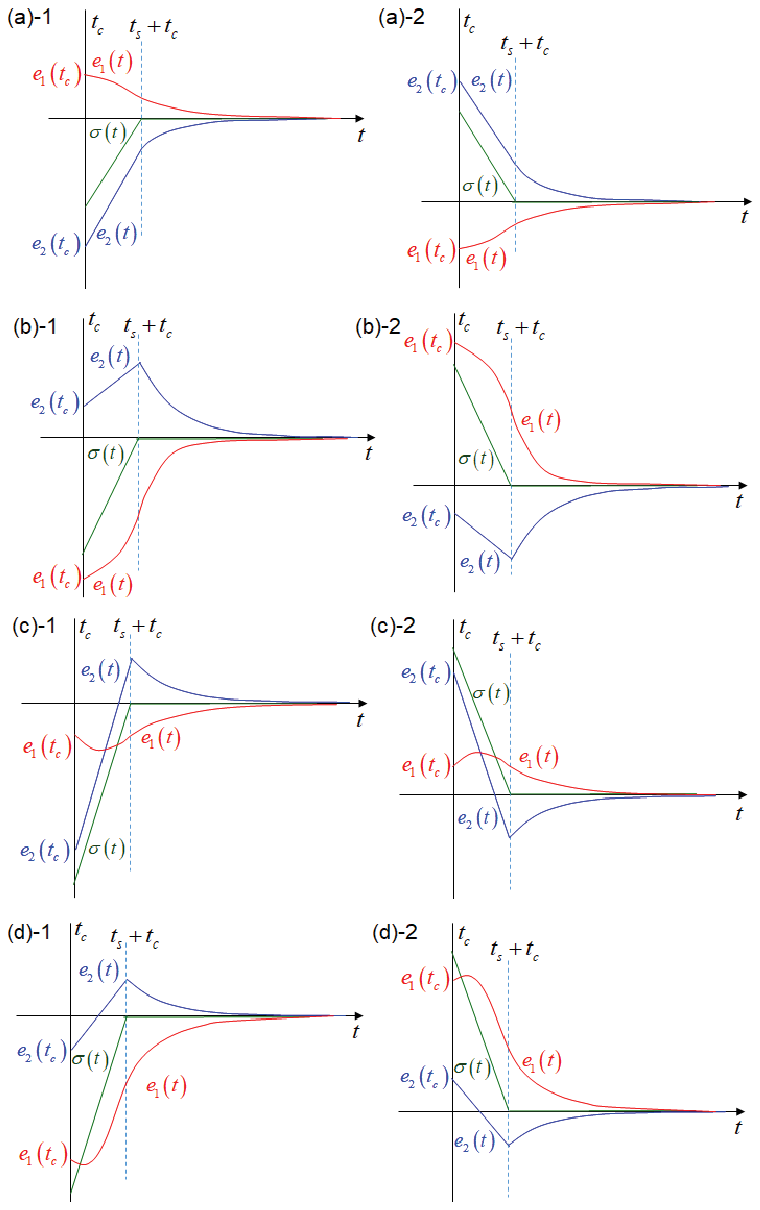}
\caption{Arranged trajectories of $e_{1}\left( t\right) $, $e_{2}\left(
t\right) $ and $\protect\sigma \left( t\right) $ for $e_{1}\left( t\right) $
non-overshooting convergence.}
\end{figure}

\emph{Trajectories arrangement of sliding variables:} In general, for the
different zones of $e_{1}\left( t_{c}\right) $ and $e_{2}\left( t_{c}\right) 
$ shown in Fig. 13, the parameters $k_{1}$ and $k_{2}$ will be determined to
make sliding variables $e_{1}\left( t\right) $, $e_{2}\left( t\right) $ and
function $\sigma \left( t\right) $ generate the desired convergent
trajectories with $e_{1}\left( t\right) $ non-overshooting, shown in Fig. 14:

1) Fig. 14 (a)-1 and (a)-2 are for the zones II-2 and IV-2 in Fig. 13,
respectively;

2) Fig. 14 (b)-1 and (b)-2 are for the zones IV-1 and II-1 in Fig. 13,
respectively;

3) Fig. 14 (c)-1 and (c)-2 are for the zones III-1 and I-1 in Fig. 13,
respectively; and Fig. 13 (d)-1 and (d)-2 are for zones III-2 and I-2 in
Fig. 13, respectively.\ 

\bigskip

\emph{Conditions on robust non-overshooting stability}

\emph{1) For zone II-2: }$-e_{2}\left( t_{c}\right) >e_{1}\left(
t_{c}\right) >0$

From Figure 13, for zone II-2, we know that the corresponding symmetrical
zone is IV-2.

\begin{figure}[tbp]
\centering
\includegraphics[width=5.00in]{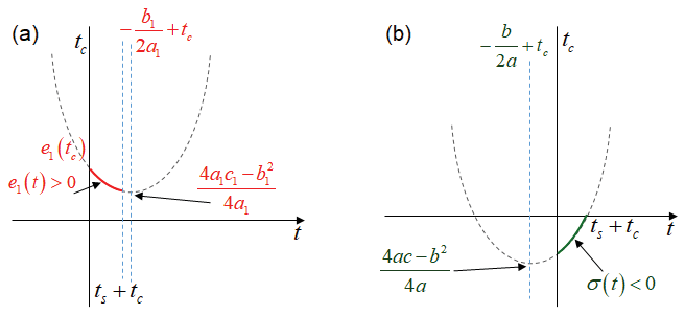}
\caption{Arranged trajectories of $e_{1}\left( t\right) $ and $\protect%
\sigma \left( t\right) $ for $e_{1}\left( t\right) $ non-overshooting
convergence for $t\in \left[ t_{c},t_{c}+t_{s}\right) $ in range II-2: $%
\emph{-}e_{2}\left( t_{c}\right) >e_{1}\left( t_{c}\right) >0$.}
\end{figure}

For zone II-2, we can get the conditions of non-overshooting convergence for 
$t\in \left[ t_{c},t_{c}+t_{s}\right) $:

1) $\bar{k}_{2}>k_{1}\left\vert e_{2}\left( t\right) \right\vert $
[finite-time stability to get convergence law $\dot{e}_{1}\left( t\right)
=-k_{1}e_{1}\left( t\right) $];

2) the sliding function $\sigma \left( t\right) =e_{2}\left( t\right)
+k_{1}e_{1}\left( t\right) <0$ for $t\in \left[ t_{c},t_{c}+t_{s}\right) $,
and $\sigma \left( t_{c}+t_{s}\right) =e_{2}\left( t_{c}+t_{s}\right)
+k_{1}e_{1}\left( t_{c}+t_{s}\right) =0$;

3) $e_{1}\left( t\right) >0$ for $t\in \left[ t_{c},t_{c}+t_{s}\right) $ [ $%
e_{1}\left( t\right) $ does not go beyond zero];

4) $e_{2}\left( t\right) <0$ for $t\in \left[ t_{c},t_{c}+t_{s}\right) $ [ $%
e_{2}\left( t\right) $ does not go beyond zero].

In the following, we will determine $k_{1}$ and $k_{2}$ to satisfy these
conditions.

From (90), for $e_{1}\left( t\right) $ (a segment of a\ parabola) for $t\in %
\left[ t_{c},t_{c}+t_{s}\right) $, we get its axis of symmetry

\begin{equation}
-\frac{b_{1}}{2a_{1}}=-\frac{e_{2}\left( t_{c}\right) }{\bar{k}_{2}}>0
\end{equation}%
because $e_{2}\left( t_{c}\right) <0$. In order to make $e_{1}\left(
t\right) >0$ for $t\in \left[ t_{c},t_{c}+t_{s}\right) $ (See $e_{1}\left(
t\right) $ in Figure 15(a)), the vertex needs to satisfy

\begin{equation}
\frac{4a_{1}c_{1}-b_{1}^{2}}{4a_{1}}=\frac{2\bar{k}_{2}e_{1}\left(
t_{c}\right) -e_{2}^{2}\left( t_{c}\right) }{2\bar{k}_{2}}>0
\end{equation}%
Because $e_{1}\left( t_{c}\right) >0$ and $e_{2}\left( t_{c}\right) <0$, for
(99), the positive $\bar{k}_{2}$ should satisfy

\begin{equation}
\bar{k}_{2}>\frac{e_{2}^{2}\left( t_{c}\right) }{2e_{1}\left( t_{c}\right) }
\end{equation}%
Therefore, from $\bar{k}_{2}$ in condition (100), we can get that $%
e_{1}\left( t\right) >0$ for $t\in \left[ t_{c},t_{c}+t_{s}\right) $.

From (91), for the sliding function $\sigma \left( t\right) =e_{2}\left(
t\right) +k_{1}e_{1}\left( t\right) $ (a segment of a\ parabola) for $t\in %
\left[ t_{c},t_{c}+t_{s}\right) $, we get its axis of symmetry

\begin{equation}
-\frac{b}{2a}=-\frac{\bar{k}_{2}+k_{1}e_{2}\left( t_{c}\right) }{k_{1}\bar{k}%
_{2}}<0
\end{equation}%
because of the finite-time convergence condition

\begin{equation}
\bar{k}_{2}>-k_{1}e_{2}\left( t_{c}\right)
\end{equation}%
In order to make $\sigma \left( t\right) =e_{2}\left( t\right)
+k_{1}e_{1}\left( t\right) <0$ for $t\in \left[ t_{c},t_{c}+t_{s}\right) $
(See $\sigma \left( t\right) $ in Figure 15(b)), the vertex needs to satisfy

\begin{equation}
\frac{4ac-b^{2}}{4a}=\frac{2k_{1}\bar{k}_{2}\left( e_{2}\left( t_{c}\right)
+k_{1}e_{1}\left( t_{c}\right) \right) -\left( \bar{k}_{2}+k_{1}e_{2}\left(
t_{c}\right) \right) ^{2}}{2k_{1}\bar{k}_{2}}<0
\end{equation}%
In order to get the relation in (103), we can let $\sigma \left(
t_{c}\right) =e_{2}\left( t_{c}\right) +k_{1}e_{1}\left( t_{c}\right) <0$,
i.e., $k_{1}e_{1}\left( t_{c}\right) <-e_{2}\left( t_{c}\right) $.
Therefore, we get the condition on $k_{1}$, as follows:

\begin{equation}
k_{1}<\frac{-e_{2}\left( t_{c}\right) }{e_{1}\left( t_{c}\right) }
\end{equation}%
For $e_{2}\left( t\right) $ in (89), we get the time instant when $%
e_{2}\left( t\right) =0$ as follows:

\begin{equation}
\left. t\right\vert _{e_{2}\left( t\right) =0}=\frac{-e_{2}\left(
t_{c}\right) }{\bar{k}_{2}}+t_{c}
\end{equation}%
Comparing the settling time $t_{c}+t_{s}$ in (97) and $\left. t\right\vert
_{e_{2}\left( t\right) =0}$ in (105), we can get

\begin{equation}
t_{s}+t_{c}<\left. t\right\vert _{e_{2}\left( t\right) =0}
\end{equation}%
Then, it follows that%
\begin{equation}
e_{2}\left( t\right) <0\text{ for }t\in \left[ t_{c},t_{c}+t_{s}\right)
\end{equation}

We know that zones II-2 and IV-2 have the same convergence performance
because of the odd function property in the sliding mode. Therefore, combing
(100), (102) and (104), for the zones II-2 (where, $-e_{2}\left(
t_{c}\right) >e_{1}\left( t_{c}\right) >0$) and IV-2 (where, $e_{2}\left(
t_{c}\right) >-e_{1}\left( t_{c}\right) >0$), we get the non-overshooting
convergence conditions, when $e_{1}\left( t_{c}\right) e_{2}\left(
t_{c}\right) <0$ and $\left\vert e_{1}\left( t_{c}\right) \right\vert
<\left\vert e_{2}\left( t_{c}\right) \right\vert $, as follows:

\begin{eqnarray}
k_{1} &\in &\left( 0,\frac{\left\vert e_{2}\left( t_{c}\right) \right\vert }{%
\left\vert e_{1}\left( t_{c}\right) \right\vert }\right) \\
\bar{k}_{2} &>&\max \left\{ k_{1}\left\vert e_{2}\left( t_{c}\right)
\right\vert ,\frac{e_{2}^{2}\left( t_{c}\right) }{2\left\vert e_{1}\left(
t_{c}\right) \right\vert }\right\}
\end{eqnarray}%
Furthermore, considering the disturbance $d\left( t\right) $, the conditions
of parameters selection, when $e_{1}\left( t_{c}\right) e_{2}\left(
t_{c}\right) <0$ and $\left\vert e_{1}\left( t_{c}\right) \right\vert
<\left\vert e_{2}\left( t_{c}\right) \right\vert $, are expressed as follows:

\begin{eqnarray}
k_{1} &\in &\left( 0,\frac{\left\vert e_{2}\left( t_{c}\right) \right\vert }{%
\left\vert e_{1}\left( t_{c}\right) \right\vert }\right) \\
k_{2} &>&\max \left\{ k_{1}\left\vert e_{2}\left( t_{c}\right) \right\vert
+L_{d},\frac{e_{2}^{2}\left( t_{c}\right) }{2e_{1}\left( t_{c}\right) }%
+L_{d}\right\}
\end{eqnarray}%
Therefore, when $k_{1}$ and $k_{2}$ are selected from (110) and (111), the
sliding mode is non-overshooting stable for $t\in \left[ t_{c},t_{c}+t_{s}%
\right) $, and $\sigma \left( t\right) =e_{2}\left( t\right)
+k_{1}e_{1}\left( t\right) =0$ holds for $t\geq t_{c}+t_{s}$. Then, the
linear convergence law $\dot{e}_{1}\left( t\right) =-k_{1}e_{1}\left(
t\right) $ makes $\underset{t\rightarrow \infty }{\lim }e_{1}\left( t\right)
=0$ without overshoot. In addition, from $\sigma \left( t\right)
=e_{2}\left( t\right) +k_{1}e_{1}\left( t\right) =0$ for $t\geq t_{c}+t_{s}$%
, we get $\underset{t\rightarrow \infty }{\lim }e_{2}\left( t\right) =%
\underset{t\rightarrow \infty }{\lim }\left[ -k_{1}e_{1}\left( t\right) %
\right] =0$.

In general, when $e_{1}\left( t_{c}\right) $ and $e_{2}\left( t_{c}\right) $%
\ are in zones II-2 and IV-2, the system is exponentially stable, and no
overshoot exists for the variable $e_{1}\left( t\right) $. This confirms the
convergence curves of sliding variables $e_{1}\left( t\right) $, $%
e_{2}\left( t\right) $ and function $\sigma \left( t\right) $ described in
Figures 14 (a)-1 and (a)-2 for the zones II-2 and IV-2, respectively.

\bigskip

\emph{2) For zone IV-1: }$-e_{1}\left( t_{c}\right) \geq e_{2}\left(
t_{c}\right) >0$

From Figure 13, for zone IV-1, we know that its corresponding symmetrical
zone is II-1.

For zone IV-1, we can get the conditions of non-overshooting convergence for 
$t\in \left[ t_{c},t_{c}+t_{s}\right) $:

1) $\bar{k}_{2}>k_{1}\left\vert e_{2}\left( t\right) \right\vert $
[finite-time stability to get convergence law $\dot{e}_{1}\left( t\right)
=-k_{1}e_{1}\left( t\right) $]

2)$\ $the sliding function $\sigma \left( t\right) =e_{2}\left( t\right)
+k_{1}e_{1}\left( t\right) <0$ for $t\in \left[ t_{c},t_{c}+t_{s}\right) $,
and $\sigma \left( t_{c}+t_{s}\right) =e_{2}\left( t_{c}+t_{s}\right)
+k_{1}e_{1}\left( t_{c}+t_{s}\right) =0$;

3) $e_{1}\left( t\right) <0$ for $t\in \left[ t_{c},t_{c}+t_{s}\right) $ [ $%
e_{1}\left( t\right) $ does not go beyond zero];

4) $e_{2}\left( t\right) >0$ for $t\in \left[ t_{c},t_{c}+t_{s}\right) $ [ $%
e_{2}\left( t\right) $ does not go beyond zero].

In the following, we will determine $k_{1}$ and $k_{2}$ to satisfy these
conditions.

From (90), for $e_{1}\left( t\right) $ (a segment of a\ parabola) for $t\in %
\left[ t_{c},t_{c}+t_{s}\right) $, we get its axis of symmetry

\begin{equation}
-\frac{b_{1}}{2a_{1}}=-\frac{e_{2}\left( t_{c}\right) }{\bar{k}_{2}}<0
\end{equation}%
because $e_{2}\left( t_{c}\right) >0$. Also, its vertex satisfies

\begin{equation}
\frac{4a_{1}c_{1}-b_{1}^{2}}{4a_{1}}=\frac{2\bar{k}_{2}e_{1}\left(
t_{c}\right) -e_{2}^{2}\left( t_{c}\right) }{2\bar{k}_{2}}<0
\end{equation}%
because $e_{1}\left( t_{c}\right) <0$ (See $e_{1}\left( t\right) $ in Figure
16(a)).

\begin{figure}[tbp]
\centering
\includegraphics[width=5.00in]{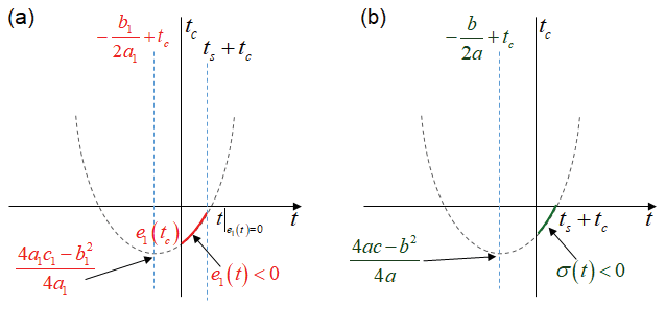}
\caption{Arranged trajectories of $e_{1}\left( t\right) $ and $\protect%
\sigma \left( t\right) $ for $e_{1}\left( t\right) $ non-overshooting
convergence for $t\in \left[ t_{c},t_{c}+t_{s}\right) $ in range IV-1: $%
-e_{1}\left( t_{c}\right) \geq e_{2}\left( t_{c}\right) >0$.}
\end{figure}

From (91), for the sliding function $\sigma \left( t\right) =e_{2}\left(
t\right) +k_{1}e_{1}\left( t\right) $ for $t\in \left[ t_{c},t_{c}+t_{s}%
\right) $, its axis of symmetry satisfies

\begin{equation}
-\frac{b}{2a}=-\frac{\bar{k}_{2}+k_{1}e_{2}\left( t_{c}\right) }{k_{1}\bar{k}%
_{2}}<0
\end{equation}%
because of the finite-time stability condition:

\begin{equation}
\bar{k}_{2}>k_{1}\left\vert e_{2}\left( t_{c}\right) \right\vert
\end{equation}%
In order to make $\sigma \left( t\right) =e_{2}\left( t\right)
+k_{1}e_{1}\left( t\right) <0$ for $t\in \left[ t_{c},t_{c}+t_{s}\right) $
(See $\sigma \left( t\right) $ in Figure 16(b)), the vertex needs to satisfy

\begin{equation}
\frac{4ac-b^{2}}{4a}=\frac{2k_{1}\bar{k}_{2}\left( e_{2}\left( t_{c}\right)
+k_{1}e_{1}\left( t_{c}\right) \right) -\left( \bar{k}_{2}+k_{1}e_{2}\left(
t_{c}\right) \right) ^{2}}{2k_{1}\bar{k}_{2}}<0
\end{equation}%
To satisfy (116), we can make $\sigma \left( t_{c}\right) =e_{2}\left(
t_{c}\right) +k_{1}e_{1}\left( t_{c}\right) <0$ i.e., $-k_{1}e_{1}\left(
t_{c}\right) >e_{2}\left( t_{c}\right) $. Therefore, we get the condition on 
$k_{1}$, as follows:

\begin{equation}
k_{1}>\frac{e_{2}\left( t_{c}\right) }{-e_{1}\left( t_{c}\right) }
\end{equation}%
Comparing the settling time $t_{c}+t_{s}$ in (97) and $\left. t\right\vert
_{e_{1}\left( t\right) =0}$ in (94), we get

\begin{equation}
t_{c}+t_{s}<\left. t\right\vert _{e_{1}\left( t\right) =0}
\end{equation}%
Therefore, we get

\begin{equation}
e_{1}\left( t\right) <0\text{ for }t\in \left[ t_{c},t_{c}+t_{s}\right)
\end{equation}%
For $e_{2}\left( t\right) $, when $t=t_{c}+t_{s}$, from (89) and (97), we get

\begin{equation}
\left. e_{2}\left( t\right) \right\vert _{t=t_{c}+t_{s}}=e_{2}\left(
t_{c}\right) +\bar{k}_{2}t_{s}=-\frac{\bar{k}_{2}}{k_{1}}+\sqrt{%
e_{2}^{2}\left( t_{c}\right) +\left( \frac{\bar{k}_{2}}{k_{1}}\right) ^{2}-2%
\bar{k}_{2}e_{1}\left( t_{c}\right) }
\end{equation}%
For $t\in \left[ t_{c},t_{c}+t_{s}\right) $, considering the finite-time
convergence condition, we need

\begin{equation}
\bar{k}_{2}>k_{1}\max \left\{ \left\vert e_{2}\left( t\right) \right\vert
\right\} =k_{1}\left\vert \left. e_{2}\left( t\right) \right\vert
_{t=t_{c}+t_{s}}\right\vert
\end{equation}%
Combining (120) and (121), we get

\begin{equation}
\bar{k}_{2}>k_{1}\left\vert \left. e_{2}\left( t\right) \right\vert
_{t=t_{c}+t_{s}}\right\vert =k_{1}\left\vert -\frac{\bar{k}_{2}}{k_{1}}+%
\sqrt{e_{2}^{2}\left( t_{c}\right) +\left( \frac{\bar{k}_{2}}{k_{1}}\right)
^{2}-2\bar{k}_{2}e_{1}\left( t_{c}\right) }\right\vert
\end{equation}%
Therefore, $\bar{k}_{2}$ should satisfy

\begin{equation}
\bar{k}_{2}>\frac{k_{1}^{2}}{3}\left[ \left\vert e_{1}\left( t_{c}\right)
\right\vert +\sqrt{e_{1}^{2}\left( t_{c}\right) +3\left( \frac{e_{2}\left(
t_{c}\right) }{k_{1}}\right) ^{2}}\right]
\end{equation}%
We know that zones IV-1 and II-1 have the same convergence performance
because of the odd function property in the sliding mode. Therefore, combing
(115), (117) and (123), for the zones IV-1 (where, $-e_{1}\left(
t_{c}\right) \geq e_{2}\left( t_{c}\right) >0$) and II-1 (where, $%
e_{1}\left( t_{c}\right) \geq -e_{2}\left( t_{c}\right) >0$), we get the
non-overshooting convergence conditions, when $e_{1}\left( t_{c}\right)
e_{2}\left( t_{c}\right) <0$ and $\left\vert e_{1}\left( t_{c}\right)
\right\vert \geq \left\vert e_{2}\left( t_{c}\right) \right\vert $, as
follows:

\begin{eqnarray}
k_{1} &\in &\left( \frac{\left\vert e_{2}\left( t_{c}\right) \right\vert }{%
\left\vert e_{1}\left( t_{c}\right) \right\vert },\infty \right) \\
\bar{k}_{2} &>&\max \left\{ k_{1}\left\vert e_{2}\left( t_{c}\right)
\right\vert ,\frac{k_{1}^{2}}{3}\left[ \left\vert e_{1}\left( t_{c}\right)
\right\vert +\sqrt{e_{1}^{2}\left( t_{c}\right) +3\left( \frac{e_{2}\left(
t_{c}\right) }{k_{1}}\right) ^{2}}\right] \right\}
\end{eqnarray}%
Furthermore, considering the disturbance $d\left( t\right) $, the conditions
of parameters selection, when $e_{1}\left( t_{c}\right) e_{2}\left(
t_{c}\right) <0$ and $\left\vert e_{1}\left( t_{c}\right) \right\vert \geq
\left\vert e_{2}\left( t_{c}\right) \right\vert $, are expressed as follows:

\begin{eqnarray}
k_{1} &\in &\left( \frac{\left\vert e_{2}\left( t_{c}\right) \right\vert }{%
\left\vert e_{1}\left( t_{c}\right) \right\vert },\infty \right) \\
k_{2} &>&\max \left\{ k_{1}\left\vert e_{2}\left( t_{c}\right) \right\vert
+L_{d},\frac{k_{1}^{2}}{3}\left[ \left\vert e_{1}\left( t_{c}\right)
\right\vert +\sqrt{e_{1}^{2}\left( t_{c}\right) +3\left( \frac{e_{2}\left(
t_{c}\right) }{k_{1}}\right) ^{2}}\right] +L_{d}\right\} ,
\end{eqnarray}%
Therefore, the sliding mode is non-overshooting stable for $t\in \left[
t_{c},t_{c}+t_{s}\right) $, and $\sigma \left( t\right) =0$ holds for $t\geq
t_{c}+t_{s}$. Then, the linear convergence law $\dot{e}_{1}\left( t\right)
=-k_{1}e_{1}\left( t\right) $ makes $\underset{t\rightarrow \infty }{\lim }%
e_{1}\left( t\right) =0$ without overshoot. In addition, from $e_{2}\left(
t\right) +k_{1}e_{1}\left( t\right) =0$ for $t\geq t_{c}+t_{s}$, we get $%
\underset{t\rightarrow \infty }{\lim }e_{2}\left( t\right) =0$.

In general, for $e_{1}\left( t_{c}\right) $ and $e_{2}\left( t_{c}\right) $
in zones IV-1 and II-1, when $k_{1}$ and $k_{2}$ are selected from (126) and
(127), the system is exponentially stable, and no overshoot exists for the
variable $e_{1}\left( t\right) $. This confirms the convergence curves of
sliding variables $e_{1}\left( t\right) $, $e_{2}\left( t\right) $ and
function $\sigma \left( t\right) $ described in Fig. 14 (b)-1 and (b)-2 for
the zones IV-1 and II-1, respectively.

\bigskip

\emph{3) For zone III, i.e., III-1: }$-e_{2}\left( t_{c}\right) \geq
-e_{1}\left( t_{c}\right) \geq 0$\emph{\ and III-2: }$-e_{1}\left(
t_{c}\right) >-e_{2}\left( t_{c}\right) \geq 0$

From Figure 13, we know that the corresponding symmetrical zone of III is
zone I; and the corresponding symmetrical zone of III-2 is zone I-2.

\begin{figure}[tbp]
\centering
\includegraphics[width=5.00in]{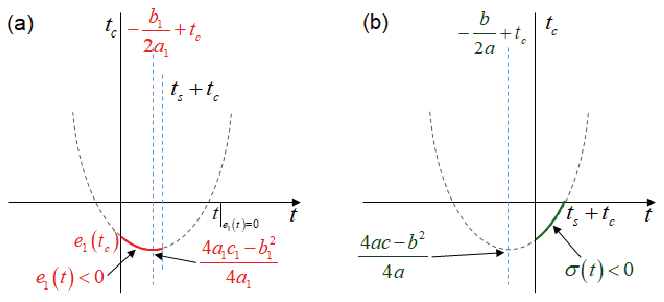}
\caption{Arranged trajectories of $e_{1}\left( t\right) $ and $\protect%
\sigma \left( t\right) $ for $e_{1}\left( t\right) $ non-overshooting
convergence for $t\in \left[ t_{c},t_{c}+t_{s}\right) $ in range III-1: $%
-e_{2}\left( t_{c}\right) \geq -e_{1}\left( t_{c}\right) \geq 0$\ and range
III-2: $-e_{1}\left( t_{c}\right) >-e_{2}\left( t_{c}\right) \geq 0$.}
\end{figure}

For zone III, we can get the conditions of non-overshooting convergence for $%
t\in \left[ t_{c},t_{c}+t_{s}\right) $:

1) $\bar{k}_{2}>k_{1}\left\vert e_{2}\left( t\right) \right\vert $
[finite-time stability to get convergence law $\dot{e}_{1}\left( t\right)
=-k_{1}e_{1}\left( t\right) $];

2) the sliding function $\sigma \left( t\right) =e_{2}\left( t\right)
+k_{1}e_{1}\left( t\right) <0$ for $t\in \left[ t_{c},t_{c}+t_{s}\right) $,
and $\sigma \left( t_{c}+t_{s}\right) =e_{2}\left( t_{c}+t_{s}\right)
+k_{1}e_{1}\left( t_{c}+t_{s}\right) =0$;

3) $e_{1}\left( t\right) <0$ [ $e_{1}\left( t\right) $ does not go beyond
zero].

In the following, we will determine $k_{1}$ and $k_{2}$ to satisfy these
conditions.

From (90), for $e_{1}\left( t\right) $ (a segment of a\ parabola) for $t\in %
\left[ t_{c},t_{c}+t_{s}\right) $, we get its axis of symmetry

\begin{equation}
-\frac{b_{1}}{2a_{1}}=-\frac{e_{2}\left( t_{c}\right) }{\bar{k}_{2}}\geq 0
\end{equation}%
because $e_{2}\left( t_{c}\right) \leq 0$. Also, its vertex satisfies

\begin{equation}
\frac{4a_{1}c_{1}-b_{1}^{2}}{4a_{1}}=\frac{2\bar{k}_{2}e_{1}\left(
t_{c}\right) -e_{2}^{2}\left( t_{c}\right) }{2\bar{k}_{2}}<0
\end{equation}%
because $e_{1}\left( t_{c}\right) \leq 0$ (See $e_{1}\left( t\right) $ in
Figure 17(a)).

From (91), for $\sigma \left( t\right) =e_{2}\left( t\right)
+k_{1}e_{1}\left( t\right) $ (the segment of a\ parabola) for $t\in \left[
t_{c},t_{c}+t_{s}\right) $, we get its axis of symmetry:

\begin{equation}
-\frac{b}{2a}=-\frac{\bar{k}_{2}+k_{1}e_{2}\left( t_{c}\right) }{k_{1}\bar{k}%
_{2}}<0
\end{equation}%
because of the finite-time convergence condition

\begin{equation}
\bar{k}_{2}>k_{1}\left\vert e_{2}\left( t_{c}\right) \right\vert
\end{equation}%
In order to make $\sigma \left( t\right) =e_{2}\left( t\right)
+k_{1}e_{1}\left( t\right) <0$ for $t\in \left[ t_{c},t_{c}+t_{s}\right) $
(See Figure 17(b)), its vertex needs to satisfy

\begin{equation}
\frac{4ac-b^{2}}{4a}=\frac{2k_{1}\bar{k}_{2}\left( e_{2}\left( t_{c}\right)
+k_{1}e_{1}\left( t_{c}\right) \right) -\left( \bar{k}_{2}+k_{1}e_{2}\left(
t_{c}\right) \right) ^{2}}{2k_{1}\bar{k}_{2}}<0
\end{equation}%
To satisfy (132), it should be that $\sigma \left( t_{c}\right) =e_{2}\left(
t_{c}\right) +k_{1}e_{1}\left( t_{c}\right) <0$. Therefore, for zones III,
we get the $k_{1}$ condition as

\begin{equation}
k_{1}>0
\end{equation}%
Comparing the settling time $t_{c}+t_{s}$ in (97) and $\left. t\right\vert
_{e_{1}\left( t\right) =0}$ in (94), we get

\begin{equation}
t_{c}+t_{s}<\left. t\right\vert _{e_{1}\left( t\right) =0}
\end{equation}%
Therefore, we get

\begin{equation}
e_{1}\left( t\right) <0\text{ for }t\in \left[ t_{c},t_{c}+t_{s}\right)
\end{equation}%
For $e_{2}\left( t\right) $, when $t=t_{c}+t_{s}$, from (89) and (97), we get

\begin{equation}
\left. e_{2}\left( t\right) \right\vert _{t=t_{c}+t_{s}}=e_{2}\left(
t_{c}\right) -\bar{k}_{2}t_{s}=-\frac{\bar{k}_{2}}{k_{1}}+\sqrt{%
e_{2}^{2}\left( t_{c}\right) +\left( \frac{\bar{k}_{2}}{k_{1}}\right) ^{2}-2%
\bar{k}_{2}e_{1}\left( t_{c}\right) }
\end{equation}%
For $t\in \left[ t_{c},t_{c}+t_{s}\right) $, considering the finite-time
convergence condition, we need

\begin{equation}
\bar{k}_{2}>k_{1}\max \left\{ \left\vert e_{2}\left( t\right) \right\vert
\right\} =k_{1}\left\vert \left. e_{2}\left( t\right) \right\vert
_{t=t_{c}+t_{s}}\right\vert
\end{equation}%
Combining (136) and (137), we get

\begin{equation}
\bar{k}_{2}>k_{1}\left\vert \left. e_{2}\left( t\right) \right\vert
_{t=t_{c}+t_{s}}\right\vert =k_{1}\left\vert -\frac{\bar{k}_{2}}{k_{1}}+%
\sqrt{e_{2}^{2}\left( t_{c}\right) +\left( \frac{\bar{k}_{2}}{k_{1}}\right)
^{2}-2\bar{k}_{2}e_{1}\left( t_{c}\right) }\right\vert
\end{equation}%
Therefore, $\bar{k}_{2}$ should satisfy

\begin{equation}
\bar{k}_{2}>\frac{k_{1}^{2}}{3}\left[ \left\vert e_{1}\left( t_{c}\right)
\right\vert +\sqrt{e_{1}^{2}\left( t_{c}\right) +3\left( \frac{e_{2}\left(
t_{c}\right) }{k_{1}}\right) ^{2}}\right]
\end{equation}

We know that zones III and I have the same convergence performance because
of the odd function property in the sliding mode. Therefore, combing (131),
(133) and (139), for the zones III and I, we get the non-overshooting
convergence conditions:

\begin{eqnarray}
k_{1} &\in &\left( 0,\infty \right) \\
\bar{k}_{2} &>&\max \left\{ k_{1}\left\vert e_{2}\left( t_{c}\right)
\right\vert ,\frac{k_{1}^{2}}{3}\left[ \left\vert e_{1}\left( t_{c}\right)
\right\vert +\sqrt{e_{1}^{2}\left( t_{c}\right) +3\left( \frac{e_{2}\left(
t_{c}\right) }{k_{1}}\right) ^{2}}\right] \right\}
\end{eqnarray}%
Furthermore, considering the disturbance $d\left( t\right) $, the conditions
of parameters selection are expressed as follows:

\begin{eqnarray}
k_{1} &\in &\left( 0,\infty \right) \\
k_{2} &>&\max \left\{ k_{1}\left\vert e_{2}\left( t_{c}\right) \right\vert
+L_{d},\frac{k_{1}^{2}}{3}\left[ \left\vert e_{1}\left( t_{c}\right)
\right\vert +\sqrt{e_{1}^{2}\left( t_{c}\right) +3\left( \frac{e_{2}\left(
t_{c}\right) }{k_{1}}\right) ^{2}}\right] +L_{d}\right\}
\end{eqnarray}

Therefore, the sliding mode is non-overshooting stable for $t\in \left[
t_{c},t_{c}+t_{s}\right) $, and $\sigma \left( t\right) =0$ holds for $t\geq
t_{c}+t_{s}$. Then, the linear convergence law $\dot{e}_{1}\left( t\right)
=-k_{1}e_{1}\left( t\right) $ makes $\underset{t\rightarrow \infty }{\lim }%
e_{1}\left( t\right) =0$ without overshoot. In addition, from $e_{2}\left(
t\right) +k_{1}e_{1}\left( t\right) =0$ for $t\geq t_{c}+t_{s}$, we get $%
\underset{t\rightarrow \infty }{\lim }e_{2}\left( t\right) =0$.

In general, for $e_{1}\left( t_{c}\right) $ and $e_{2}\left( t_{c}\right) $
in zones III and I, when $k_{1}$ and $k_{2}$ are selected from (142) and
(143), the system is exponentially stable, and no overshoot exists for the
variable $e_{1}\left( t\right) $. This confirms the convergence curves of
sliding variables $e_{1}\left( t\right) $, $e_{2}\left( t\right) $ and
function $\sigma \left( t\right) $ described in Fig. 13 (c)-1 and (c)-2 for
the zones III-1 and I-1, and Fig. 13 (d)-1 and (d)-2 for zones III-2 and
I-2, respectively.

Finally, combing parameter selection conditions (110)-(111), (126)-(127),
and (142)-(143) in the different zones, we can get the parameter conditions
(19) and (20) for non-overshooting stable system in Theorem 3.1.

\bigskip

\textbf{C. Determination of }$e_{1c}$\textbf{\ and }$e_{2c}$\textbf{\ for
bounded system gain and without overshoot}

From the expression of $k_{2}$ in (20), in order to make the system gain
bounded, we need to reduce $\left\vert e_{1}\left( t_{c}\right) \right\vert $%
, $\left\vert e_{2}\left( t_{c}\right) \right\vert $ and $\frac{%
e_{2}^{2}\left( t_{c}\right) }{2\left\vert e_{1}\left( t_{c}\right)
\right\vert }$. For the first subsystem, according to the robust and
non-overshooting reachability, we get $e_{1c}=e_{1}(t_{c})$ and $%
e_{2c}=e_{2}(t_{c})$ when $t=t_{c}$. Therefore, we can select the bounded $%
e_{1c}$ and $e_{2c}$ for bounded $\left\vert e_{1}\left( t_{c}\right)
\right\vert $ and $\left\vert e_{2}\left( t_{c}\right) \right\vert $.
Furthermore, we hope to get the fast convergence as the mode in the zones
II-1 and IV-1, and the up-bound of $k_{2}$ from the system gain limitation
is considered. In order to get the non-overshooting stability with the
bounded system gain, we select

\begin{eqnarray}
e_{1c} &\in &\left( 0,k_{2M}-L_{d}\right)  \notag \\
e_{2c} &\in &\left( e_{1c},\sqrt{\left( k_{2M}-L_{d}\right) e_{1c}}\right]
\end{eqnarray}%
where, $k_{2M}$ is the up-bound of $k_{2}$. In fact, due to $%
e_{1c}=e_{1}(t_{c})$ and $e_{2c}=e_{2}(t_{c})$ when $t=t_{c}$, we get the
parameter $k_{1}\in \left( 0,\frac{e_{2c}}{e_{1c}}\right) $ for the second
subsystem. Therefore, there exists $\beta _{11}\in (0,1)$ such that $%
k_{1}=\beta _{11}\frac{e_{2c}}{e_{1c}}$. In (20), we know that

\begin{equation}
k_{1}e_{2c}\leq k_{2M}-L_{d}
\end{equation}%
From $k_{1}=\beta _{11}\frac{e_{2c}}{e_{1c}}$ and (145), we can get

\begin{equation}
\beta _{11}\frac{e_{2c}^{2}}{e_{1c}}\leq k_{2M}-L_{d}
\end{equation}%
In (20), we also know that

\begin{equation}
\frac{e_{2c}^{2}}{2e_{1c}}\leq k_{2M}-L_{d}
\end{equation}%
Combing (146) and (147), we can select

\begin{equation}
\max \left\{ \beta _{11},\frac{1}{2}\right\} \frac{e_{2c}^{2}}{e_{1c}}\leq
k_{2M}-L_{d}
\end{equation}%
Because the system gain limitation $0<k_{2}\leq k_{2M}$, i.e., $k_{2M}$ is
the maximum implementation of $k_{2}$. Because $\beta _{11}\in (0,1)$, we
get $\max \left\{ \beta _{11},\frac{1}{2}\right\} <1$. Therefore, from $%
e_{1c}<e_{2c}$, for (148), we select

\begin{equation}
e_{1c}<e_{2c}\leq \sqrt{\left( k_{2M}-L_{d}\right) e_{1c}}
\end{equation}%
i.e., $e_{1c}^{2}<\left( k_{2M}-L_{d}\right) e_{1c}$ and $e_{1c}<e_{2c}\leq 
\sqrt{\left( k_{2M}-L_{d}\right) e_{1c}}$. Then, we get (144).

This concludes the proof. $\blacksquare $

\bigskip

\emph{Proof of Theorem 3.2:}

(i) Firstly, we consider $\left\vert e_{1}\left( t\right) \right\vert
>e_{1c} $.

Case one: If $e_{1}\left( t\right) >e_{1c}$, for (24), we get

\begin{equation}
\frac{d\left( e_{2}\left( t\right) +e_{2c}\right) }{dt}=-k_{c}\text{tanh}%
\left[ \rho _{c}\left( e_{2}\left( t\right) +e_{2c}\right) \right] +d(t)
\end{equation}%
A Lyapunov function candidate is selected as

\begin{equation}
V_{c}=\frac{1}{2}\left( e_{2}\left( t\right) +e_{2c}\right) ^{2}
\end{equation}%
Then, taking the derivative for $V_{c}$, we get

\begin{eqnarray}
\dot{V}_{c} &=&\left( e_{2}\left( t\right) +e_{2c}\right) \left\{ -k_{c}%
\text{tanh}\left[ \rho _{c}\left( e_{2}\left( t\right) +e_{2c}\right) \right]
+d(t)\right\}  \notag \\
&\leq &-\left( k_{c}\left\vert \text{tanh}\left[ \rho _{c}\left( e_{2}\left(
t\right) +e_{2c}\right) \right] \right\vert -L_{d}\right) \left\vert
e_{2}\left( t\right) +e_{2c}\right\vert  \notag \\
&=&-\sqrt{2}\left( k_{c}\left\vert \text{tanh}\left[ \rho _{c}\left(
e_{2}\left( t\right) +e_{2c}\right) \right] \right\vert -L_{d}\right) V_{c}^{%
\frac{1}{2}}
\end{eqnarray}%
From (152), if $\left\vert \text{tanh}\left[ \rho _{c}\left( e_{2}\left(
t\right) +e_{2c}\right) \right] \right\vert >\frac{L_{d}}{k_{c}}$, then $%
\dot{V}_{c}\left( t\right) <0$, and $\left\vert e_{2}\left( t\right)
+e_{2c}\right\vert $ decreases, and we get

\begin{equation}
\left\vert \text{tanh}\left[ \rho _{c}\left( e_{2}\left( t\right)
+e_{2c}\right) \right] \right\vert \leq \frac{L_{d}}{k_{c}}
\end{equation}%
i.e.,

\begin{equation}
\left\vert 1-\frac{2}{\text{e}^{2\rho _{c}\left( e_{2}\left( t\right)
+e_{2c}\right) }+1}\right\vert \leq \frac{L_{d}}{k_{c}}
\end{equation}%
Therefore, there exists a finite time $t_{c1}>0$, for $t\geq t_{c1}$, such
that

\begin{equation}
\left\vert e_{2}\left( t\right) +e_{2c}\right\vert \leq \frac{1}{2\rho _{c}}%
\ln \frac{k_{c}+L_{d}}{k_{c}-L_{d}}
\end{equation}%
Because $\rho _{c}\gg \frac{1}{2}\ln \frac{k_{c}+L_{d}}{k_{c}-L_{d}}$, we
get $\left\vert e_{2}\left( t\right) +e_{2c}\right\vert \ll 1$. There exists 
$\left\vert O\left( 1/\rho \right) \right\vert \leq \frac{1}{2\rho _{c}}\ln 
\frac{k_{c}+L_{d}}{k_{c}-L_{d}}$, such that $e_{2}\left( t\right)
=-e_{2c}+O\left( 1/\rho \right) $. From $\dot{e}_{1}=e_{2}$, we get $\dot{e}%
_{1}=-e_{2c}+O\left( 1/\rho _{c}\right) $ for $t\geq t_{c1}$. Therefore,
there exists a finite time $t_{c2}>t_{c1}>0$, for $t\geq t_{c2}$, such that $%
e_{1}\left( t\right) \leq e_{1c}$.

We know that $k_{c}>L_{d}$. Therefore, there exists a finite time $t_{c1}>0$%
, for $t\geq t_{c1}$, such that $e_{2}\left( t\right) =-e_{2c}$. According
to $\dot{e}_{1}=e_{2}$, we get that $\dot{e}_{1}\left( t\right) =-e_{2c}$
for $t\geq t_{c1}$. Therefore, there exists a finite time $t_{c}>0$, for $%
t\geq t_{c}$, such that $e_{1}\left( t\right) \leq e_{1c}$.

Case two: If $e_{1}\left( t\right) <-e_{1c}$, for (24), we get

\begin{equation}
\frac{d\left( e_{2}\left( t\right) -e_{2c}\right) }{dt}=-k_{c}\text{tanh}%
\left[ \rho _{c}\left( e_{2}\left( t\right) -e_{2c}\right) \right] +d(t)
\end{equation}%
Similar method to case $e_{1}\left( t\right) >e_{1c}$, when we select the
Lyapunov function candidate as $V_{c}=\frac{1}{2}\left( e_{2}\left( t\right)
-e_{2c}\right) ^{2}$, there exists a finite time $t_{c}>0$, for $t\geq t_{c}$%
, such that $e_{1}\left( t\right) \geq -e_{1c}$.

Combining cases one and two, we can get that $\left\vert e_{1}\left(
t\right) \right\vert \leq e_{1c}$ for $t\geq t_{c}$.

\bigskip

In the following, we discuss the sliding mode system when $\left\vert
e_{1}\left( t\right) \right\vert \leq e_{1c}$.

For the smoothed sliding mode (24), a Lyapunov function candidate is
selected as

\begin{equation}
V_{2}\left( t\right) =\frac{1}{2}\left[ e_{2}\left( t\right)
+k_{1}e_{1}\left( t\right) \right] ^{2}
\end{equation}%
Then, taking the derivative for $V_{2}\left( t\right) $, we get

\begin{eqnarray}
\dot{V}_{2} &=&\left[ e_{2}\left( t\right) +k_{1}e_{1}\left( t\right) \right]
\left\{ -k_{2}\text{tanh}\left[ \rho \left( e_{2}\left( t\right)
+k_{1}e_{1}\left( t\right) \right) \right] -d(t)+k_{1}e_{2}\left( t\right)
\right\}  \notag \\
&\leq &-k_{2}\left[ e_{2}\left( t\right) +k_{1}e_{1}\left( t\right) \right]
\left\vert \text{tanh}\left[ \rho (e_{2}\left( t\right) +k_{1}e_{1}\left(
t\right) )\right] \right\vert \text{sign}\left[ e_{2}\left( t\right)
+k_{1}e_{1}\left( t\right) \right]  \notag \\
&&+(k_{1}\left\vert e_{2}\left( t\right) \right\vert +L_{d})\left\vert
e_{2}\left( t\right) +k_{1}e_{1}\left( t\right) \right\vert  \notag \\
&=&-\left( k_{2}\left\vert \text{tanh}\left[ \rho (e_{2}\left( t\right)
+k_{1}e_{1}\left( t\right) )\right] \right\vert -k_{1}\left\vert e_{2}\left(
t\right) \right\vert -L_{d}\right) \left\vert e_{2}\left( t\right)
+k_{1}e_{1}\left( t\right) \right\vert  \notag \\
&=&-\left( k_{2}\left\vert \text{tanh}\left[ \rho (e_{2}\left( t\right)
+k_{1}e_{1}\left( t\right) )\right] \right\vert -k_{1}\left\vert e_{2}\left(
t\right) \right\vert -L_{d}\right) V^{\frac{1}{2}}\left( t\right)
\end{eqnarray}

From (158), if $\left\vert \text{tanh}\left[ \rho (e_{2}\left( t\right)
+k_{1}e_{1}\left( t\right) )\right] \right\vert >\frac{k_{1}\left\vert
e_{2}\left( t\right) \right\vert +L_{d}}{k_{2}}$, then $\dot{V}_{2}\left(
t\right) <0$, and $\left\vert e_{2}\left( t\right) +k_{1}e_{1}\left(
t\right) \right\vert $ decreases. We know that, the function $\left\vert 
\text{tanh}\left[ \rho (e_{2}\left( t\right) +k_{1}e_{1}\left( t\right) )%
\right] \right\vert $ is the the monotonically increasing function about $%
\left\vert e_{2}\left( t\right) +k_{1}e_{1}\left( t\right) \right\vert $.
Therefore, $\left\vert e_{2}\left( t\right) +k_{1}e_{1}\left( t\right)
\right\vert $ decreases until

\begin{equation}
\left\vert \text{tanh}\left[ \rho (e_{2}\left( t\right) +k_{1}e_{1}\left(
t\right) )\right] \right\vert \leq \frac{k_{1}\left\vert e_{2}\left(
t\right) \right\vert +L_{d}}{k_{2}}
\end{equation}%
We define

\begin{equation}
\max \left\{ \left\vert e_{2}\left( t\right) \right\vert \right\} =\max
\left\{ \left\vert e_{2}\left( t_{c}\right) \right\vert ,\frac{k_{1}}{3}%
\left[ \left\vert e_{1}\left( t_{c}\right) \right\vert +\sqrt{%
e_{1}^{2}\left( t_{c}\right) +3\left( \frac{e_{2}\left( t_{c}\right) }{k_{1}}%
\right) ^{2}}\right] \right\} \overset{\text{define}}{=}e_{2\max }
\end{equation}%
For (159), from (160), we get

\begin{equation}
\left\vert 1-\frac{2}{\text{e}^{2\rho \left[ e_{2}\left( t\right)
+k_{1}e_{1}\left( t\right) \right] }+1}\right\vert \leq \frac{k_{1}e_{2\max
}+L_{d}}{k_{2}}
\end{equation}%
Function $e_{2}+k_{1}e_{1}<0$ holds for $t\in \left[ t_{c},t_{c}+t_{s}%
\right) $ in the zones (II-2, IV-1, III-1 and III-2). Therefore, (161) can
be expressed by

\begin{equation}
\frac{2}{\text{e}^{2\rho \left[ e_{2}\left( t\right) +k_{1}e_{1}\left(
t\right) \right] }+1}-1\leq \frac{k_{1}e_{2\max }+L_{d}}{k_{2}}
\end{equation}%
Then, it follows that

\begin{equation}
\left\vert e_{2}\left( t\right) +k_{1}e_{1}\left( t\right) \right\vert \leq 
\frac{1}{2\rho }\ln \frac{k_{2}+k_{1}e_{2\max }+L_{d}}{k_{2}-k_{1}e_{2\max
}-L_{d}}
\end{equation}%
From the relation $\dot{e}_{1}\left( t\right) =e_{2}\left( t\right) $ in the
sliding mode (24), for (163), there exists a function $\beta \left( t\right) 
$, where $\left\vert \beta \left( t\right) \right\vert \leq \frac{1}{2\rho }%
\ln \frac{k_{2}+k_{1}e_{2\max }+L_{d}}{k_{2}-k_{1}e_{2\max }-L_{d}}$, such
that the following convergence law holds:

\begin{equation}
\dot{e}_{1}\left( t\right) =-k_{1}e_{1}\left( t\right) +\beta \left( t\right)
\end{equation}%
The solution to the convergence law (164) is

\begin{equation}
e_{1}\left( t\right) =\left( \int_{t_{c}}^{t}\beta \left( \tau \right) \text{%
e}^{k_{1}\tau }d\tau \right) \text{e}^{-k_{1}t}
\end{equation}%
Therefore, we get

\begin{equation}
\left\vert e_{1}\left( t\right) \right\vert \leq \left\vert \beta \left(
\tau \right) \right\vert \left( \int_{t_{c}}^{t}\text{e}^{k_{1}\tau }d\tau
\right) \text{e}^{-k_{1}t}\leq \frac{1}{2\rho k_{1}}\ln \frac{%
k_{2}+k_{1}e_{2\max }+L_{d}}{k_{2}-k_{1}e_{2\max }-L_{d}}\left( 1-\text{e}%
^{-k_{1}\left( t-t_{c}\right) }\right)
\end{equation}%
Then, it follows that

\begin{equation}
\underset{t\rightarrow \infty }{\lim }\left\vert e_{1}\left( t\right)
\right\vert \leq \frac{1}{2\rho k_{1}}\ln \frac{k_{2}+k_{1}e_{2\max }+L_{d}}{%
k_{2}-k_{1}e_{2\max }-L_{d}}
\end{equation}%
Because $\rho \gg \frac{1}{2k_{1}}\ln \frac{k_{2}+k_{1}e_{2\max }+L_{d}}{%
k_{2}-k_{1}e_{2\max }-L_{d}}$ is selected, the up-bound of $\underset{%
t\rightarrow \infty }{\lim }\left\vert e_{1}\left( t\right) \right\vert $ is
sufficiently small, and $\frac{1}{2\rho k_{1}}\ln \frac{k_{2}+k_{1}e_{2\max
}+L_{d}}{k_{2}-k_{1}e_{2\max }-L_{d}}\ll 1$ holds. For $e_{2}\left( t\right) 
$, from (163) and (166), we have

\begin{eqnarray}
\left\vert e_{2}\left( t\right) \right\vert &=&\left\vert e_{2}\left(
t\right) +k_{1}e_{1}\left( t\right) -k_{1}e_{1}\left( t\right) \right\vert
\leq \left\vert e_{2}\left( t\right) +k_{1}e_{1}\left( t\right) \right\vert
+k_{1}\left\vert e_{1}\left( t\right) \right\vert  \notag \\
&\leq &\frac{1}{2\rho }\ln \frac{k_{2}+k_{1}e_{2\max }+L_{d}}{%
k_{2}-k_{1}e_{2\max }-L_{d}}+\frac{1}{2\rho }\ln \frac{k_{2}+k_{1}e_{2\max
}+L_{d}}{k_{2}-k_{1}e_{2\max }-L_{d}}\left( 1-\text{e}^{-k_{1}\left(
t-t_{c}\right) }\right)
\end{eqnarray}%
Therefore, we get

\begin{equation}
\underset{t\rightarrow \infty }{\lim }\left\vert e_{2}\left( t\right)
\right\vert \leq \frac{1}{\rho }\ln \frac{k_{2}+k_{1}e_{2\max }+L_{d}}{%
k_{2}-k_{1}e_{2\max }-L_{d}}
\end{equation}%
Because $\rho \gg \ln \frac{k_{2}+k_{1}e_{2\max }+L_{d}}{k_{2}-k_{1}e_{2\max
}-L_{d}}$ is selected, the up-bound of $\underset{t\rightarrow \infty }{\lim 
}\left\vert e_{2}\left( t\right) \right\vert $ is sufficiently small, and $%
\frac{1}{\rho }\ln \frac{k_{2}+k_{1}e_{2\max }+L_{d}}{k_{2}-k_{1}e_{2\max
}-L_{d}}\ll 1$ holds.

(ii) Specially, when $\rho $ is selected large enough, i.e., $\rho
\rightarrow +\infty $, the sliding variable up-bounds in (167) and (169)
approach to zero, and

\begin{equation*}
\underset{t\rightarrow \infty }{\lim }\underset{\rho \rightarrow +\infty }{%
\lim }e_{1}\left( t\right) =0\text{ and }\underset{t\rightarrow \infty }{%
\lim }\underset{\rho \rightarrow +\infty }{\lim }e_{2}\left( t\right) =0
\end{equation*}%
From $\underset{\rho \rightarrow +\infty }{\lim }$tanh$\left( \rho \cdot
x\right) =$sign$(x)$, the tanh-function-based sliding mode (24) becomes the
ideal 2-sliding mode (18). This concludes the proof. $\blacksquare $

\bigskip

\emph{Proof of Theorem 5.1:}

Define $e_{1}\left( t\right) =x_{d}\left( t\right) -x_{1}\left( t\right) $
and $e_{2}\left( t\right) =\dot{x}_{d}\left( t\right) -x_{2}\left( t\right) $%
. Then, the error system is

\begin{eqnarray}
\dot{e}_{1}\left( t\right) &=&e_{2}\left( t\right)  \notag \\
\dot{e}_{2}\left( t\right) &=&-h(t)-u(t)+\ddot{x}_{d}\left( t\right) +\delta
\left( t\right)
\end{eqnarray}%
The desired stable sliding mode (18) in Theorem 3.1 is selected, where, $%
d(t)=\ddot{x}_{d}\left( t\right) +\delta \left( t\right) $. In order to turn
the error system (170) into the sliding mode (18), we select

\begin{eqnarray}
\dot{e}_{2}\left( t\right) &=&-h(t)-u(t)+\ddot{x}_{d}\left( t\right)
+d\left( t\right)  \notag \\
&=&\left\{ 
\begin{array}{l}
-k_{c}\text{sign}\left[ e_{2}\left( t\right) +e_{2c}\text{sign}\left(
e_{1}\left( t\right) \right) \right] +\ddot{x}_{d}\left( t\right) +\delta
\left( t\right) ,\text{ if }\left\vert e_{1}\left( t\right) \right\vert
>e_{1c}; \\ 
-k_{2}\text{sign}\left[ e_{2}\left( t\right) +k_{1}e_{1}\left( t\right) %
\right] +\ddot{x}_{d}\left( t\right) +\delta \left( t\right) ,\text{ if }%
\left\vert e_{1}\left( t\right) \right\vert \leq e_{1c}%
\end{array}%
\right.
\end{eqnarray}%
Therefore, we get the controller as follows:

\begin{equation}
u(t)=\left\{ 
\begin{array}{l}
k_{c}\text{sign}\left[ e_{2}\left( t\right) +e_{2c}\text{sign}\left(
e_{1}\left( t\right) \right) \right] -h(t),\text{ if }\left\vert e_{1}\left(
t\right) \right\vert >e_{1c} \\ 
k_{2}\text{sign}\left[ e_{2}\left( t\right) +k_{1}e_{1}\left( t\right) %
\right] -h(t),\text{ if }\left\vert e_{1}\left( t\right) \right\vert \leq
e_{1c}%
\end{array}%
\right.
\end{equation}%
Thus, for the uncertain system (39), when the controller (40) is selected,
the system is stable, and $x_{1}$ tracking $x_{d}\left( t\right) $ is
non-overshooting. This concludes the proof. $\blacksquare $

\bigskip

\emph{Proof of Theorem 5.2:}

Define $e_{1}\left( t\right) =x_{d}\left( t\right) -x_{1}\left( t\right) $
and $e_{2}\left( t\right) =\dot{x}_{d}\left( t\right) -x_{2}\left( t\right) $%
. Then, the error system is

\begin{eqnarray}
\dot{e}_{1}\left( t\right) &=&e_{2}\left( t\right)  \notag \\
\dot{e}_{2}\left( t\right) &=&-h(t)-u(t)+\ddot{x}_{d}\left( t\right) +\delta
\left( t\right)
\end{eqnarray}

The desired stable sliding mode (24) in Theorem 3.2 is selected, where, $%
d(t)=\ddot{x}_{d}\left( t\right) +\delta \left( t\right) $. In order to turn
the error system (173) into the sliding mode (24), we select

\begin{eqnarray}
\dot{e}_{2} &=&-h(t)-u(t)+\ddot{x}_{d}\left( t\right) +d\left( t\right) 
\notag \\
&=&\left\{ 
\begin{array}{l}
-k_{c}\text{tanh}\left[ \rho _{c}\left( e_{2}\left( t\right) +e_{2c}\text{%
sign}\left( e_{1}\left( t\right) \right) \right) \right] +\ddot{x}_{d}\left(
t\right) +\delta \left( t\right) ,\text{ if }\left\vert e_{1}\left( t\right)
\right\vert >e_{1c}; \\ 
-k_{2}\text{tanh}\left[ \rho (e_{2}\left( t\right) +k_{1}e_{1}\left(
t\right) )\right] +\ddot{x}_{d}\left( t\right) +\delta \left( t\right) ,%
\text{ if }\left\vert e_{1}\left( t\right) \right\vert \leq e_{1c}%
\end{array}%
\right.
\end{eqnarray}%
Therefore, we get the controller as follows:

\begin{equation}
u(t)=\left\{ 
\begin{array}{l}
k_{c}\text{tanh}\left[ \rho _{c}\left( e_{2}\left( t\right) +e_{2c}\text{sign%
}\left( e_{1}\left( t\right) \right) \right) \right] -h(t),\text{ if }%
\left\vert e_{1}\left( t\right) \right\vert >e_{1c}; \\ 
k_{2}\text{tanh}\left[ \rho (e_{2}\left( t\right) +k_{1}e_{1}\left( t\right)
)\right] -h(t),\text{ if }\left\vert e_{1}\left( t\right) \right\vert \leq
e_{1c}%
\end{array}%
\right.
\end{equation}%
Thus, for the uncertain system (39), when the controller (44) is selected,
the system is stable, and $x_{1}$ tracking $x_{d}\left( t\right) $ is
non-overshooting. This concludes the proof. $\blacksquare $

\bigskip

\end{document}